\documentclass[prb,showpacs,preprintnumbers,amssymb,printfigures,twocolumn,superscriptaddress,aps]{revtex4}
\usepackage{natbib}
\usepackage{graphicx}
\usepackage{tikz} 
\graphicspath{{figures/}}
\usepackage{bm}
\usepackage[T1]{fontenc}
\usepackage{lmodern}
 \usepackage{textcomp}
\usepackage{amstext}
\usepackage[Symbol]{upgreek}
\usepackage{subfigure}
\usepackage{booktabs}
\usepackage{dcolumn}
\usepackage{color}
\usepackage{enumerate}
\usepackage{latexsym,bm}
\usepackage{longtable}
\usepackage[colorlinks=true, allcolors=blue]{hyperref}
\usepackage{amsmath,empheq}
\makeatletter
\def\l@subsection#1#2{}
\def\l@subsubsection#1#2{}
\makeatother

\begin{document}

\preprint{manuscript-multi-Weyl Semimetals-Liu}
\title{Fermi arcs of topological surface states in multi-Weyl Semimetals}

\author{Y. C. Liu}
\thanks{liuyachao@xaut.edu.cn (Y. C. Liu).}
\affiliation{Department of Applied Physics, Xi'an University of Technology, Xi'an 710054, China}
\affiliation{National Institute for Materials Science, Tsukuba, 305-0044, Japan}

\author{V. Wang}
\affiliation{Department of Applied Physics, Xi'an University of Technology, Xi'an 710054, China}
\email{orcid.org/0000-0002-9499-1823}

\author{J. B. Lin}
\affiliation{National Institute for Materials Science, Tsukuba, 305-0044, Japan}
\email{https://orcid.org/0000-0003-0769-9857}

\author{J. Nara}
\affiliation{National Institute for Materials Science, Tsukuba, 305-0044, Japan}
\thanks{nara.jun@nims.go.jp(J.Nara).}

 \date{\today}

\begin{abstract}
The Fermi arcs of topological surface state in the three-dimensional multi-Weyl semimetals are investigated systematically by a continuum model. We calculated the energy spectra and wave function for bulk quadratic- and cubic-Weyl semimetal with a single Weyl point. The Fermi arcs pattern of topological surface state in multi-Weyl semimetals are derived analytically. We demonstrate that the number of the Fermi arc emitting from any Weyl point is always equal to its chirality. In addition, the topological Lifshitz phase transition of Fermi arcs with respect to boundary condition parameter is revealed clearly, and the critical point for Lifshitz transition in multi-Weyl semimetals is determined. Our theoretical results provide explicit relations between boundary condition parameter and Fermi arcs in multi-Weyl semimetals, which may lead to an effective modulation of topological surface states by boundary modification in future experimental research.
\end{abstract}
\maketitle

\section{Introduction}\label{sec:intro}

The conception of symmetry breaking originating at condensed matter physics renovated the quantum field theory in particle physics profoundly. Similarly, the topological phases emerging in condensed matter physics demonstrate vividly the important concepts in particle physics, such as topological charges and quantum anomalies, and even flourish and enrich the concept quite non-trivially\cite{Kane2010RMP,Zhang2011RMP}. We thus have good reason to expect that the interplay between them should further motivate novel developments which may go beyond themselves. 

Topological phases are classified theoretically by dimensions and discrete symmetries\cite{Schnyder2008prb,Kitaev2009mg}. The discrete symmetries are important for topological classification because of the robustness against small deformation in resultant phenomena, which even works well at the continuum limit where details of lattice structures have been erased. The hallmark of the topological phase is the gapless edge modes, which arise from the so-called bulk-edge correspondence \cite{Jackiw1976prd,hatsugai1993}.

Both the form of boundary condition and the value of parameter have significant effects on the surface states of topological phase. This was not realized in the research of topological insulator because of the adoption of open boundary condition, in which case there is no parameter can be tuned. On the other hand, 3-dimensional (3D) Weyl semimetal has been observed experimentally\cite{Xu2015Science,Huang2015NC,Weng2015PRX} in 2015 after its explicit predictions based on both topological argument and first principle calculations\cite{Murakami2007PRB,Murakami2007NJP,Wan2011,Yang2011,Burkov2011PRL,Xu2011PRL,Burkov2011PRB}. The distinctive mark of the surface states in 3D Weyl semimetal is the Fermi arc on surface Brillouin zone\cite{Armitage2018weyl}. Surely, the existence of topological surface states for a given 3D Weyl semimetal can be explained by the topological number of bulk theory, but how the boundary condition acts on the wave-function of zero-energy surface state and the pattern of Fermi arcs, has not been investigated thoroughly in the literature\cite{isaev2011PRB,okugawa2014dispersion}. One significant reason is that the open boundary extensively used in analytical model of topological insulator is not suitable for surface states in 3D Weyl semimetal. 

Until 2016, Witten pointed out that the generic boundary condition for continuum model of 3D Weyl semimetal must have matrix form with a simple angle parameter\cite{Witten2016}. As an example, he also derived the wave function of the surface state in a special angle parameter. Then in 2017 Hashimoto et.al. systematically studied the generic boundary condition of 3D Weyl semimetal in the continuum limit\cite{Hashimoto2017}. They also obtained the generic boundary condition with a single real parameter for 3D Weyl semimetal both in continuum and lattice models. They demonstrated how a generic surface term in the Lagrangian affects the surface states of a Weyl fermion, especially the shape of Fermi arc connecting two Weyl points. Besides, Devizorova et.al. pointed out the key role of inter-valley interaction in the formation of Fermi arcs in Weyl semimetals\cite{Volkov2017PRB}, which adds indeed another parameter into the generic boundary condition and increase the degree of model freedom.

In fact, there are not only usual Dirac and Weyl points of $\pm1$ chirality, but also there are exotic Dirac and Weyl points of higher chirality which could exist in condensed matter physics under the protection of crystalline symmetry, such as multi-Weyl topological semimetal with partial non-linear dispersion, spin-1 excitations with threefold degeneracy and spin-3/2 Rarita-Schwinger-Weyl fermions with quadruple degeneracy \cite{Bernevig2012PRL,Zhang2016PRB,Bradlyn2016,Tang2017}. The so called multi-Weyl semimetals are quite amazing\cite{Huang2017topological,Ahn2017optical,Dantas2018magnetotransport,Yang2019topological,Dantas2020non-Abelian,Menon2020}. However, so far as the authors are informed, the Fermi arc of topological surface states in multi-Weyl semimetal has not been investigated analytically in literature due to their nonlinear dispersion relation near multi-Weyl points. We thus aim to fill the gap in this paper by extending the theory of Witten and Hashimoto\cite{Witten2016,Hashimoto2017} to describe multi-Weyl semimetals. In general, the boundary condition for continuum model of multi-Weyl semimetal should include derivative term of spinor wave function when the quadratic or cubic momentum terms present in Hamiltonian\cite{Volkov2015JETP}. However, the generic boundary condition deduced for linear Weyl semimetal remains valid for multi-Weyl semimetals. 

Our study is divided into two parts: first, the reformulation of the theory obtained by Hashimoto et.al and generalization of the boundary condition to double flat boundary with Lagrangian formulation; second, the systematical study of Fermi arcs of surface states in linear-Weyl and multi-Weyl semimetals. The paper is organized as follows. In Sec.~\ref{sec:bc_3DWeyl}, we study the 3D Weyl semimetals and their generic boundary conditions in the continuum limit. The relations between energy dispersions, wave functions of edge states and the boundary conditions have been derived. In Sec.~\ref{sec:generic}, we solve the eigenequation in linear-Weyl semimetal and obtain the dispersion relations of bulk and surface states as well as the wave function of surface states. Besides, we reformulate the Fermi arc of surface states in a complex function formalism for easy generalization from linear-Weyl to multi-Weyl semimetals. 

In Sec.~\ref{sec:bc_single nodes}, we first generalize the theory of linear-Weyl semimetal to multi-Weyl semimetal with single Weyl point. The emphasis is put on the situation for quadratic- and cubic-Weyl semimetals, which can be stabilized in crystal materials with high order point group symmetry. Sec.~\ref{sec:FS_pair nodes} discusses the surface states in multi-Weyl semimetals with single and double pairs of Weyl nodes; analyze the evolution of their Fermi arcs varying with the boundary angle parameter. The emphasis is put on the situation for quadratic- and cubic-Weyl semimetals with two pairs of Weyl nodes, which obey the Nielsen-Ninomiya theorem. In Appendix.~\ref{sec:lattice}, we re-derive the orthogonal boundary condition for spinor within a lattice model which offers a complementary explanation to the results obtained in the continuum limit.


\section{Boundary condition for 3D Weyl semimetals}\label{sec:bc_3DWeyl}


In this paper, we are following the theory of Weyl semimetal presented by Witten\cite{Witten2016} and Hashimoto et.al.\cite{Hashimoto2017}, derive the boundary condition for continuum model of Weyl semimetal with some modification, and generalize it to the multi-Weyl systems with single and couple boundary surfaces.

For Weyl semimetal in 3D space, the Hamiltonian in continuum limit near a Weyl point is generically given by 
\begin{align}\label{Ham}
	\mathcal{H}=p_i\sigma_i=p_1\sigma_1+p_2\sigma_2+p_3\sigma_3,
\end{align}
where $p_i$ and $\sigma_i$ are components of 3D momentum and Pauli operator acting in spin or orbital space. The Weyl point is set at the origin of the 3D momentum space.
Since the Hamiltonian \eqref{Ham} is linear to the momenta, the boundary condition is a linear combination of the wave-function and can be always expressed in the form $(M+1)\psi\Big|_{S}=0$. The energy eigenstates of the semi-infinite system with a single boundary condition at $x^3=0$ can be described by
\begin{empheq}[left=\empheqlbrace]{align}
		&\label{HE} \mathcal{H}\psi=\epsilon\psi\\
		\label{BC} &(M+1)\psi\Big|_{x^3=0}=0,
\end{empheq}
where the system is put in the spatial region $x^3\geq 0$. The $\epsilon$ is the energy eigenvalue and $M$ is a $2\times2$ complex constant matrix.

The Hamiltonian and boundary condition above are suitable to depict a simple two-band system with a double degeneracy at the Weyl point. It has been verified that the model captures the topological nature and is identical to the Hamiltonian of a 3D Weyl fermion. 
The boundary condition implies that the two components of $\psi$ are related to each other at the boundary by the matrix $M$, which still reserves the arbitrariness in the choice of the boundary parameter. In physics, we could imagine different boundary conditions, such as a slicing of the material, hydrogen termination or oxidization and even deposition of light metal atoms on boundary surfaces. However, Hashimoto et.al.\cite{Hashimoto2017} have demonstrated that the generic boundary condition \eqref{BC} depicted by the arbitrary matrix $M$ is parameterized only by a single real angle $\theta$ in $[0,2\pi)$.

\subsection{Generic boundary condition and its parameterization}\label{sec:bc_cont}

In this subsection we re-derive the boundary condition matrix $M$ in \eqref{BC} and parametrize it. It turns out that $M$ in the boundary condition \eqref{BC} is determined by the Hermiticity of the Hamiltonian \eqref{Ham} and the requirement that the combined matrix $M+1$ should have a zero eigenvalue.

\subsubsection{constraint of Hermiticity on boundary condition}

Hamiltonian \eqref{Ham}should be Hermitian, which gives rise to a constraint on the boundary condition.\cite{Witten2016} This constraint is
\begin{align}
	\langle \mathcal{H}\psi_1 | \psi_2 \rangle= \langle \psi_1 | \mathcal{H} \psi_2 \rangle,
\end{align}
which remains valid for arbitrary normalizable $\psi_1$ and $\psi_2$.
If the inner product above is written as an integral, a surface term arises and it must be zero:
\begin{align} 
	[(\sigma_3\psi_1)^{\dagger}\psi_2]|_{x^3=0}=0.
\end{align} 
The boundary condition \eqref{BC} must be consistent with this equation, which demands 
\begin{align}
	\frac{1}{2}[((\sigma_3M+M^{\dagger}\sigma_3)\psi_1)^{\dagger}\psi_2]|_{x^3=0}=0.
\end{align}
This is fulfilled for any choice of $\psi_1$ and $\psi_2$ only if 
\begin{align}\label{MSSM}
	M^{\dagger}\sigma_3=-\sigma_3M,
\end{align}
which limits the form of M and partly removes the arbitrariness of boundary condition.

In general, $M$ is a $2\times 2$ complex matrix and should have 4 complex degrees of freedom (d.o.f), that is, 8 real parameters. However, the boundary condition \eqref{BC} actually implies that the eigenvalues of operator $M$ cannot be anything but $\pm1$, which means that $M$ should also be Hermitian($M^{\dagger}=M$). Thus one can express $M$ as:
\begin{align}\label{M2}
	M=A_\mu \sigma_\mu=A_0 \sigma_0 + A_i \sigma_i,
\end{align}
with only four real coefficients $A_0$ and $ A_i$, where $\sigma_0$ is identity matrix and $\sigma_i$ are three Pauli matrix.

Then equation \eqref{MSSM} could be expressed as
\begin{align}\label{MSSM2}
	M\sigma_3+\sigma_3M=\{M,\sigma_3\}=0.
\end{align}
Substituting \eqref{M2} into \eqref{MSSM2} , we obtain
\begin{align}
	0&=M\sigma_3+\sigma_3M
	\nonumber \\
	&=A_{\mu}\sigma_{\mu}\sigma_3+\sigma_3\sigma_{\mu}A_{\mu} \nonumber
	\\
	&=2A_0\sigma_3+2A_3\sigma_0,
\end{align}
which removes further two real d.o.f of $M$, i.e., 
 \begin{align}
 	A_0=A_3=0. 
 \end{align}
Therefore, a boundary condition matrix in terms of two real parameters is left,
\begin{align} \label{AABB}
M = A_1 \sigma_1 + A_2 \sigma_2.
\end{align}
You might be inclined to think that it has been simplified enough, but unexpectedly, matrix $M$ has a single real parameter actually. To see that, let us consider its eigenvalue-problem.

\subsubsection{parameterization of $M$}

The boundary condition \eqref{BC} can be regarded as an eigenequation of matrix $M$. If equation \eqref{AABB} is substituted into the determinant of \eqref{BC}
\begin{align}
	\text{det}(M-\lambda)=0,
\end{align}
one obtains
\begin{align}
	\lambda_{\pm}=\pm\sqrt{A_1^2+A_2^2}.
\end{align}
The boundary condition demands $M$ shoud have a real eigenvalue $-1$, which means:
\begin{align}	
	A_1^2+A_2^2=1.
\end{align}
The generic boundary condition matrix thus could be rewritten as
\begin{align}\label{MS}
	M=\cos\theta \sigma_1+\sin\theta\sigma_2,
\end{align}
with $A_1=\cos\theta$ and $A_2=\sin\theta$. Consequently, $M$ is parametrized only by a single real angular parameter with $0\leq\theta<2\pi$.
This boundary condition indeed points to a direction for Pauli vector on the surface of 3D Weyl semimetal, which is obvious when we reformulate it as 
\begin{align}
M=\vec{\sigma}\cdot\vec{m}=(\sigma_1\:\sigma_2\:\sigma_3)\cdot(\cos\theta\:\sin \theta\:0),
\end{align}
where $\theta$ is the included angle between the unit vector $\vec{m}$ and the $x_1$ axis. Thus, the boundary condition means that the wave function of surface state is the one eigenstate of Pauli projection operator $M=\vec{\sigma}\cdot\vec{m}$ with eigenvalue $-1$. 
The boundary condition can be also formulated by angle parameter $\theta$ as 
\begin{align}
	\left(\begin{array}{cc}
	1 & e^{-i\theta} 
	\\ 
	e^{i\theta} &1 
	\end{array}\right)
	\psi\Big|_{x^3=0}=0.
\end{align}
Note that 
\begin{align}
\left(\begin{array}{cc}
	1 & e^{-i\theta} 
	\\ 
	e^{i\theta} & 1 
	\end{array}\right)
	=
	\left(\begin{array}{c}
	1 \\
	e^{i\theta} 
	\end{array}\right)
	\left(\begin{array}{cc}1 & e^{-i\theta} 
	\end{array}\right),
\end{align}
the boundary condition can be recasted to the following simpler form
\begin{align}\label{bc'}
	\left(\begin{array}{cc}1 & e^{-i\theta} 
	\end{array}\right)
	\psi\Big|_{x^3=0}=0.
\end{align}
We thereby conclude that the generic boundary condition is dictated by a single real parameter. In addition, the equation \eqref{bc'} also tells us that, at the boundary, two components of the fermion need to have the identical magnitude, and the relative phase between them is determined 
by $\theta$. This is true for both edge modes and bulk modes. 

If we set the boundary condition with the wave function of surface state as the other eigenstate of $M=\vec{\sigma}\cdot\vec{m}$ with eigenvalue $+1$, then the boundary condition in term of $\theta$ becomes
\begin{align}\label{bc'n}
	\left(\begin{array}{cc}1 & -e^{-i\theta} 
	\end{array}\right)
	\psi\Big|_{x^3=0}=0.
\end{align}
This form of boundary condition will naturally present in a Lagrangian formulation which describes the Weyl semi-metal materials with double parallel surfaces boundary.

\subsection{Lagrangian formalism}

Lagrangian formalism permit the natural derivation of the boundary condition \eqref{MSSM}. Let us first consider a generic theory for a Weyl semimetal in 1+3 spacetime dimensions without boundary. 

We chose the metric convention as $\eta_{\mu\nu}=\mbox{diag}(+,-,-,-)_{\mu\nu}$, then the bulk Lagrangian (for a right-handed Weyl fermion) can be written as 
\begin{align}
{\cal L} = \frac{i}{2} \psi^\dagger \sigma^\mu\overleftrightarrow{\partial}_{\mu}\psi
\label{eq:Lag2}
\end{align}
where $\sigma^\mu=(\sigma_0, \sigma_1,\sigma_2,\sigma_3)$.
The Euler equation is a Dirac equation:
\begin{align}
\sigma^\mu \partial_\mu \psi=0.
\end{align}
Rewrite it as
\begin{align}
\left[i\sigma_0\partial_0 +i\sigma_i \partial_i\right]\psi
=0,
\end{align}
where $i=1,2,3$. The corresponding Hamiltonian is, 
\begin{align}
{\cal H} = p_1 \sigma_1 + p_2 \sigma_2 + p_3 \sigma_3,
\label{eq:hamil2}
\end{align}
which is the standard Hamiltonian near a Weyl point.

\subsubsection{single flat surface boundary}

To study surface states, we should add boundary conditions to the system. First consider a single flat surface boundary. In this situation, a surface term should be introduced to the above Lagrangian. Then the total action becomes
\begin{align}
S = \int_{x^3\geq 0} \!\!\!\!d^3x \; 
\frac{i}{2} \psi^\dagger \sigma^\mu\overleftrightarrow{\partial}_{\mu}\psi+ \frac12 \int_{x^3=0} \! d^2x \;
\psi^\dagger N \psi \, .
\end{align}
The first bulk term is the Weyl Lagrangian of half-infinite system; the second term is the boundary surface Lagrangian with a Hermitian matrix $N$. We point out that this surface term captures the essential physics and could describe effectively the influence of ideal surface or interface, as well as the surface oxidization, reconstruction, hydrogenation and even metal atom adsorption. Generally speaking, $N$ should be a function of 2D surface momentum and coordinates. In the simplest case for ideal surface, however, it can be a constant Hermitian operator.

Performing variations $\psi\to\psi + \delta\psi$ and $\psi^\dagger \to \psi^\dagger+\delta\psi^\dagger$ and considering the arbitrariness of $\delta\psi$ and $\delta\psi^\dagger$, we get
\begin{align}
\psi^\dagger\Big|_{x^3=0} (-i\sigma_3 +N)= 0, \quad 
(i\sigma_3 +N)\psi\Big|_{x^3=0}= 0. 
\label{boudagger}
\end{align}
These two equations are complex-conjugate to each other. Using $\sigma_{3}\psi$ to right multiply the former and plus the latter left multiplied by $\psi^\dagger\sigma_{3}$ , one obtains
\begin{align}
\label{NS3}
N \sigma_3 + \sigma_3 N =\{N,\sigma_3\}= 0.
\end{align}
Equation \eqref{NS3} and the Hermiticity of $N$ mean that it can be formulated as
\begin{align}\label{NS}
N = B_1 \sigma_1 + B_2 \sigma_2,
\end{align}
just like that for $M$.
The non-triviality of boundary condition demands $det(N+i\sigma_{3})=0$, which means that
$B_1^2+B_2^2-1=0$. $N$ therefore can also be expressed with a angular parameter $\phi$ as
\begin{align}\label{NP}
N=\cos\phi\,\sigma_1+\sin\phi\,\sigma_2. 
\end{align}
Similar to $M$, $N$ can also be expressed in term of a projector vector as
\begin{align}
N=\vec{\sigma}\cdot\vec{n}=(\sigma_1\:\sigma_2\:\sigma_3)\cdot(\cos\phi\:\sin\phi\:0).
\end{align}
Since $N$ and $M$ have the same form, let us determine their relation. Left multiply $-i\sigma_3$ to the second equation in \eqref{boudagger} gives
\begin{align}
		\label{BC2} &[(-i\sigma_3\,N)+1]\psi\Big|_{x^3=0}=0.
\end{align}
Comparing to \eqref{BC} and considering \eqref{MS} as well as \eqref{NP}, one obtains
\begin{align}\label{TP}
		\cos\theta=-\sin\phi,\,\sin\theta=\cos\phi\,,
\end{align}
which means $\theta=\phi+\pi/2$ and $\vec{n}\perp\vec{m}$.

So far, we have indicated that the boundary condition is dictated by a boundary ''mass'' term manifested as the Hermitian matrix $N$, which is determined by an angular parameter and is equivalent to \eqref{BC}. The subsequent work is to explore the two parallel boundary cases, which are also typical for realistic materials. 

\subsubsection{double parallel flat surfaces boundary}

Presume the two parallel boundary surfaces are still along $x^3$ direction, with one surface boundary at $x^3=0$ and the other at $x^3=L$. In such situation, the total action becomes
\begin{align}
S &= \int_{0\leq\,x^3\leq\,L} \!\!\!\!d^3x \; 
\frac{i}{2} \psi^\dagger \sigma^\mu\overleftrightarrow{\partial}_{\mu}\psi+ \frac12 \int_{x^3=0} \! d^2x \;
\psi^\dagger N_0 \psi \,\nonumber\\
&+ \frac12 \int_{x^3=L} \! d^2x\;
\psi^\dagger N_L \psi \, ,
\end{align}
where the first is bulk term in the limited space; the second and third terms are the surface Lagrangian for boundaries at $x^3=0$ and $x^3=L$ with Hermitian matrices $N_0$ and $N_L$. 

After the variations $\psi\to\psi + \delta \psi$ and $\psi^\dagger \to \psi^\dagger+\delta\psi^\dagger$, one can obtain the boundary condition at the surface $x^3=0$ 
\begin{align}\label{BCN0}
- i\psi^\dagger \sigma_3 + \psi^\dagger N _0= 0 , \quad 
i\sigma_3 \psi + N_0\psi = 0 ,
\end{align}
which is the same as the single flat boundary case \eqref{boudagger}. While the boundary condition at the surface $x^3=L$ is 
\begin{align}\label{BCNL}
 i\psi^\dagger \sigma_3 + \psi^\dagger N _L= 0 , \quad 
-i\sigma_3 \psi + N_L\psi = 0 ,
\end{align}
which is different from that condition at $x^3=0$ with an extra minus sign in $\sigma_3$ term. We comment that this minus sign will have significance on the relative direction of Fermi arcs on the two surface Brillouin zones. However, the $N_L$ is still satisfies the same anti-commutation with $\sigma_3$ as \eqref{NS3}, thus it processes the same form as $N$ and can be characterized by a single angle parameter. 

Let us consider the simplest case where $N_L=\cos\phi\,\sigma_1+\sin\phi\,\sigma_2=N$, which means that the physical structure and environment on the two surfaces are the same as each other. In this case, the spinor wave function $\psi$ should satisfy the following boundary conditions 
\begin{align}\label{BCN0LN}
[(-i\sigma_3\,N)+1]\psi\Big|_{x^3=0}=0 , \quad 
[(i\sigma_3\,N)+1]\psi\Big|_{x^3=L}=0 .
\end{align}
In term of M, the boundary conditions become
\begin{align}\label{BCM0L}
[M+1]\psi\Big|_{x^3=0}=0 , \quad 
[M-1]\psi\Big|_{x^3=L}=0 .
\end{align}
One can find that the boundary conditions for the two identical parallel surfaces have the same boundary operator $M$, just as expected, but with different eigenvalues which is unexpected more or less.

\section{Surface states in linear-Weyl semimetal} \label{sec:generic}

One may expect the existence of the topological surface modes arising from the non-trivial topology of Weyl Semimetal. In this subsection we look for surface state solution of linear-Weyl semimetal. With the generic boundary condition \eqref{bc'}, the dispersion relation and the wave function of the surface states have been obtained by Hashimoto et.al\cite{Hashimoto2017}. Here we reformulate them in term of parameter $\theta$ for convenient generalization to multi-Weyl semimetal cases.

\subsection{Wave function of surface states}

The surface modes solution to eigenequation \eqref{HE} can be solved with a two-component wave function ansatz
\begin{align}
	\psi=
	\left(\begin{array}{c} 
		\chi \\
		\eta
	\end{array}\right),
\end{align}
the eigenequation \eqref{HE} is written as
\begin{align}\label{HE'}
	\left(\begin{array}{cc}
	-i\partial_3-\epsilon & p_1-ip_2 
	\\
	p_1+ip_2 & i\partial_3-\epsilon
	\end{array}\right)
	\left(\begin{array}{c}\chi \\\eta\end{array}\right)
	=0.
\end{align}
This equation is equivalent to two independent second-order differential equations:
\begin{align}
 \left(p_1^2+p_2^2-\epsilon^2-\partial_3^2\right)
 \begin{pmatrix}
 \chi \\ \eta
 \end{pmatrix} = 0.
\end{align}
To obtain the modes localized at the boundary, we need 
\begin{align}\label{al}
	\alpha^2\equiv p_1^2+p_2^2-\epsilon^2>0,
\end{align}
then the corresponding solution becomes
\begin{align} \label{EM}
 \begin{pmatrix}
 \chi \\ \eta
 \end{pmatrix}
 = e^{- \alpha x^3}
 \begin{pmatrix}
 \chi_0 \\ \eta_0
 \end{pmatrix}
\end{align} 
where $\chi_0$ and $\eta_0$ have no dependence on $x^3$. This is the general edge mode up to a normalization factor, and the components $\chi_0$ and $\eta_0$ will be further determined by the boundary condition \eqref{bc'} up to a phase factor.
\begin{align} \label{EM2}
 \begin{pmatrix}
 \chi_0 \\ \eta_0
 \end{pmatrix}
 \propto
 \begin{pmatrix}
 1 \\ -e^{i\theta}
 \end{pmatrix}
\end{align} 

We have already utilized most of the information and are left with normalization condition only, with which we can determine the wavefunction completely. To obtain the wave function of the surface states.
 we substitute \eqref{EM} to the normalization condition
\begin{align}
	\int^\infty_0 dx^3~\psi^{\dagger} \psi=1,
\end{align}
and we get a constraint
\begin{align}\label{NC}
	|\chi_0|^2+|\eta_0|^2=2\alpha.
\end{align}
Combined \eqref{EM2}with \eqref{NC}, they are determined up to an irrelevant overall phase:
\begin{align}
 \begin{pmatrix}
 \chi_0 \\ \eta_0
 \end{pmatrix}
 = \sqrt{\alpha}
 \begin{pmatrix}
 1 \\ -e^{i\theta} 
 \end{pmatrix}.
\end{align}
So the general edge mode wave function is 
\begin{align}
\label{edgegeneral}
	\psi(x^3)&=\sqrt{\alpha}~\text{exp}(-\alpha x^3)
	\left(\begin{array}{c}1 \\ -e^{i\theta}
	\end{array}\right),
\end{align}
where $\alpha=p_1\sin{\theta}-p_2\cos{\theta}$.
Note that the edge modes exist only in a limited region of the momentum space, since we 
need to require $\alpha>0$. The linear inequality $\alpha>0$ specifies a half of the momentum space,
only in which the dispersion exists.

In the limit $\alpha = 0$, that is, on the line $p_1\sin{\theta}-p_2\cos{\theta}=0$ in the momentum
space, the edge mode approaches a non-normalizable mode, which is a constant wave function in the $x^3$ space.
It corresponds to $p_3=0$ bulk mode, whose dispersion is $\epsilon = \pm \sqrt{p_1^2+p_2^2}$. 

In fact, the edge dispersion \eqref{EDR1} is identical to that under the condition $\alpha=0$. Therefore we have a consistent picture
for any value of $\theta$: when the edge mode approaches a non-normalizable state in the momentum space, it is consistently and continuously absorbed into the bulk modes.

\subsection{Dispersion relation of surface states}
Combining the results from eigenvalue equation \eqref{HE} and boundary condition \eqref{BC} for surface eigenmodes and
substituting equations \eqref{EM} and \eqref{EM2} into equation \eqref{HE'}, we get an independent equation:
\begin{align}\label{HE"}
	(i\alpha-\epsilon) -(p_1-ip_2)\,e^{i\theta}=0.
\end{align}
The real and imaginary parts of the left part in above equation equalling to zero respectively gives rise to 
\begin{align}\label{EDR1}
	\epsilon=-p_1\cos{\theta}-p_2\sin{\theta}=-\vec{p}\cdot\vec{m},
\end{align}
\begin{align}\label{EDR-2}
	\alpha=p_1\sin{\theta}-p_2\cos{\theta}=\vec{p}\cdot\vec{n}>0.
\end{align}
The first equation \eqref{EDR1} is the dispersion relation of the edge states and is linear with respect to $p_1$ and $p_2$; the second equation \eqref{EDR2} give the relation of localization factor with the surface momentum $p_1$ and $p_2$. These two equations can be written into a compact form:
\begin{align}
	\left(\begin{array}{c}\epsilon \\\alpha\end{array}\right)
	=
	-\left(\begin{array}{cc}\cos{\theta} & \sin{\theta} \\ -\sin{\theta} & \cos{\theta}\end{array}\right)
	\left(\begin{array}{c}p_1 \\p_2\end{array}\right).
	\label{rotea}
\end{align}
Equation \eqref{rotea} indicates that the effect of boundary is rotating the momenta $(p_1,p_2)$ into $(\epsilon,\alpha)$, the energy and the inverse of edge mode decay width (penetration depth). For any certain $p_1$ and $p_2$, the pair $(\epsilon, \alpha)$ can be regard as a vector rotating around the origin by $\theta$. When $|\epsilon|$ becomes large, $\alpha$ becomes small, then the penetration depth is large. On the other hand, when $|\epsilon|$ is small, $\alpha$ becomes large and then the penetration depth is reduced. Thus we conclude that the penetration depth is inversely proportional to the energy approximately.
 
\subsection{Complex function formalism and Fermi arc}
Taking complex conjugate of \eqref{HE"} and define complex function $\omega=\epsilon+i\alpha$ and complex momentum $p=p_1+ip_2$, we then obtain a more compact and easier generalization form: 
\begin{align}
\omega=p\,e^{-i(\theta+\pi)}=p\,e^{-i(\phi+3\pi/2)},\,(\Im(\omega)>0),
\end{align}
where $\theta=\phi+\pi/2$ as indicated in \eqref{TP}. If we rewrite $p=|p|e^{i\arg(p)}$, then 
\begin{align}
\omega=|p|e^{i[\arg(p)-(\theta+\pi)]},\,(\Im(\omega)>0),
\end{align}
which means $\omega$ has the same modulus with p but rotate $\theta+\pi$ clockwise under the condition that $\sin[\arg(p)-(\theta+\pi)]>0$. 
The Fermi arc is defined as the curve of the zero-energy surface state on the projected momentum plane. To obtain the Fermi arc of the linear-Weyl fermion, we demand further that $\epsilon=\Re(\omega)=0$. i.e.,
\begin{align}
&\cos(\arg(p)-(\theta+\pi))=0\cap\sin(arg(p)-(\theta+\pi))>0,
\end{align}
or more compactly,
\begin{align}
\arg(p)-\theta-\pi=\pi/2,
\end{align}
which means that the Fermi arc in this case is a ray form origin in projected momentum plane ($p_1-p_2$) along the direction with $\arg(p)=\theta-\pi/2$.


\section{Surface states in multi-Weyl semimetals with single Weyl point} \label{sec:bc_single nodes}


\subsection{General results of multi-Weyl semimetals} 

The analyses above with respect to linear-Weyl semimetals ( with the topological charge $w=1$) can be easily generalized to the multi-Weyl semimetals described by the Hamiltonian:
\begin{align}\label{Ham-m}
	\mathcal{H}=\left(\begin{array}{cc}p_3 & g^*(p) \\ g(p) & -p_3\end{array}\right)=\Re{g(p)}\,\sigma_1+\Im{g(p)}\,\sigma_2+p_3\sigma_3
\end{align}
where $g(p)=g(p_1+ip_2)$ is a complex variable function with $\Re{g(p)}$ and $\Im{g(p)}$ its real and imaginary parts, respectively. It should be pointed out that $g(p)$ needs not to be an analytical function of complex $p$. In this situation, The bulk energy dispersion 
\begin{align}
E=\pm\sqrt{[\Re{g(p)}]^2+[\Im{g(p)}]^2+p_3^2}.
\end{align}
The energy dispersion of surface states is
\begin{align}
\omega=g(p)\,e^{-i(\theta+\pi)}=g(p)\,e^{-i(\varphi+\pi/2)},\,(\Im{\omega}>0),
\end{align}
or 
\begin{align}\label{EDRG}
	\epsilon&=-\Re{g(p)}\cos{\theta}-\Im{g(p)}\sin{\theta},\\
	\alpha&=\Re{g(p)}\sin{\theta}-\Im{g(p)}\cos{\theta}>0.
\end{align}
\subsection{Quadratic-Weyl semimetal} 
For a single quadratic-Weyl node semimetal, $g(p)=p^2$, we have
the bulk energy dispersion 
\begin{align}
E=\pm\sqrt{(p_1^2+p_2^2)^2+p_3^2}.
\end{align}
The corresponding energy dispersion of surface states is
\begin{align}\label{EDR2}
	\epsilon=-(p_1^2-p_2^2)\cos{\theta}-(2p_1\,p_2)\sin{\theta},
\end{align}
\begin{align}
	\alpha=(p_1^2-p_2^2)\sin{\theta}-(2p_1\,p_2)\cos{\theta}>0.
\end{align}
The compact complex function formalism is
\begin{align}
\omega=|p|^2\,e^{i(2\arg(p)-\theta-\pi)},\,(\Im(\omega)>0),
\end{align}
which yields the argument equation of Fermi arc 
\begin{align}
2\arg(p)-\theta-\pi=\pi/2+2k\pi, (k=0,1).
\end{align}
This formula indicates that the Fermi arcs for quadratic-Weyl semimetal are two rays from the Weyl point with directional angles $\theta/2+3\pi/4$ and $\theta/2+7\pi/4$.

\subsection{Cubic-Weyl semimetal} 

For a single cubic-Weyl semimetal, $g(p)=p^3$, we have
the bulk energy dispersion 
\begin{align}
E=\pm\sqrt{(p_1^2+p_2^2)^3+p_3^2}.
\end{align}
The energy dispersion of surface states is
\begin{align}\label{EDR3}
	\epsilon=-p_1(p_1^2-3p_2^2)\cos{\theta}-p_2(3p_1^2-p_2^2)\sin{\theta},
\end{align}
\begin{align}
	\alpha=p_1(p_1^2-3p_2^2)\sin{\theta}-p_2(3p_1^2-p_2^2)\cos{\theta}>0.
\end{align}
The compact complex function formalism is
\begin{align}
\omega=|p|^3\,e^{i(3\arg(p)-\theta-\pi)},\,(\Im(\omega)>0).
\end{align}
The argument equation of Fermi arc 
\begin{align}
3\arg(p)-\theta-\pi=\pi/2+2k\pi, (k=0,1,2).
\end{align}
The Fermi arcs for single cubic-Weyl fermion are three rays from the origin with directional angles $\theta/3+\pi/2$, $\theta/3+7\pi/6$ and $\theta/3+11\pi/6$.
\subsection{multi-Weyl semimetal} 
The highest winding number of Weyl point in crystal is 4 protected by point group symmetry\cite{Weng2015PRX,Yao2021}. It is, however, still of significance to consider the surface states and Fermi arc of semimetals with $w\geqslant4$. For the general case $g(p)=p^w,(w=4,5,6,\cdots)$, the corresponding bulk energy dispersion is
\begin{align}
E=\pm\sqrt{(p_1^2+p_2^2)^w+p_3^2}.
\end{align}
The energy dispersion of surface states is
\begin{align}\label{EDR4}
	\epsilon&=-\Re{(p^w)}\cos{\theta}-\Im{(p^w)}\sin{\theta},\\
	\alpha&=\Re{(p^w)}\sin{\theta}-\Im{(p^w)}\cos{\theta}>0.
\end{align}
The compact formalism is
\begin{align}
\omega=|p|^w\,e^{i(w\arg(p)-\theta-\pi)},\,(\Im(\omega)>0).
\end{align}
The argument equation of Fermi arc 
\begin{align}
w\arg(p)-\theta-\pi=\pi/2+2k\pi, (k=0,1,2,\cdots,w-1),
\end{align}
which gives the arguments of Fermi arc rays with
\begin{align}
\arg(p)=\dfrac{\theta}{w}+\dfrac{3\pi}{2w}+\dfrac{k2\pi}{w},(k=0,1,2,\cdots,w-1).
\end{align}
Here we point out that $\alpha>0$ is for semi-infinite region $x^3\geq0$, while for $x^3\leq0$, we should take $\alpha<0$. Then the the arguments of Fermi arc rays in this case are
\begin{align}
\arg(p)=\dfrac{\theta}{w}+\dfrac{\pi}{2w}+\dfrac{k2\pi}{w},(k=0,1,2,\cdots,w-1).
\end{align}
which is the ($\pi/w$ degree) rotation image of that for positive $\alpha$. 


\section{Surface states in multi-Weyl semimetals with single and double pairs of Weyl nodes} \label{sec:FS_pair nodes}


The models above are about multi-Weyl semimetals with only one Weyl node (the zero point of $g(p)$), though providing the intuitive understanding of surface states of Weyl semimetals, are too simple to be real. In real 3D Weyl semimetal, there must be more than one Weyl nodes with the sum of their topological charges (or winding numbers) equals to zero, according to Nielsen-Ninomiya theorem\cite{Nielsen1981,Friedan1982}. Besides, since it has been observed that the Fermi arc can be manipulated by decorating potassium (K) on the surface boundary of NbAs crystal\cite{Yang2019}, thus we will discuss the more realistic Weyl semimetals with one and two pairs of Weyl nodes with winding numbers of equal magnitude but opposite sign, and investigate the evolution of Fermi arc with respective to boundary angle parameter $\theta$.

\subsection{A single-pair of Weyl nodes} 

\subsubsection{Q=(1,-1)}

\begin{figure}[htbp]
 \begin{center}
 \begin{tikzpicture}

 \node (1) at (-1.2,0) {\includegraphics[width=14em]{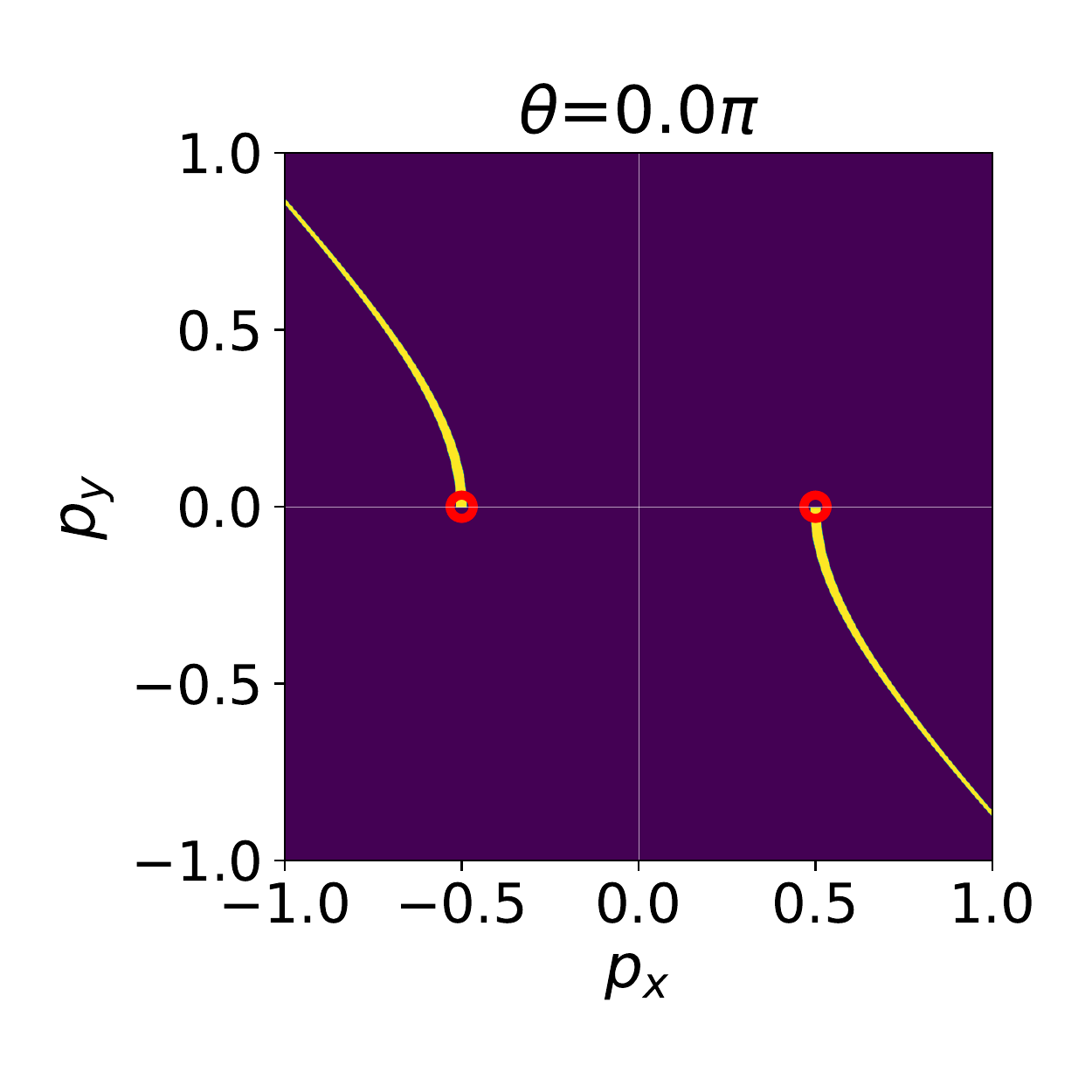}};
 \node (2) at (3.1,0) {\includegraphics[width=14em]{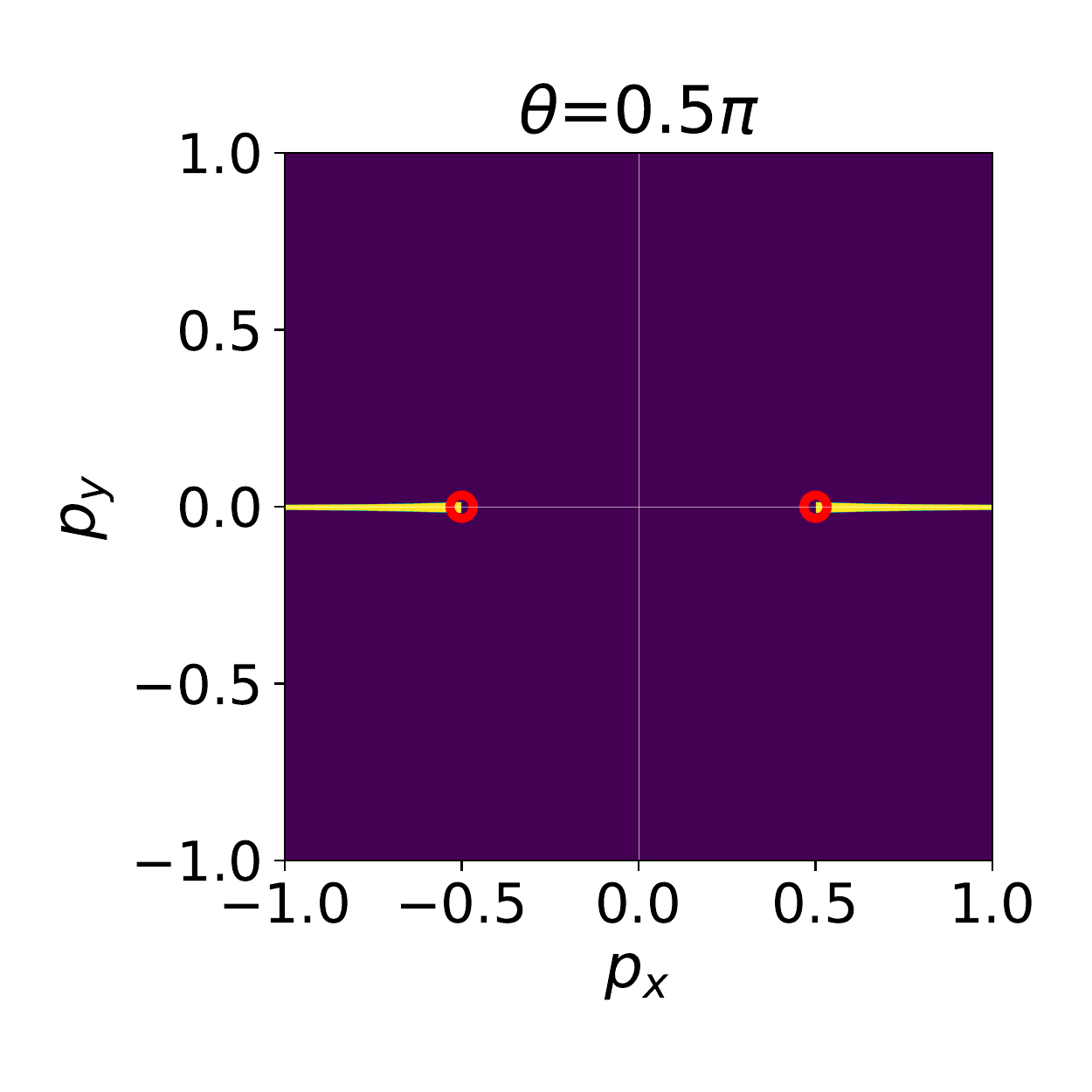}};
 \node (3) at (-1.2,-3.9) {\includegraphics[width=14em]{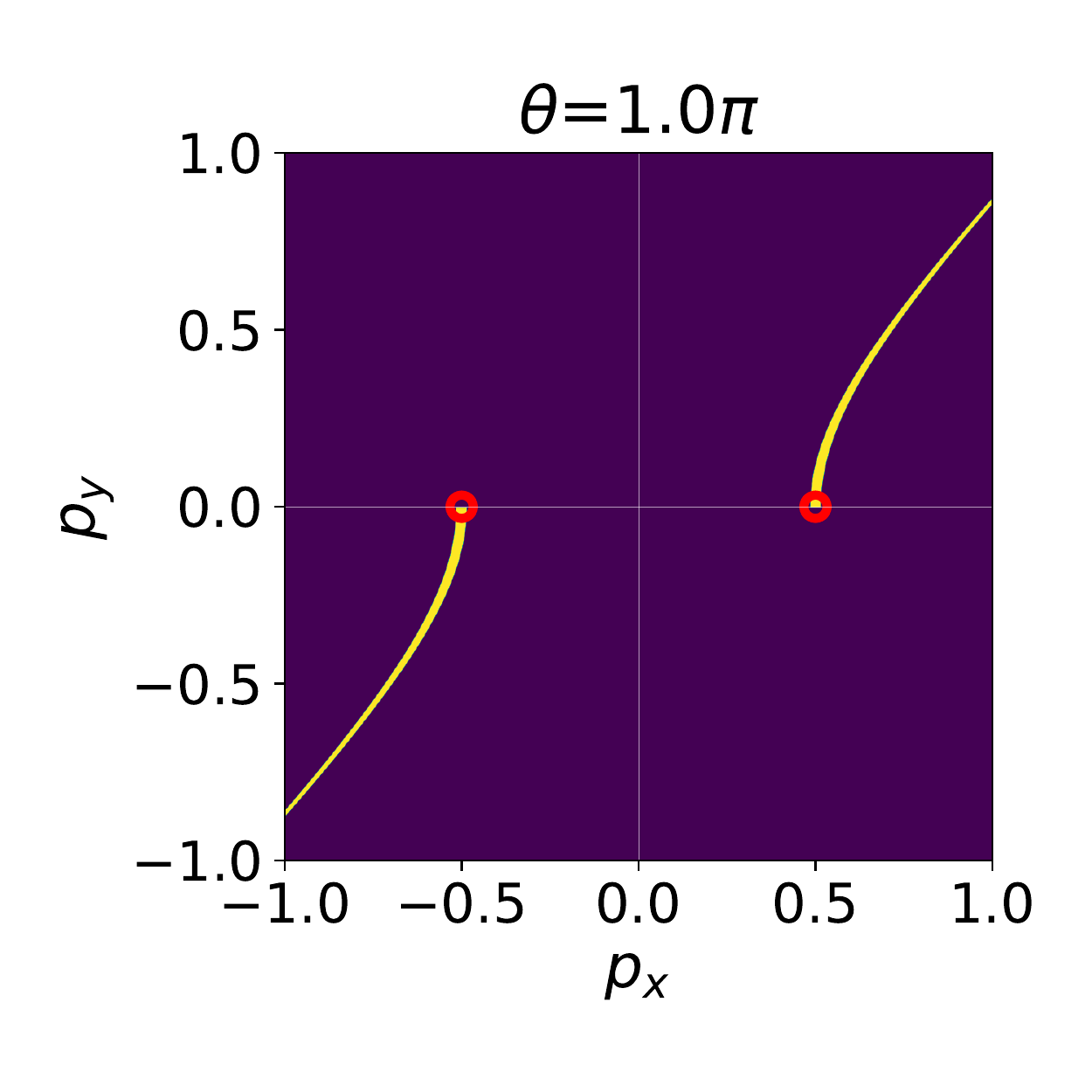}};
 \node (4) at (3.1,-3.9) {\includegraphics[width=14em]{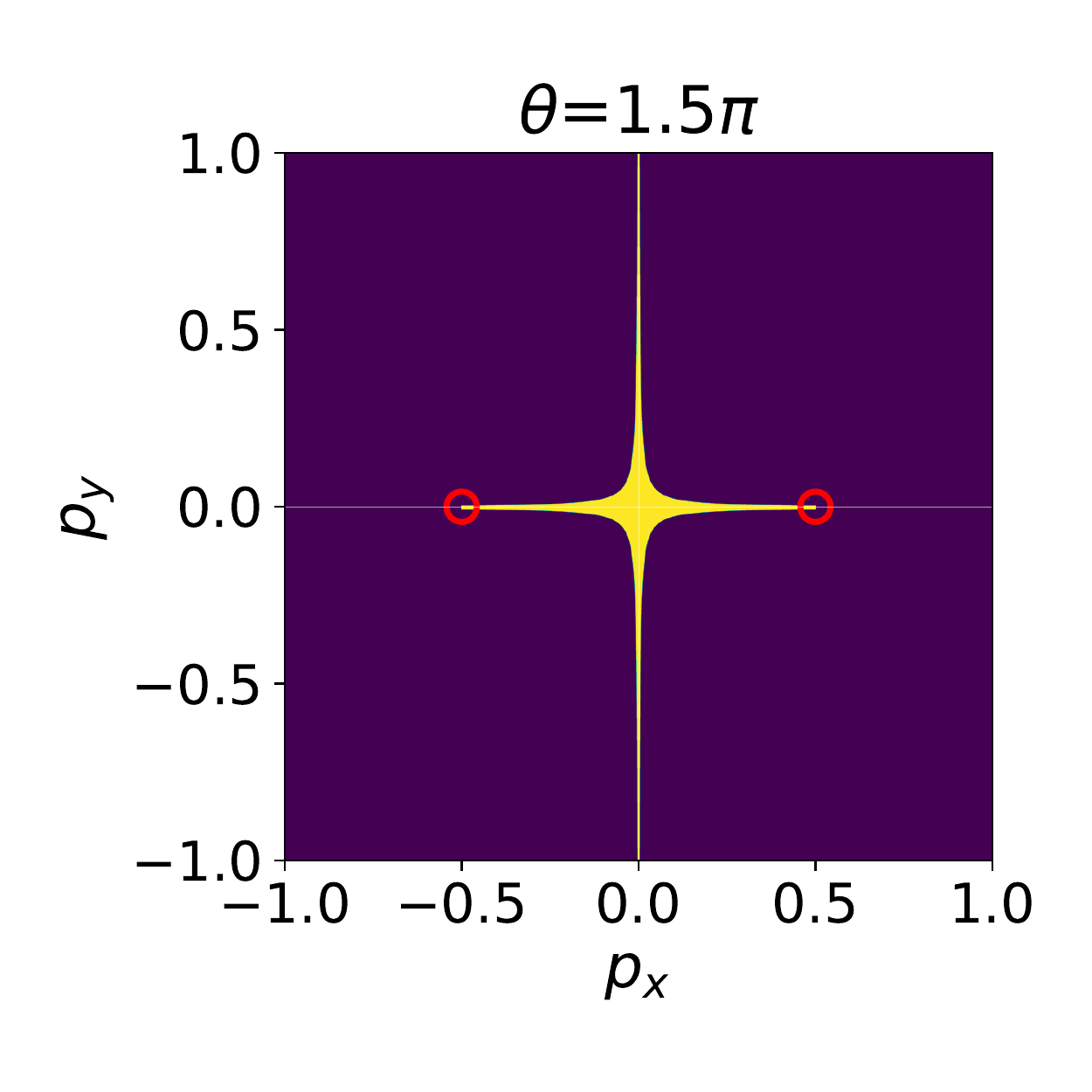}};
 \node (5) at (-1.2,-7.8) {\includegraphics[width=14em]{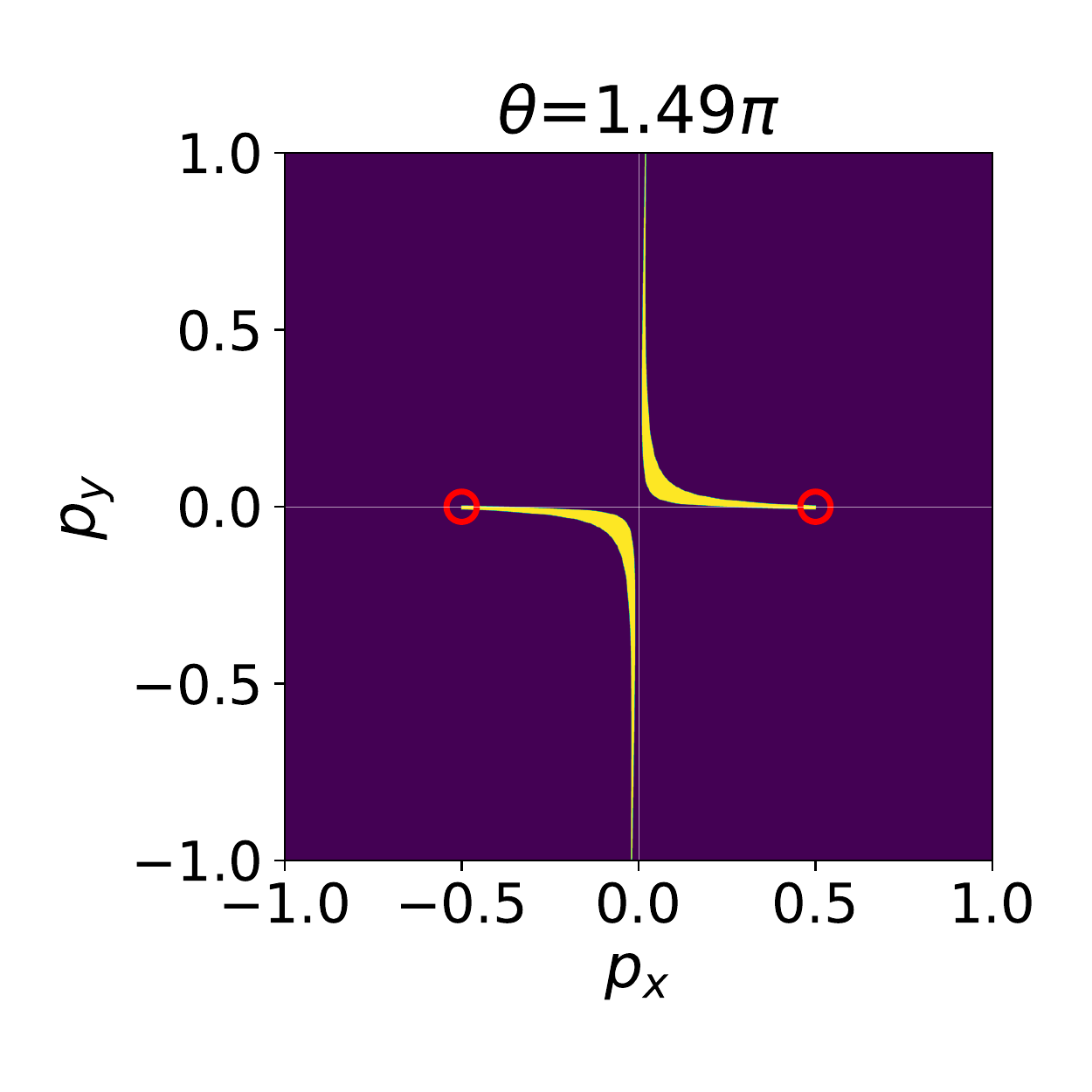}};
 \node (6) at (3.1,-7.8) {\includegraphics[width=14em]{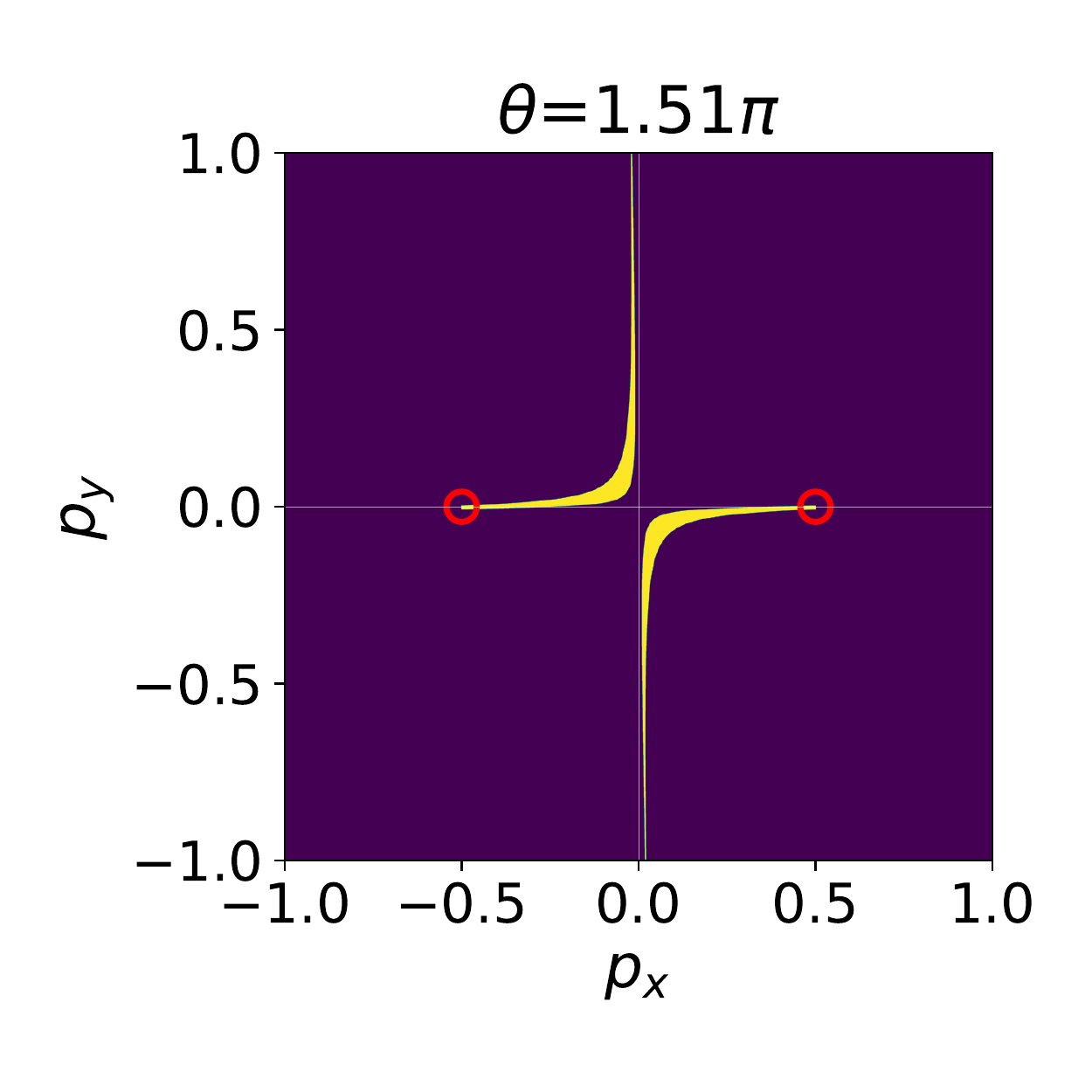}};

 \node at (-2.0,1.9) {(a)};
 \node at (2.2,1.9) {(b)};
 \node at (-2.0,-2.0) {(c)};
 \node at (2.2,-2.0) {(d)};
 \node at (-2.0,-5.9) {(e)};
 \node at (2.2,-5.9) {(f)};
 \end{tikzpicture}
 \end{center}
 \caption{The Fermi arcs of two Weyl points $Q=(1,-1)$ (sitting at $(\pm 0.5,0)$ indicated by small red circle) for the boundary condition parameter $\theta= 0,0.5\pi,1.0\pi,1.5\pi,1.49\pi,1.51\pi$ are shown successively. Here and below, the $(p_x, p_y)$ in each sub-graph denote the two-dimensional momentum of surface state, which correspond to $(p_1, p_2)$ in text.}
 \label{fig:11}
\end{figure}
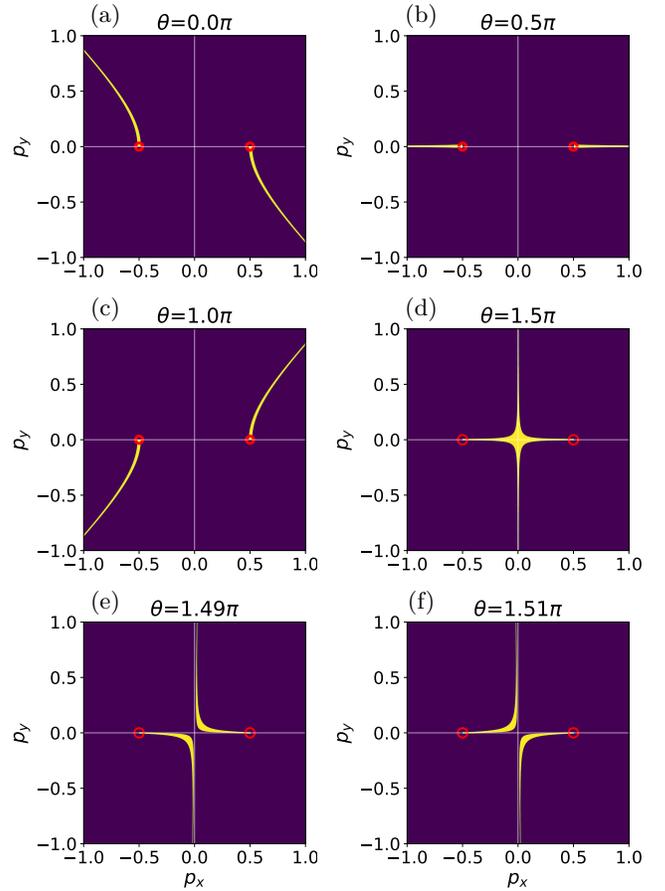

For a pair of Weyl nodes locating at $(\pm a,0)$ with topological charge $\pm1$, its $g(p)=p^2-a^2$. The bulk energy dispersion 
\begin{align}
E=\pm\sqrt{(p_1^2-p_2^2-a^2)^2+(2p_1\,p_2)^2+p_3^2}.
\end{align}
The corresponding energy dispersion of surface states is
\begin{align}\label{EDR11a}
	\epsilon&=-(p_1^2-p_2^2-a^2)\cos{\theta}-(2p_1\,p_2)\sin{\theta},
\\
	\alpha&=(p_1^2-p_2^2-a^2)\sin{\theta}-(2p_1\,p_2)\cos{\theta}>0.
\end{align}
The compact complex function formalism is
\begin{align}
\omega=|g(p)|^2\,e^{i(\arg{g(p)}-\theta-\pi)},\,(\Im(\omega)>0),
\end{align}
which gives the argument of $g(p)$. However, to obtain Fermi arc, what we need is the relation of $\arg(p)$ and $|p|$ rather than $\arg{g(p)}$. Thus we would better to begin with \eqref{EDR11a} by demanding $\epsilon=0$
 \begin{align}\label{EDR11b}
	(|p|^2\cos{2\beta}-a^2)\cos{\theta}+|p|^2\sin{2\beta}\sin{\theta}=0,
 \\
	(|p|^2\cos{2\beta}-a^2)\sin{\theta}-|p|^2\sin{2\beta}\cos{\theta}>0.
\end{align}
In above equations we have defined $\beta=\arg(p)$ and using the relation
\begin{eqnarray}
p_1^2-p_2^2=|p|^2\cos{2\beta},\quad2p_1p_2=|p|^2\sin{2\beta}.
\end{eqnarray}
The Fermi arcs in this case becomes complex with the variation of $\theta$. Let us firstly itemize four special cases.
\begin{itemize}
\item $\theta=\pi/2$, $\Rightarrow(\cos{\theta}=0,\sin{\theta}=1)$\\
 Fermi arcs: ($|p|>a,\quad\beta=0,\pi$),\\ 
which are two rays begin from $(\pm a,0)$ to$\pm\infty$ along $p_1$ axis, respectively.
\item $\theta=3\pi/2$, $\Rightarrow(\cos{\theta}=0,\sin{\theta}=-1)$\\
 Fermi arcs: ($(|p|<a\cap\cos{2\beta}=1)\cup(\cos{2\beta}=-1)$), \\
which are the line segment beginning from $(+a,0)$ to$(-a,0)$ along $p_1$ axis and the whole $p_2$ axis.
\item $\theta=0$, $\Rightarrow(\cos{\theta}=1,\sin{\theta}=0)$\\
 Fermi arcs: $(|p|^2=a^2\sec{2\beta},\,\sin{2\beta}<0)$,\\ 
which are two half parts at II and IV quadrants of each branch of the hyperbola $p_1^2-p_2^2=a^2$, with the two Weyl nodes $(\pm a,0)$ as their two vertices.
\item $\theta=\pi$, then $\cos{\theta}=-1,\sin{\theta}=0$,\\
Fermi arcs: ($|p|^2=a^2\sec{2\beta},\,\sin{2\beta}>0$),\\
which are the other half of the hyperbola $p_1^2-p_2^2=a^2$ at I and III quadrants.
\end{itemize}
The Fermi arc of $Q=(1,-1)$ for the four special cases above are shown in FIG. \ref{fig:11}(a-d). For general $\theta$, we find that Fermi arcs are still half of inclined hyperbola
\begin{align}
|p|^2=a^2\cos\theta\sec{(2\beta-\theta)}, \label{HPPL1}\\
\sin{(2\beta-\theta)}+\cos{(2\beta-\theta)}\tan{\theta}<0.\label{HPPL2}
\end{align}
The equation \eqref{HPPL1} represents a slopping hyperbola rotating $\theta/2$ counter clockwise from the hyperbola $p_1^2-p_2^2=a^2$ with vertices $a$ shorten as $a\sqrt{|\cos\theta|}$, while the inequality \eqref{HPPL2} further chooses the half of each branches of this parabola. These formulae are suit for all $\theta\in[0,2\pi)$ except for $\theta=\pi/2$ and $3\pi/2$. We thus should discuss them in two intervals of $\theta$:
\begin{itemize}
\item $-\pi/2<\theta<\pi/2$, $\Rightarrow(\cos{\theta}>0)$,\\
 Fermi arcs: 
\begin{align}
|p|^2=a^2\cos\theta\sec{(2\beta-\theta)}, \\
\cos{(2\beta-\theta)}>0\cap\sin{2\beta}<0,
\end{align} 
which are two pieces of each branch of the hyperbola with$(3\pi-2\theta)/4<\beta<\pi\cup(7\pi-2\theta)/4<\beta<2\pi$.
\item $\pi/2<\theta<3\pi/2$, $\Rightarrow(\cos{\theta}<0)$\\
 Fermi arcs: 
 \begin{align}
|p|^2=a^2\cos\theta\sec{(2\beta-\theta)}, \\
\cos{(2\beta-\theta)}<0\cap\sin{2\beta}>0,
\end{align} 
which are two pieces of each branch of the hyperbola with $(\pi<\beta<(5\pi-2\theta/4))\cup(0<\beta<(\pi-2\theta)/4)$.
\end{itemize} 

FIG. \ref{fig:11}(a,c) display the typical curves of Fermi arc belonging to the two intervals above. The Fermi arc of $Q=(1,-1)$ for the $\theta=1.49\pi$ and $\theta=1.51\pi $ which shown in FIG. \ref{fig:11}(e,f) clearly demonstrate the topological change of Fermi arc connection at $\theta=1.5\pi$ shown in FIG. \ref{fig:11}(d). 

\subsubsection{Q=(2,-2)}

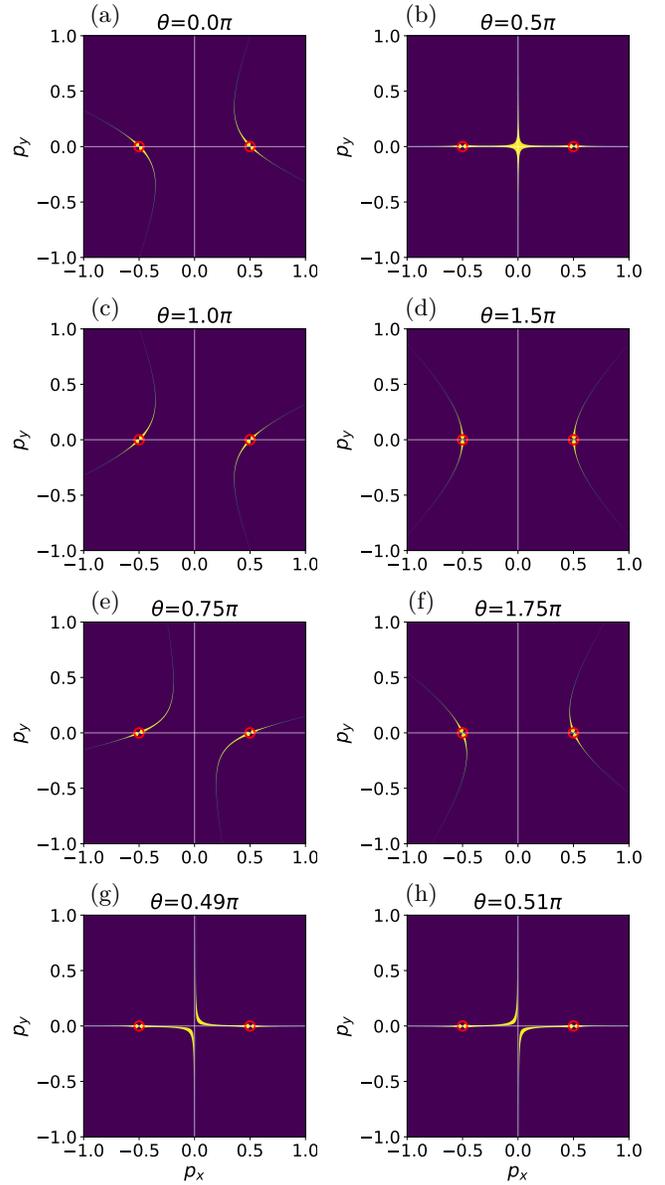
\begin{figure}[htbp]
 \begin{center}
 \begin{tikzpicture}
 \node (1) at (-1.2,0) {\includegraphics[width=14em]{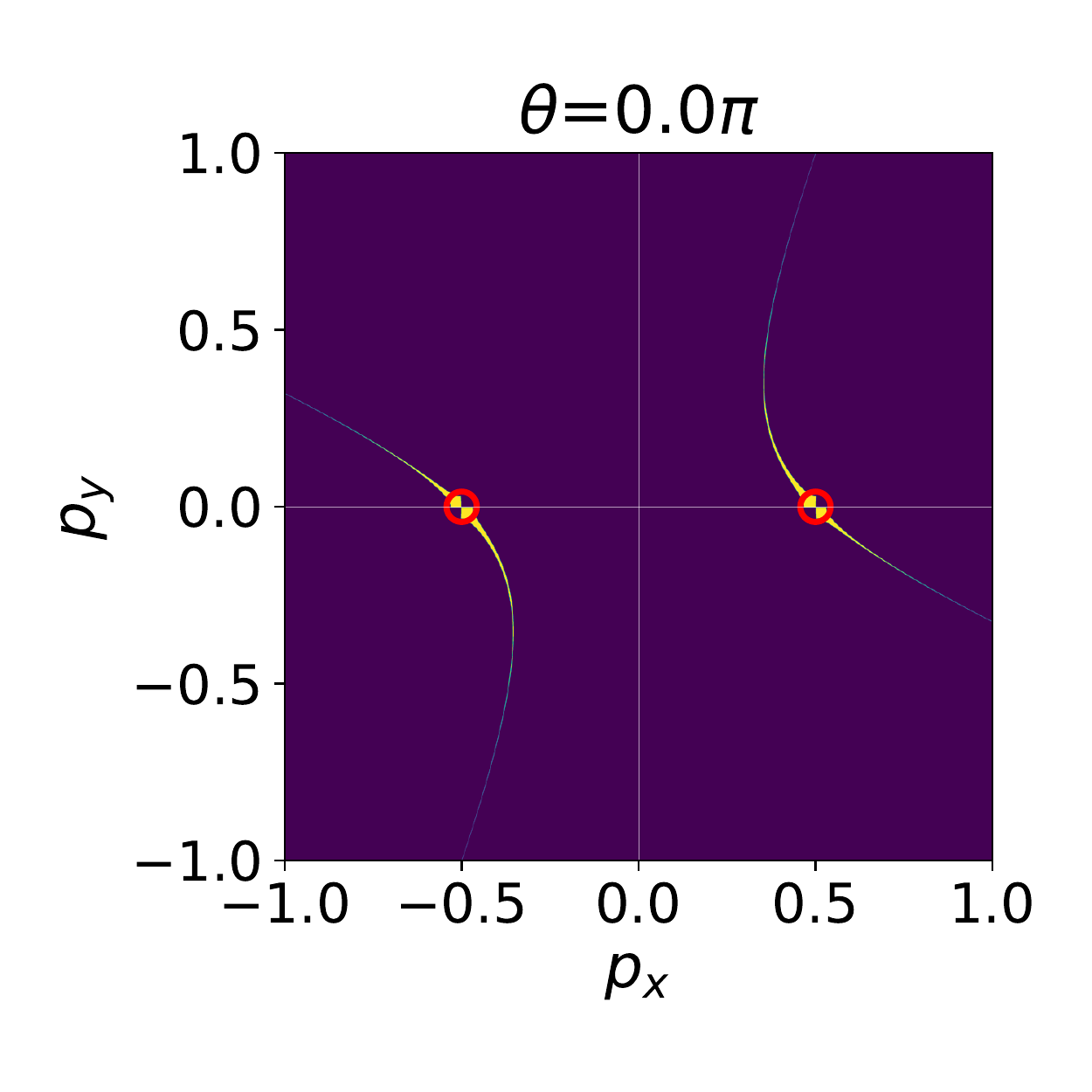}};
 \node (2) at (3.1,0) {\includegraphics[width=14em]{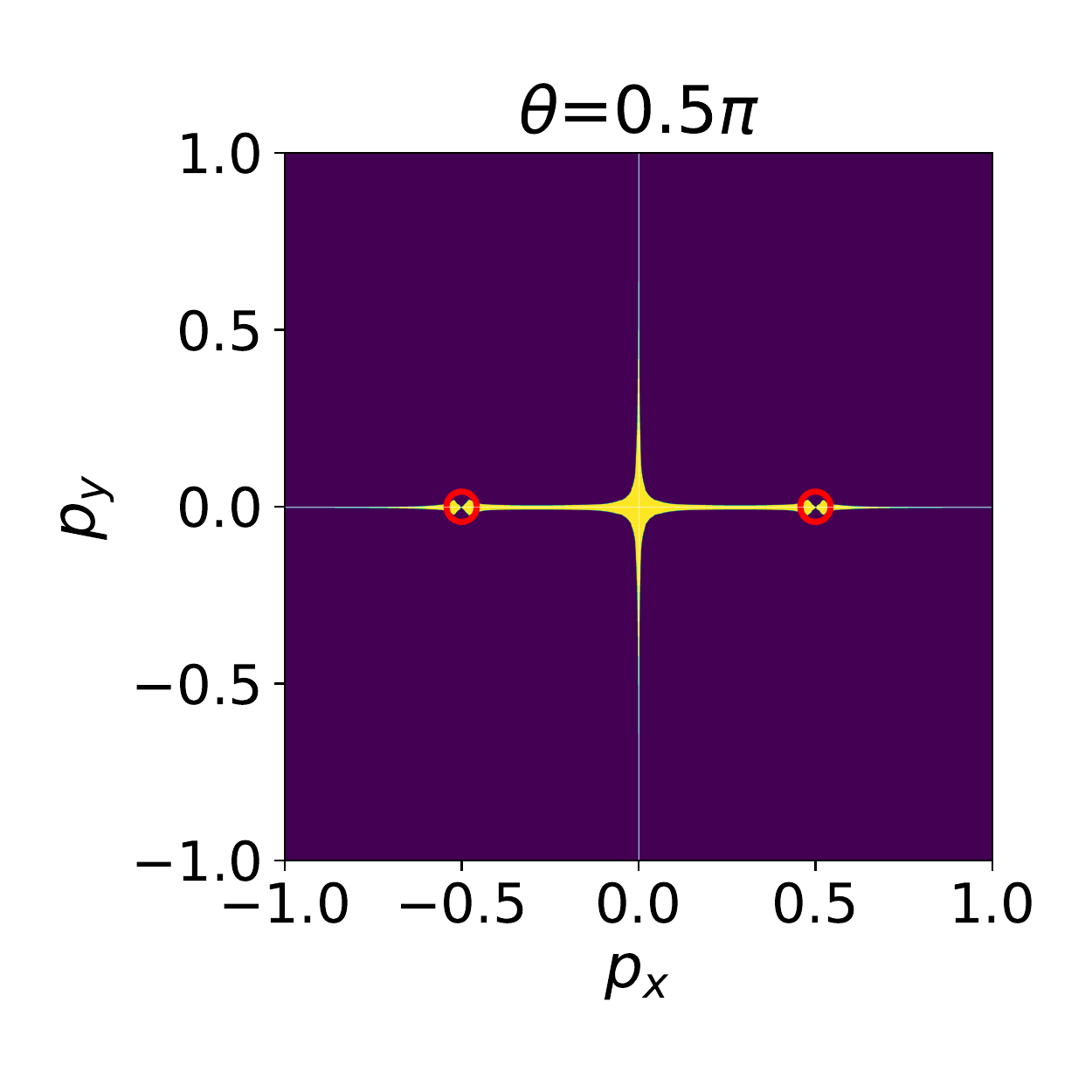}};
 \node (3) at (-1.2,-3.9) {\includegraphics[width=14em]{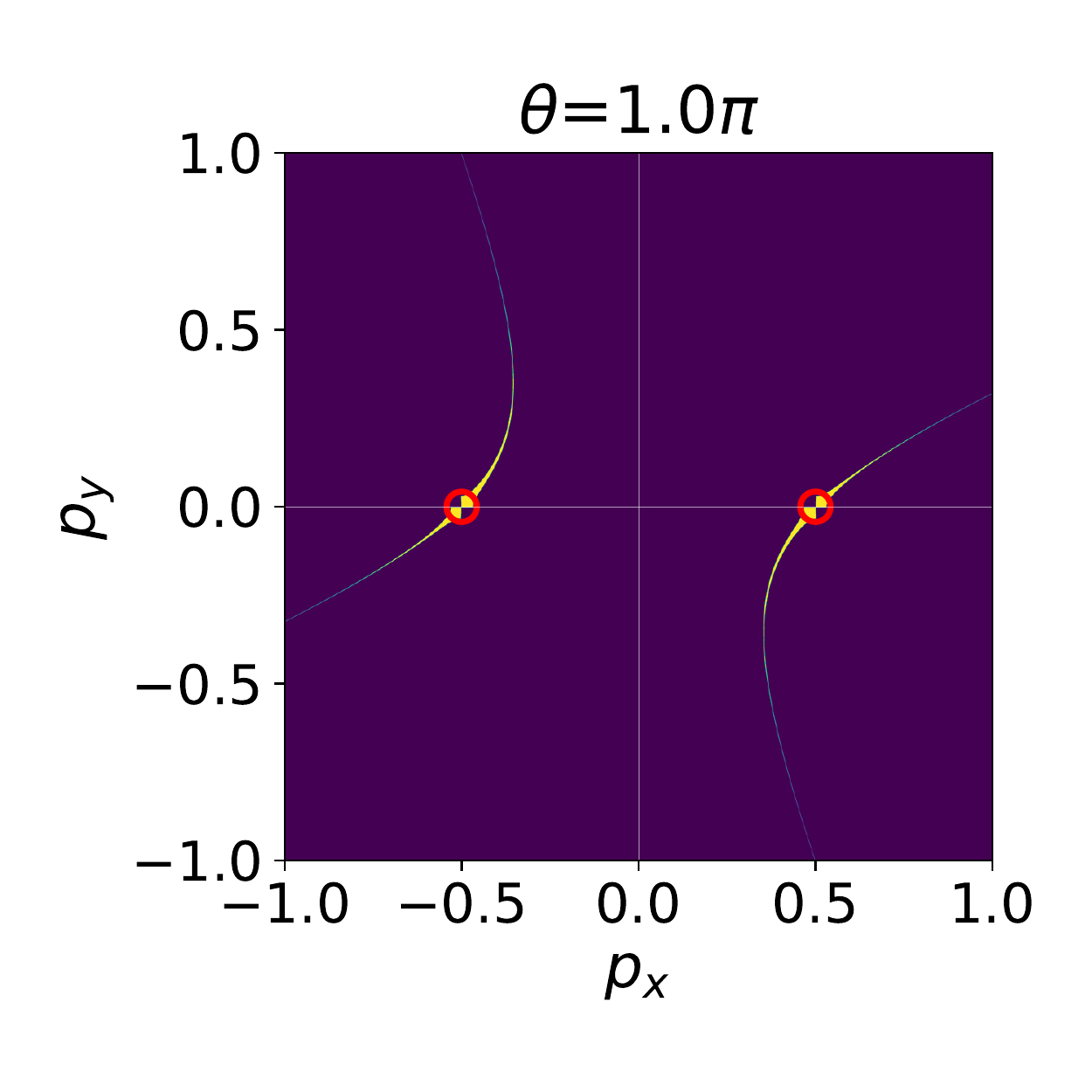}};
 \node (4) at (3.1,-3.9) {\includegraphics[width=14em]{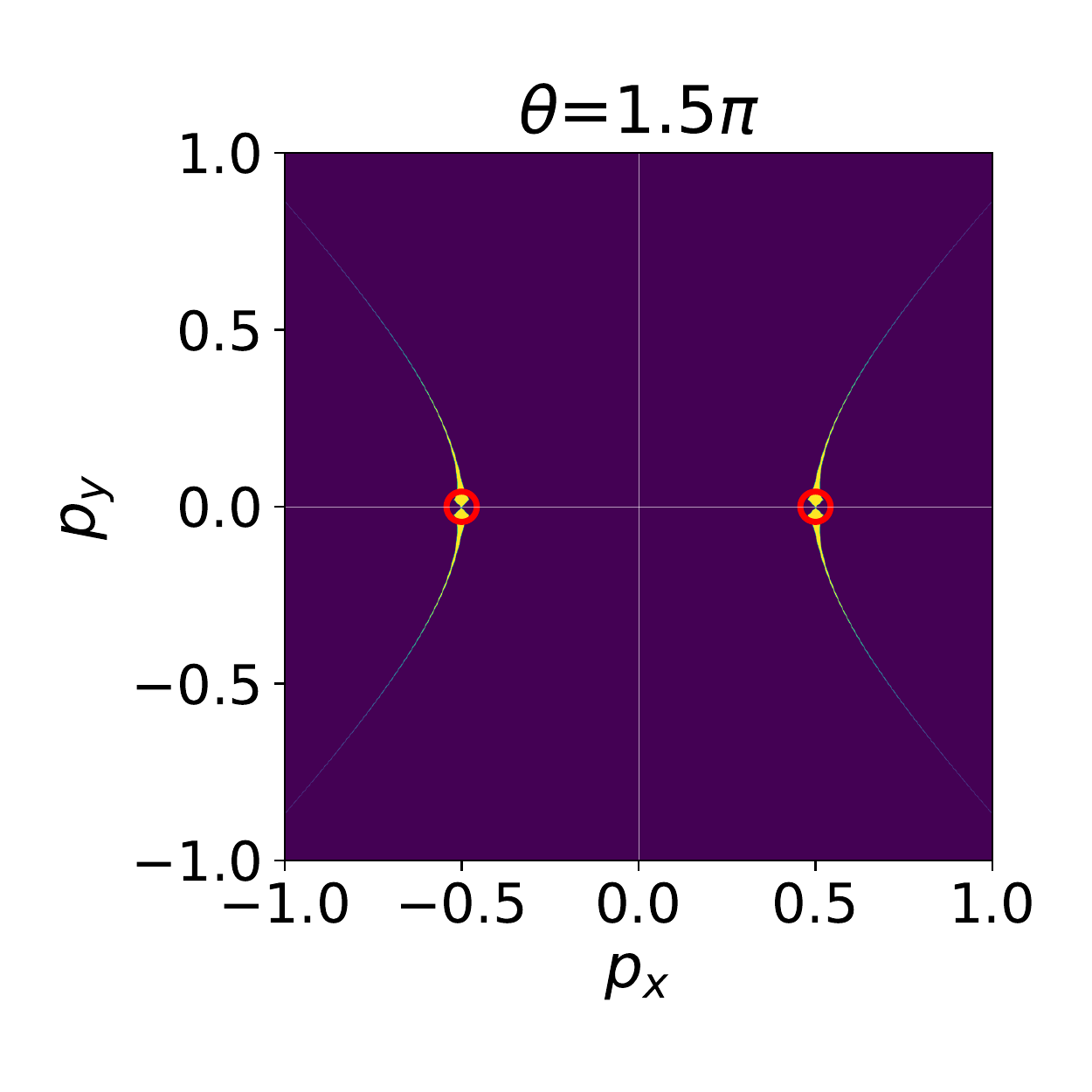}};
 \node (5) at (-1.2,-7.8) {\includegraphics[width=14em]{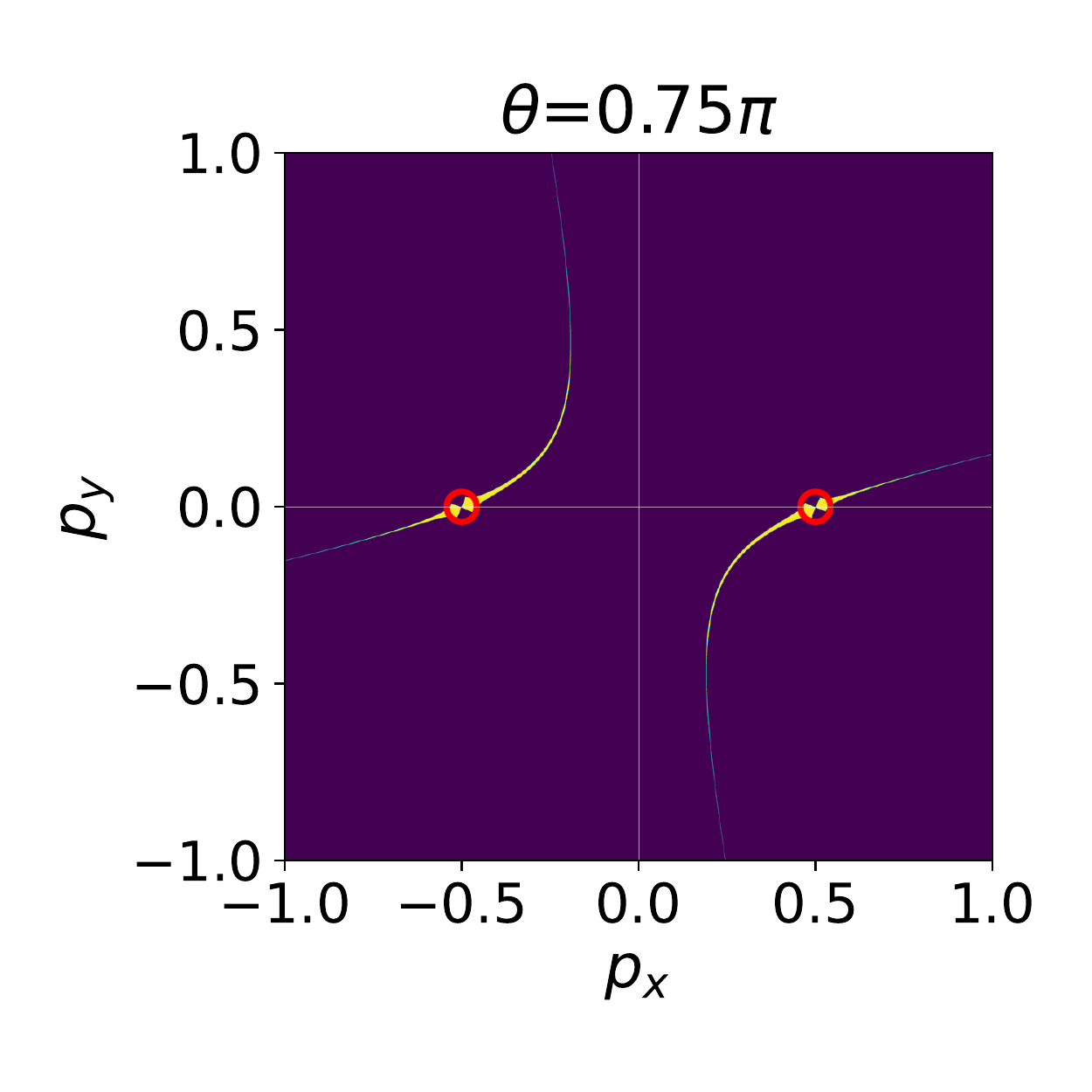}};
 \node (6) at (3.1,-7.8) {\includegraphics[width=14em]{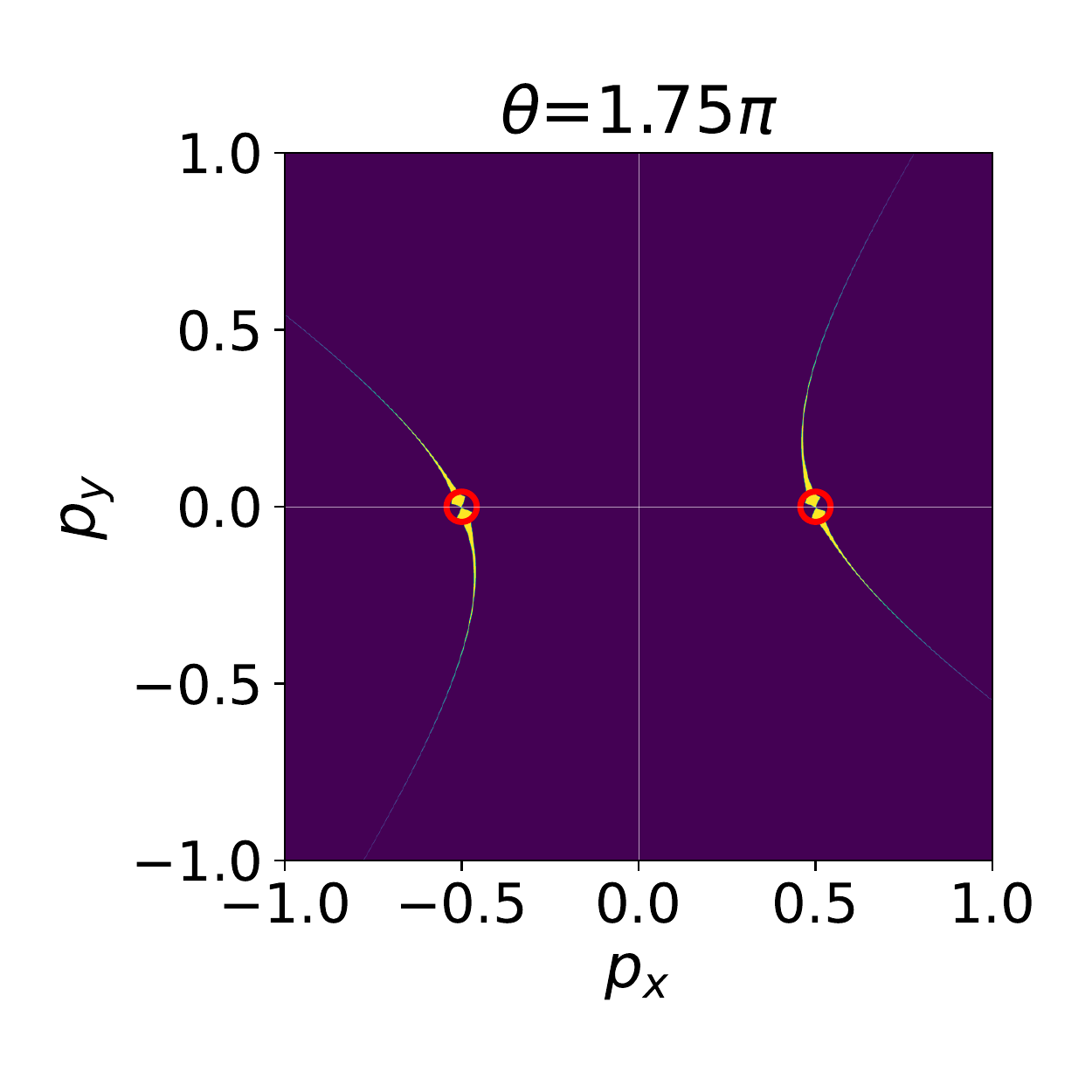}};
 \node (7) at (-1.2,-11.7) {\includegraphics[width=14em]{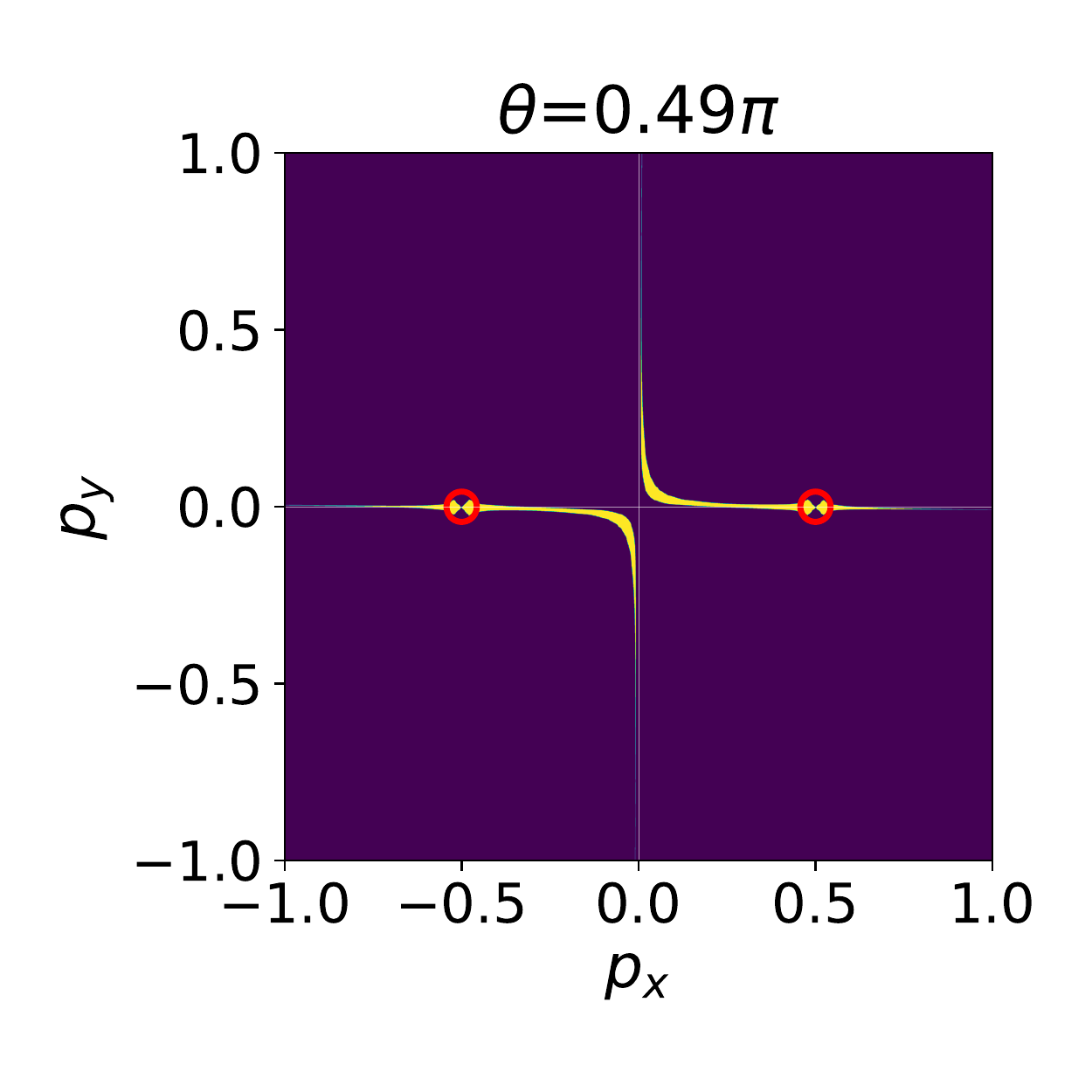}};
 \node (8) at (3.1,-11.7) {\includegraphics[width=14em]{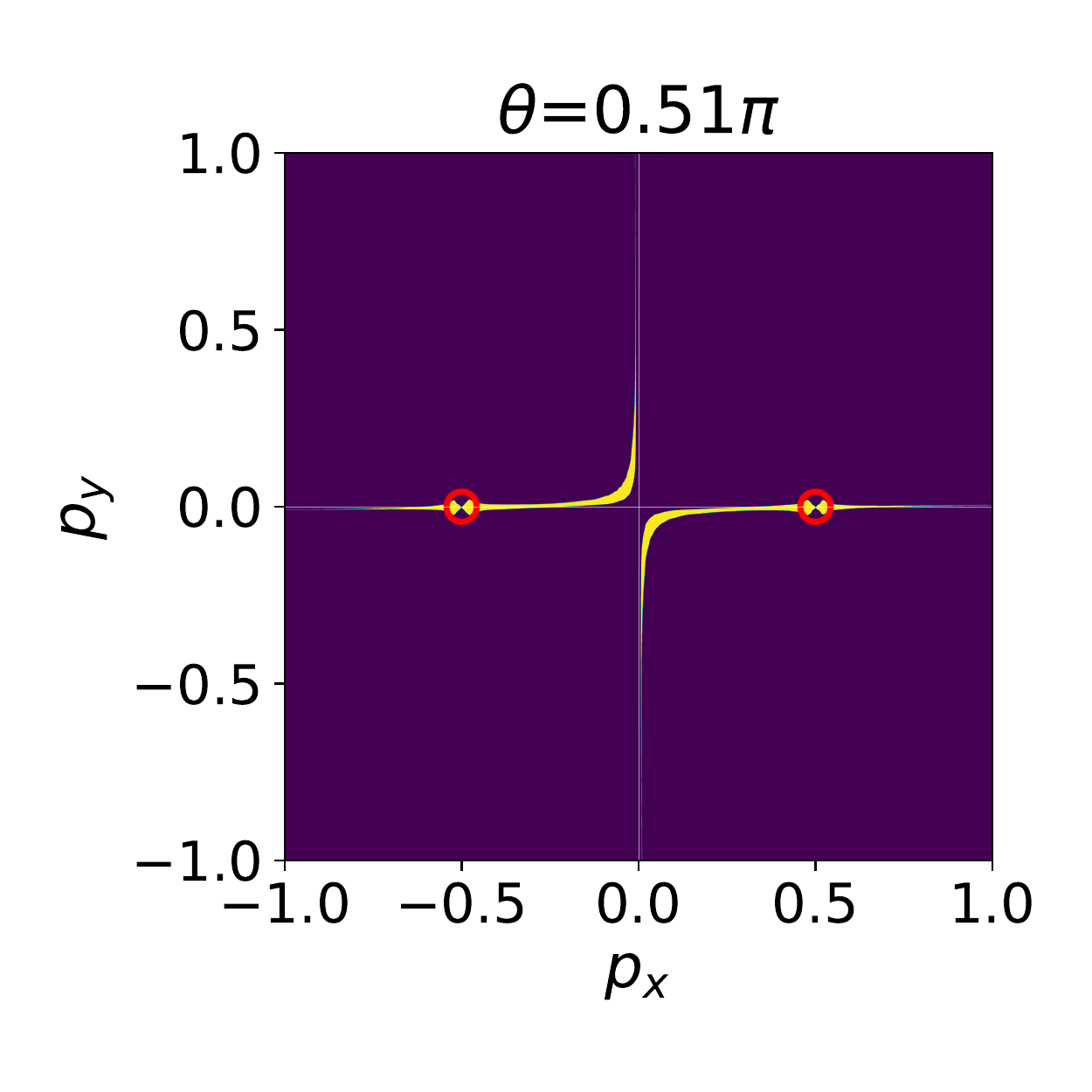}};

 \node at (-2.0,1.9) {(a)};
 \node at (2.2,1.9) {(b)};
 \node at (-2.0,-2.0) {(c)};
 \node at (2.2,-2.0) {(d)};
 \node at (-2.0,-5.9) {(e)};
 \node at (2.2,-5.9) {(f)};
 \node at (-2.0,-9.8) {(g)};
 \node at (2.2,-9.8) {(h)};
 \end{tikzpicture}
 \end{center}
 \caption{The Fermi arcs of two Weyl points $Q=(2,-2)$ (sitting at $(\pm 0.5,0)$ indicated by small red circle) for the boundary condition parameter $\theta= 0,0.5\pi,1.0\pi,1.5\pi,0.75\pi,1.75\pi,0.49\pi,0.51\pi$ are shown successively.}
 \label{fig:22}
\end{figure}

For a pair of Weyl points locating at $(\pm a,0)$ with topological charge $\pm2$, its $g(p)=(p^2-a^2)^2$. The bulk energy dispersion 
\begin{align}
E=\pm\sqrt{[(p_1^2-p_2^2-a^2)^2+(2p_1\,p_2)^2]^2+p_3^2}.
\end{align}
The corresponding energy dispersion of surface states is
\begin{align}\label{EDR11}
	\epsilon=-[(p_1^2-p_2^2-a^2)^2-(2p_1\,p_2)^2]\cos{\theta}\nonumber\\
	-4p_1p_2(p_1^2-p_2^2-a^2)\sin{\theta},
\\
	\alpha=[(p_1^2-p_2^2-a^2)^2-(2p_1\,p_2)^2]\sin{\theta}\nonumber\\
	-4p_1p_2(p_1^2-p_2^2-a^2)\cos{\theta}>0.
\end{align}
The equation of Fermi arcs is
\begin{align}\label{EDR11c}
	[(|p|^2\cos{2\beta}-a^2)^2-(|p|^2\sin{2\beta})^2]\cos{\theta}\nonumber\\ 
	+2|p|^2\sin{2\beta}(|p|^2\cos{2\beta}-a^2)\sin{\theta}=0,
 \\
	[(|p|^2\cos{2\beta}-a^2)^2-(|p|^2\sin{2\beta})^2]\sin{\theta}\nonumber\\ 
	-2|p|^2\sin{2\beta}(|p|^2\cos{2\beta}-a^2)\cos{\theta}>0,
\end{align}
where $p=|p|e^{i\beta}=\sqrt{p_1^2+p_2^2}e^{i\arg(p)}$ is used.
Let us discuss four special values of $\theta$.
\begin{itemize}
\item $\theta=\pi/2$, $\Rightarrow(\cos{\theta}=0,\sin{\theta}=1)$\\
 Fermi arcs:
\begin{align}
 (\cos{2\beta}=1\cap|p|\neq a)\cup(\cos{2\beta}=-1),
\end{align}
which are the $p_1$ axis except for the two Weyl points $\pm a$ and the whole $p_2$ axis. 
\item $\theta=3\pi/2$, $\Rightarrow(\cos{\theta}=0,\sin{\theta}=-1)$\\
 Fermi arcs: 
 \begin{align}
|p|^2=a^2\sec{2\beta}, 
\end{align} 
which is the whole hyperbola $p_1^2-p_2^2=a^2$.
\item $\theta=0$, $\Rightarrow(\cos{\theta}=1,\sin{\theta}=0)$\\
 Fermi arcs: 
\begin{align}
|p|^2=a^2\cos{(\dfrac{\pi}{4})}\sec{(2\beta-\dfrac{\pi}{4})}, 
\end{align} 
which is the whole hyperbola with $\theta=\pi/4$ in $Q=(1,-1)$ case.
\item $\theta=\pi$, then $\cos{\theta}=-1,\sin{\theta}=0$,\\
Fermi arcs: 
\begin{align}
|p|^2=a^2\cos{(-\dfrac{\pi}{4})}\sec{(2\beta+\dfrac{\pi}{4})}, 
\end{align}
which is the whole hyperbola with $\theta=-\pi/4$ in $Q=(1,-1)$ case.
\end{itemize}
The Fermi arc of $Q=(2,-2)$ for the four special cases above are shown in FIG. \ref{fig:22}(a-d). It is interesting to note that the Fermi arc with $\theta=\pi/2$ in this case is equal to that with $\theta=\pi/2$ plus $\theta=3\pi/2$ for $Q=(1,-1)$ case shown in FIG. \ref{fig:11}(b,d); while that with $\theta=3\pi/2$ in this case is equal to that with $\theta=0$ plus $\theta=\pi$ in $Q=(1,-1)$ case shown in FIG. \ref{fig:11}(a,c). For general $\theta$ in $[0,2\pi)$ except for $\pi/2$, the Fermi arcs are given by
\begin{align}
&|p|^2=a^2\cos{(\dfrac{\theta}{2}+\dfrac{\pi}{4}+k\pi)}\sec{(2\beta-\dfrac{\theta}{2}-\dfrac{\pi}{4}-k\pi)}\nonumber\\
&\cap \cos{(\dfrac{\theta}{2}+\dfrac{\pi}{4}+k\pi)}>0 ,k\in\{0,1\}.
\end{align}
FIG. \ref{fig:22}(e-h) provide the examples of Fermi arcs for $Q=(2,-2)$ in general case, in which FIG. \ref{fig:22}(g,h) demonstrate the sudden change of the connection of Fermi arcs over the $\theta=0.5\pi$.
\subsubsection{Q=(3,-3)}

For a pair of Weyl nodes locating at $(\pm a,0)$ with topological charge $\pm3$, its $g(p)=(p^2-a^2)^3$, the bulk energy dispersion is
\begin{align}
E=\pm\sqrt{[(p_1^2-p_2^2-a^2)^2+(2p_1\,p_2)^2]^3+p_3^2}.
\end{align}
The corresponding energy dispersion of surface states is
\begin{align}\label{EDR11}
	\epsilon=-(p_1^2-p_2^2-a^2)[(p_1^2-p_2^2-a^2)^3-3(2p_1\,p_2)^2]\cos{\theta}\nonumber\\ 
	-2p_1p_2[3(p_1^2-p_2^2-a^2)-(2p_1p_2)^2]\sin{\theta},\nonumber
\\
	\alpha=(p_1^2-p_2^2-a^2)[(p_1^2-p_2^2-a^2)^3-3(2p_1\,p_2)^2]\sin{\theta}\nonumber\\ 
	-2p_1p_2[3(p_1^2-p_2^2-a^2)-(2p_1p_2)^2]\cos{\theta}>0.
\end{align}
The equation of Fermi arcs is
\begin{align}\label{EDR11b1}
	(|p|^2\cos{2\beta}-a^2)[(|p|^2\cos{2\beta}-a^2)^2-3(|p|^2\sin{2\beta})^2]\cos{\theta}\nonumber\\ 
	+|p|^2\sin{2\beta}[3(|p|^2\cos{2\beta}-a^2)^2-(|p|^2\sin{2\beta})^2]\sin{\theta}=0,
 \\
	(|p|^2\cos{2\beta}-a^2)[(|p|^2\cos{2\beta}-a^2)^2-3(|p|^2\sin{2\beta})^2]\sin{\theta}\nonumber\\ 
	-|p|^2\sin{2\beta}[3(|p|^2\cos{2\beta}-a^2)^2-(|p|^2\sin{2\beta})^2]\cos{\theta}>0.
\end{align}
Let us discuss four special values of $\theta$ again.
\begin{itemize}
\item $\theta=\pi/2$, $\Rightarrow(\cos{\theta}=0,\sin{\theta}=1)$\\
 Fermi arcs:
\begin{align}
(\cos{2\beta}=1 \cap |p|> a)\cup\nonumber\\
(|p|^2=a^2\cos{(\dfrac{\pi}{6})}\sec{(2\beta\pm \dfrac{\pi}{6}})\cap(\pm\sin{(2\beta)}<0)),
\end{align}
which represent large part of $p_1$ axis with $|p|> a$ and half of the two hyperbolas $|p|^2=a^2\cos{(\dfrac{\pi}{6})}\sec{(2\beta\pm \dfrac{\pi}{6})}$. 
\item $\theta=3\pi/2$, $\Rightarrow(\cos{\theta}=0,\sin{\theta}=-1)$\\
 Fermi arcs: 
\begin{align}
[(\cos{2\beta}=1 \cap |p|< a)\cup(\cos{2\beta}=-1)] \cup\nonumber\\
(|p|^2=a^2\cos{(\dfrac{\pi}{6})}\sec{(2\beta\pm \dfrac{\pi}{6}})\cap(\pm\sin{(2\beta)}>0)),
\end{align}
which represent the small part of $p_1$ axis with $|p|< a$ and the whole $p_2$ axis as well as half pieces of the two hyperbolae $|p|^2=a^2\cos{(\dfrac{\pi}{6})}\sec{(2\beta\pm \dfrac{\pi}{6})}$. 
\item $\theta=0$, $\Rightarrow(\cos{\theta}=1,\sin{\theta}=0)$\\
 Fermi arcs:
\begin{align}
(|p|^2=a^2\sec{(2\beta)})\cap(\sin{(2\beta)}>0)\cup\nonumber\\
(|p|^2=a^2\cos{(\dfrac{\pi}{3})}\sec{(2\beta\pm \dfrac{\pi}{3}})\cap(\sin{(2\beta)}<0)),
\end{align}
which are half part of the three hyperbolas. 
\item $\theta=\pi$, $\Rightarrow(\cos{\theta}=-1,\sin{\theta}=0)$\\
 Fermi arcs: 
\begin{align}
(|p|^2=a^2\sec{(2\beta)})\cap(\sin{(2\beta)}<0)\cup\nonumber\\
(|p|^2=a^2\cos{(\dfrac{\pi}{3})}\sec{(2\beta\pm \dfrac{\pi}{3}})\cap(\sin{(2\beta)}>0)),
\end{align}
which are the other half part of the three hyperbolas. 
\end{itemize} 
The Fermi arc of $Q=(3,-3)$ for the four special cases above are shown in FIG. \ref{fig:22}(a-d). For general $\theta$, we find that Fermi arcs are
\begin{align}
&|p|^2=a^2\cos{(\dfrac{\theta+(2k+1)\pi}{3})}\sec{(2\beta-\dfrac{\theta+(2k+1)\pi}{3})},\nonumber\\
&k\in\{0,1,2\}.
\end{align}
The real Fermi arcs are only half of these three hyperbolas with the condition:
\begin{align}
&\sin{2\beta}\lessgtr 0, for \cos{(\dfrac{\theta+(2k+1)\pi}{3})}\lessgtr 0\\
&(\sin{2\beta}=0 \cap |p|\lessgtr a), for \sin{(\dfrac{\theta+(2k+1)\pi}{3})}=\mp1.
\end{align}
We find that the critical point of Lifshitz phase transition of Fermi arcs presents at $\theta=1.5\pi$($\theta=0.5\pi$) for odd (even) winding number $w$ and thus we can discuss single-pair Weyl nodes in even and odd cases seperately.
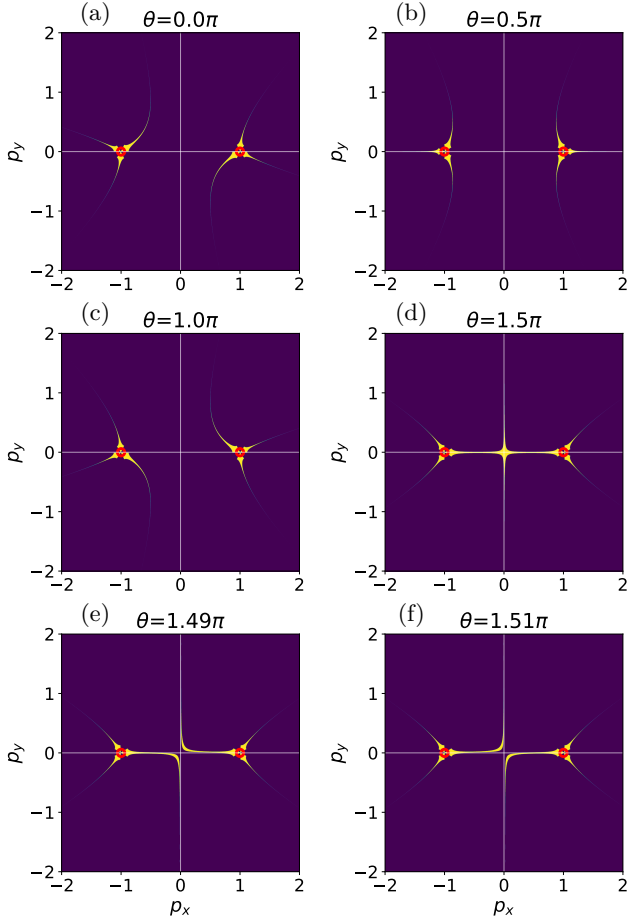
\begin{figure}[htbp]
 \begin{center}
 \begin{tikzpicture}
 \node (1) at (-1.2,0) {\includegraphics[width=14em]{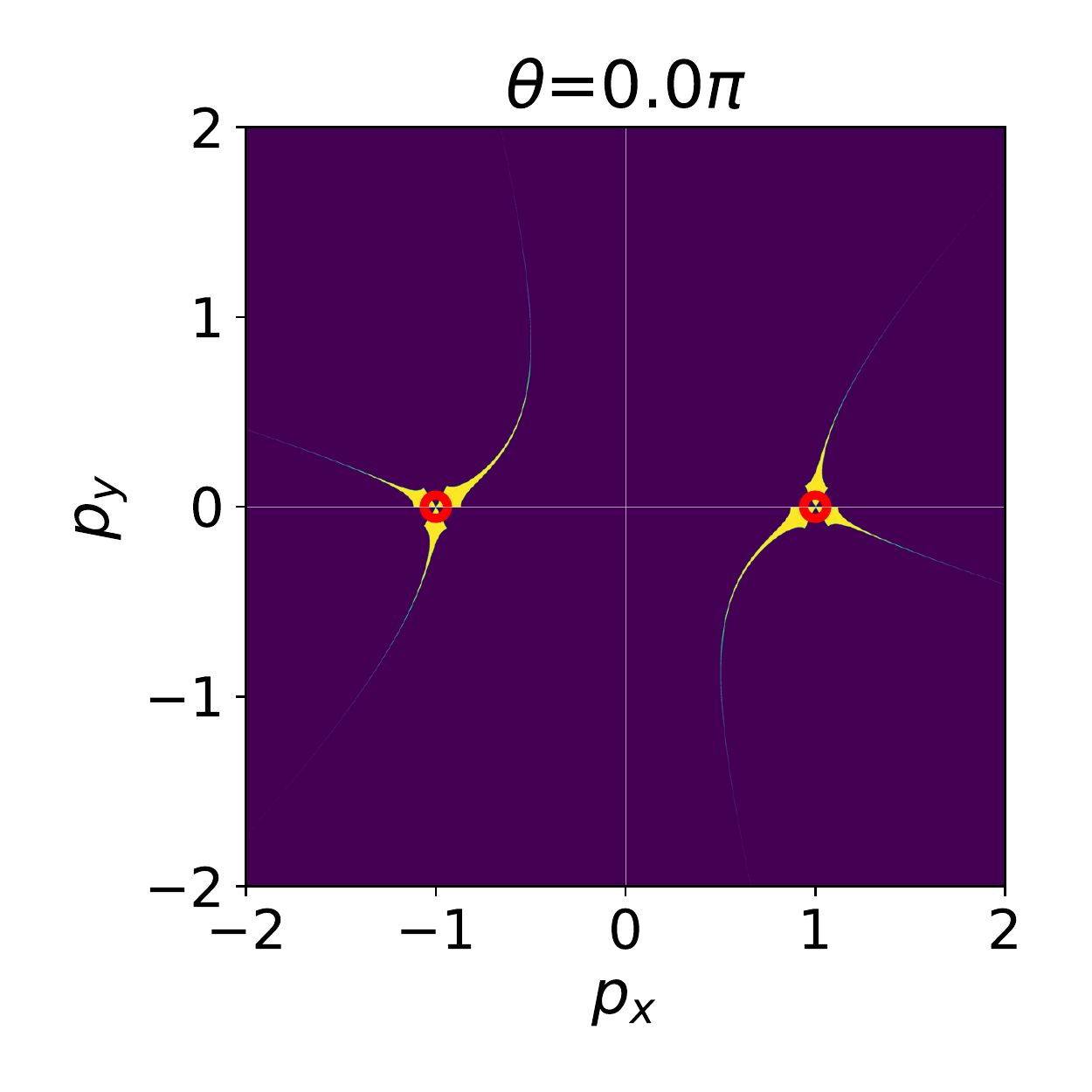}};
 \node (2) at (3.1,0) {\includegraphics[width=14em]{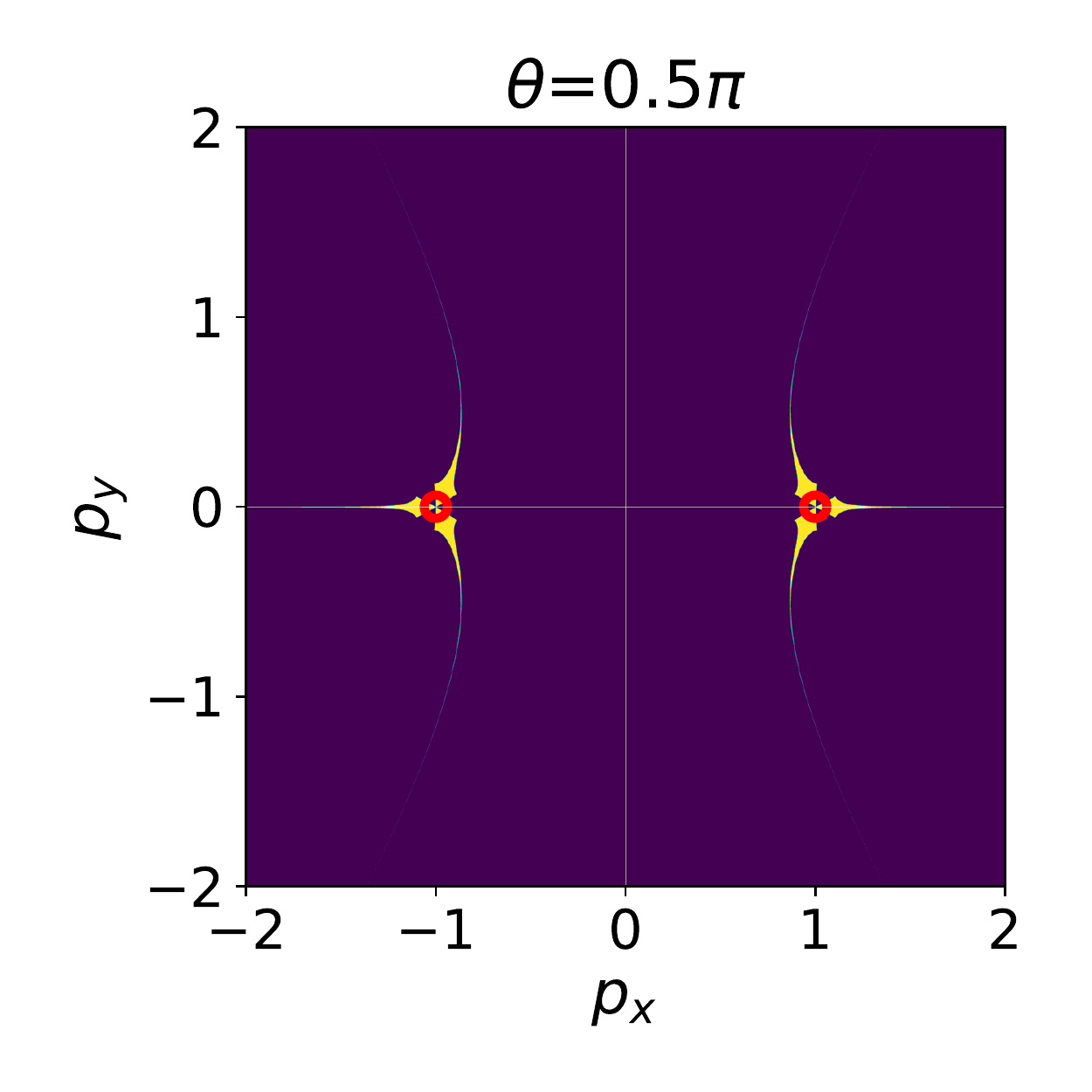}};
 \node (3) at (-1.2,-4.0) {\includegraphics[width=14em]{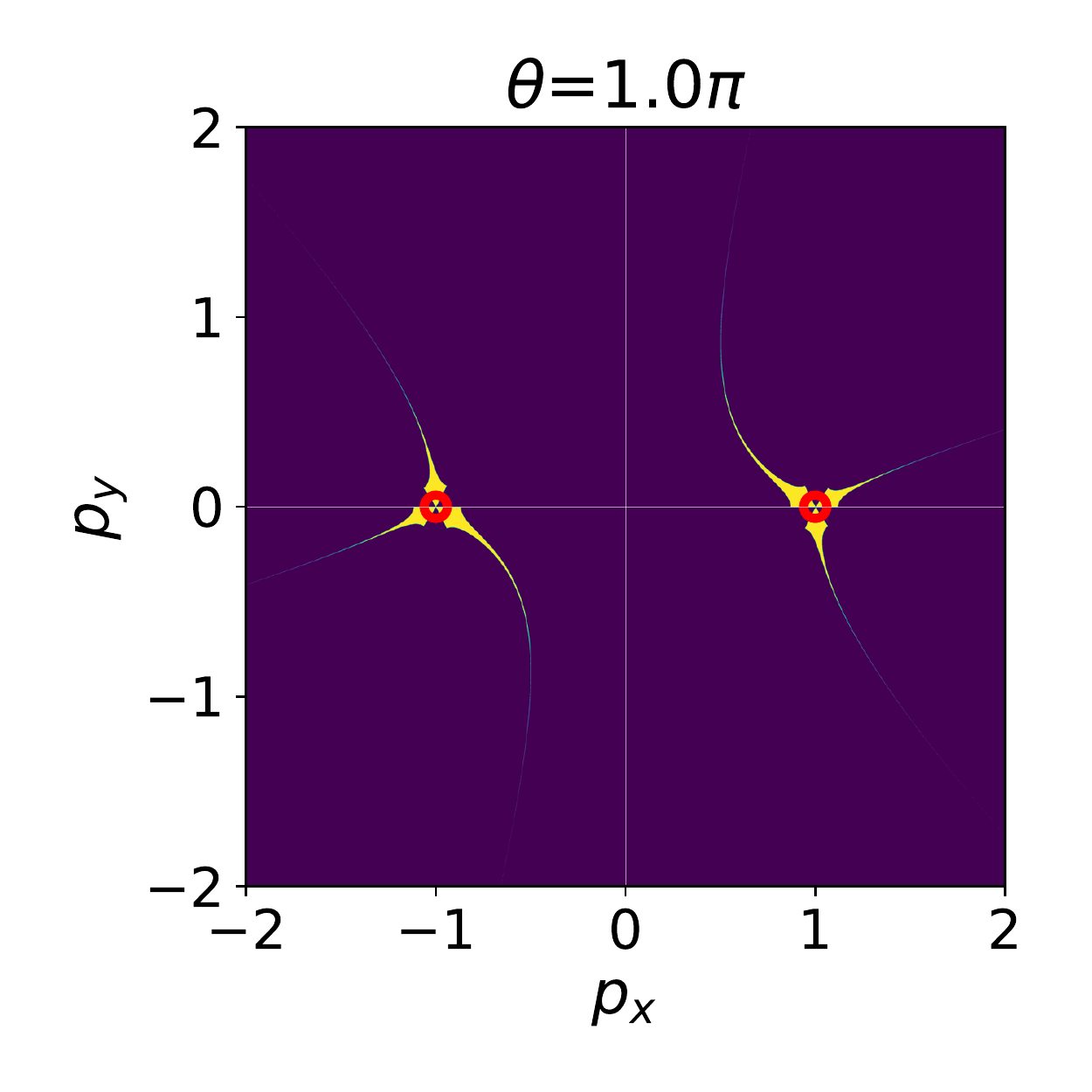}};
 \node (4) at (3.1,-4.0) {\includegraphics[width=14em]{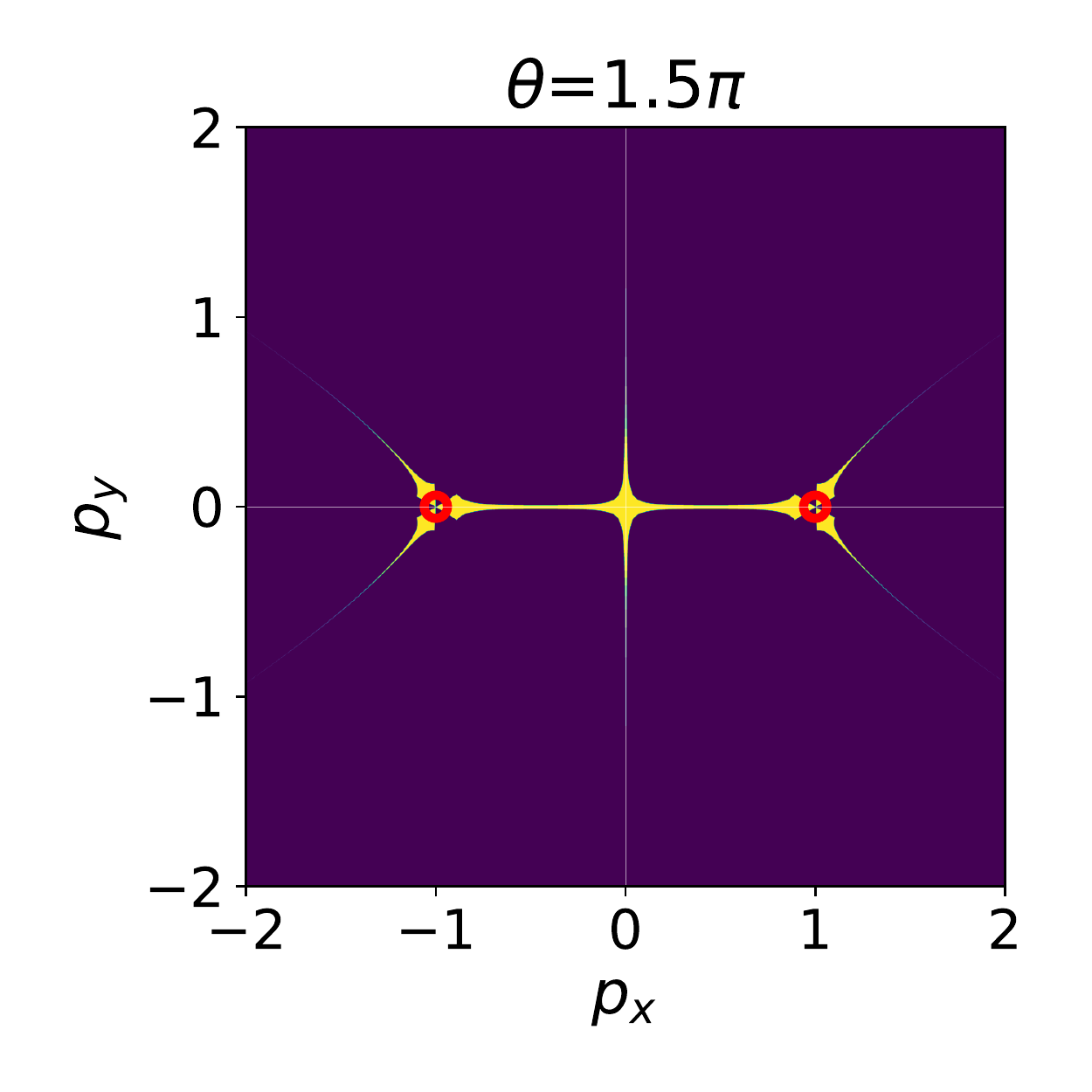}};
 \node (5) at (-1.2,-8.0) {\includegraphics[width=14em]{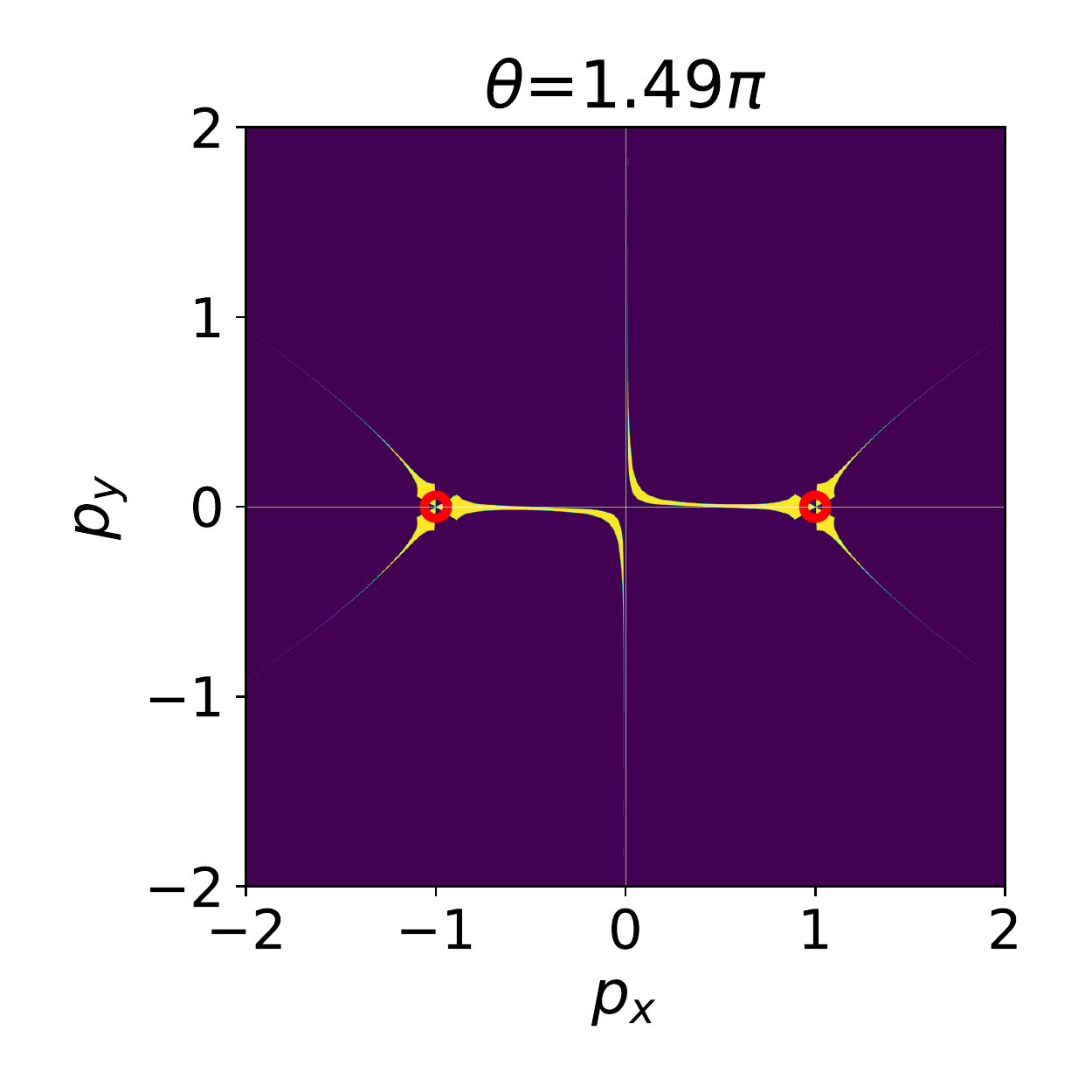}};
 \node (6) at (3.1,-8.0) {\includegraphics[width=14em]{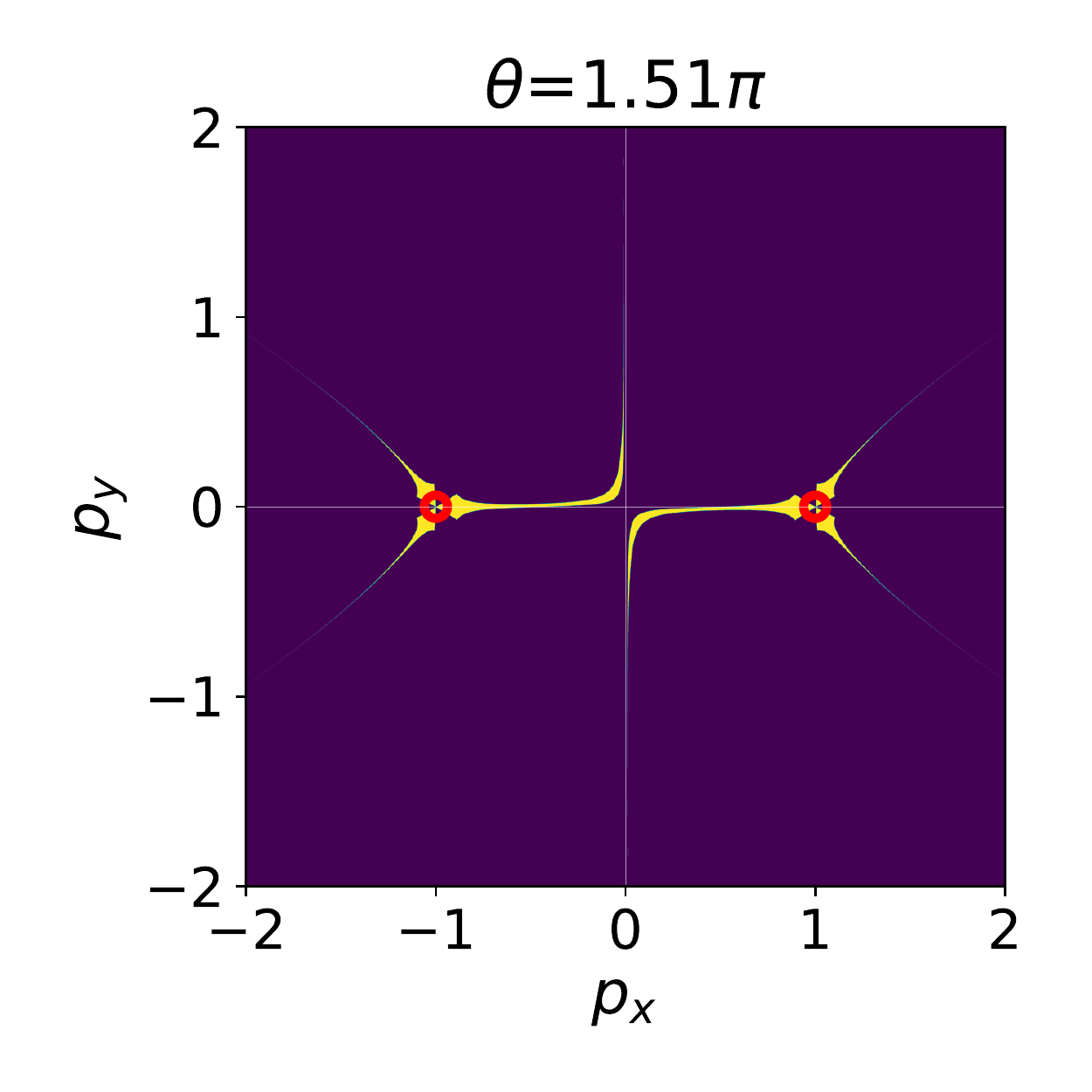}};
 \node at (-2.0,2) {(a)};
 \node at (2.2,2) {(b)};
 \node at (-2.0,-2.0) {(c)};
 \node at (2.2,-2.0) {(d)};
 \node at (-2.0,-6) {(e)};
 \node at (2.2,-6) {(f)};
 \end{tikzpicture}
 \end{center}
 \caption{The Fermi arcs of two Weyl points $Q=(3,-3)$ (sitting at $(\pm 0.5,0)$ indicated by small red circle) for the boundary condition parameter $\theta= 0,0.5\pi,1.0\pi,1.5\pi,1.49\pi,1.51\pi$ are shown successively.}
 \label{fig:33}
\end{figure}

\subsubsection{Q=($w$,$-w$)}

Notice that $g(p)=(p^2-a^2)^w $ can be regard as a function of function. If we define $f(p)=p^w$, then we have $g(p)=(p^2-a^2)^w=f(p^2-a^2)$. Thus we can obtain the Fermi arcs for Q=($w$,$-w$) case from that of multi-Weyl case with $g(p)=p^w$ and that of Q=(1,-1) case with $g(p)=p^2-a^2$. On the other hand, we have found that the Fermi arcs of single pair Weyl semimetal are generally several half or whole hyperbolas depending on the parity of the winding number $w$ of Weyl nodes. Thus we should explore the generic case by dividing even $w=2m$ and odd $w=2m+1$ cases with $m$ the non-negative integer.

\subsubsection*{a. even case: Q=($2m$,$-2m$)} 
In this case $g(p)=(p^2-a^2)^{2m}$ and the Fermi arcs in general case ($\cos{[\dfrac{\theta+(2k+\dfrac{3}{2}-m)\pi}{2m}]}\neq 0$) are intact hyperbolas.
\begin{align}
&|p|^2=a^2\cos{[\dfrac{\theta+(2k+\dfrac{3}{2}-m)\pi}{2m}]}\sec{[2\beta-\dfrac{\theta+(2k+\dfrac{3}{2}-m)\pi}{2m}]},\nonumber\\
&k\in\{0,1,2,\cdots,(2m-1)\}.
\end{align}
\subsubsection*{b. odd case: Q=($2m+1$,$-(2m+1)$)} 
In this case $g(p)=(p^2-a^2)^{2m+1}$ and the Fermi arcs are half of hyperbolas.
\begin{align}
&|p|^2=a^2\cos{[\dfrac{\theta+(2k+2-m)\pi}{2m+1}]}\sec{[2\beta-\dfrac{\theta+(2k+2-m)\pi}{2m+1}]},\nonumber\\
&k\in\{0,1,2,\cdots,(2m)\},
\end{align}
with the conditions:
\begin{align}
&\sin{2\beta}\lessgtr 0, for \cos{[\dfrac{\theta+(2k+2-m)\pi}{2m+1}]}\lessgtr 0,\\
&(\sin{2\beta}=0 \cap |p|\lessgtr a), for \sin{[\dfrac{\theta+(2k+2-m)\pi}{2m+1}]}=\mp1.
\end{align}
We find that there are two topologically different phases for single-pair Weyl semimetals: one is the phase with only single Fermi arc connected the two Weyl nodes in the projected momentum space, the other phase without any Fermi arc that connected the Weyl nodes. The condition for the Fermi arc to connect this pair of Weyl nodes is 
\begin{align}
\cos{[\dfrac{\theta+(2k+\dfrac{3}{2}-\dfrac{w}{2})\pi}{w}]}=0, k\in\{0,1,2,\cdots,(w-1)\}
\end{align}
\subsection{Double-pairs of Weyl nodes} 
To see more complex Fermi arcs pattern, we then investigate double-pairs Weyl nodes situation.
\subsubsection{$Q=(1,-1,1,-1)$}
\begin{figure}[htbp]
 \begin{center}
 \begin{tikzpicture}
 \node (1) at (-1.2,0) {\includegraphics[width=14em]{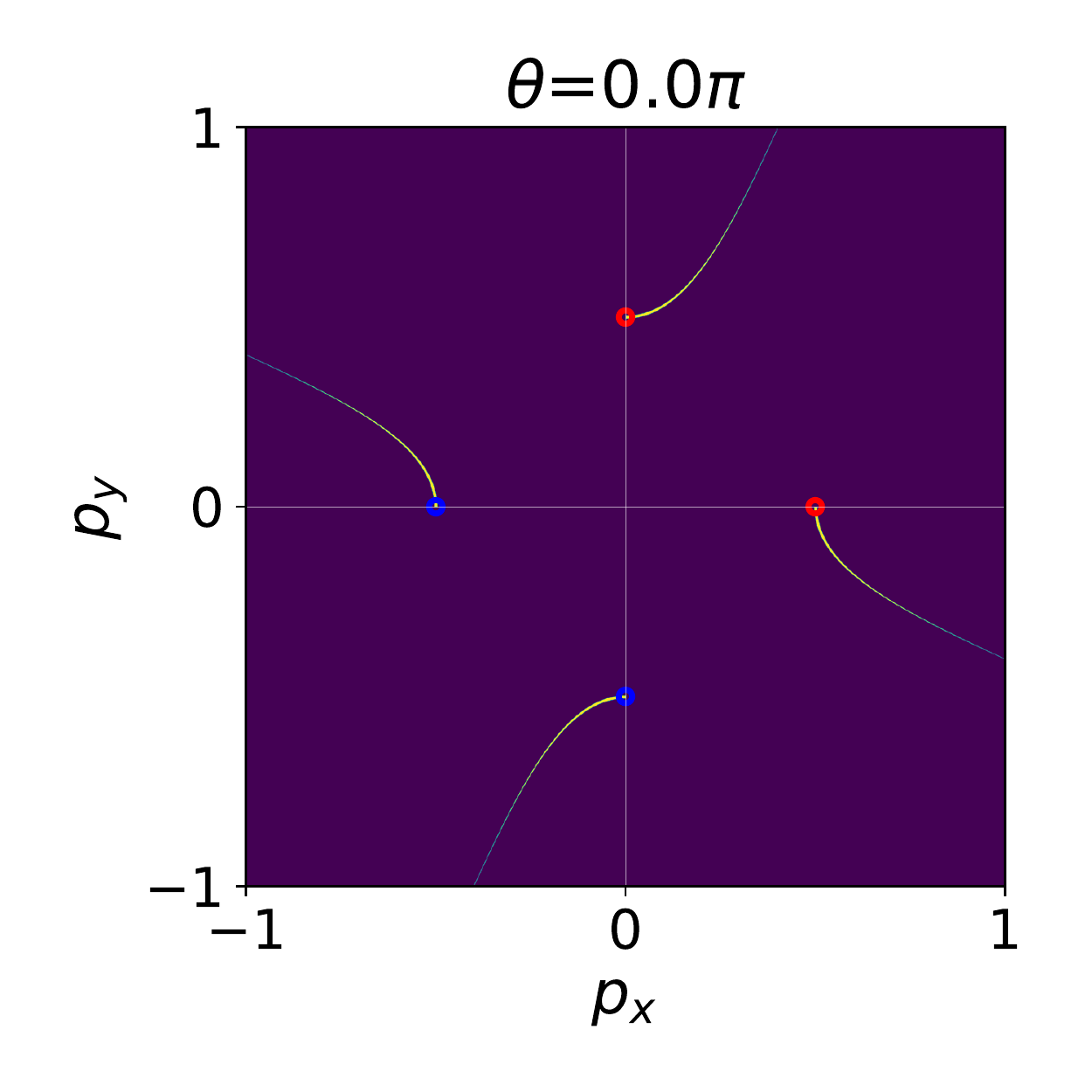}};
 \node (2) at (3.1,0) {\includegraphics[width=14em]{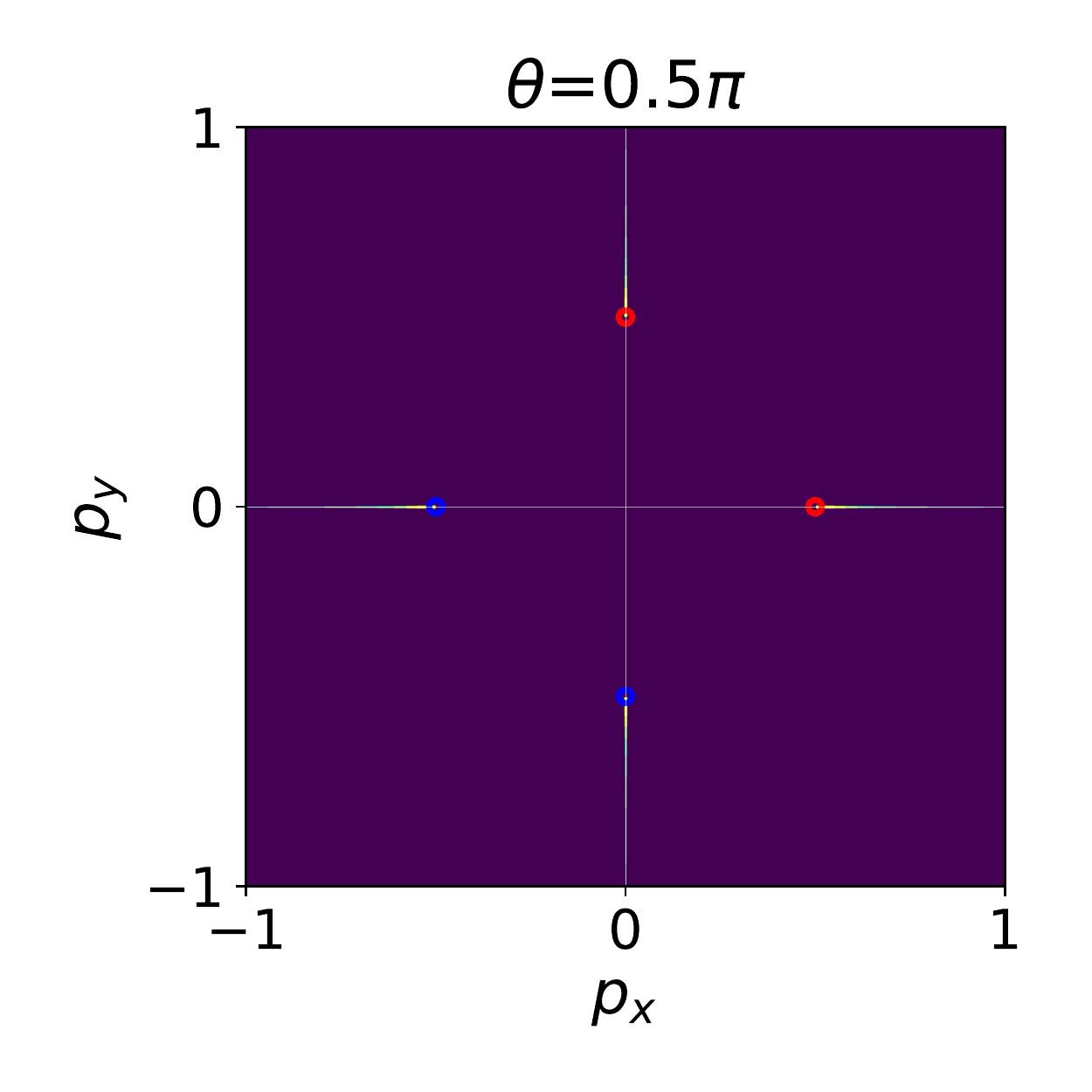}};
 \node (3) at (-1.2,-4) {\includegraphics[width=14em]{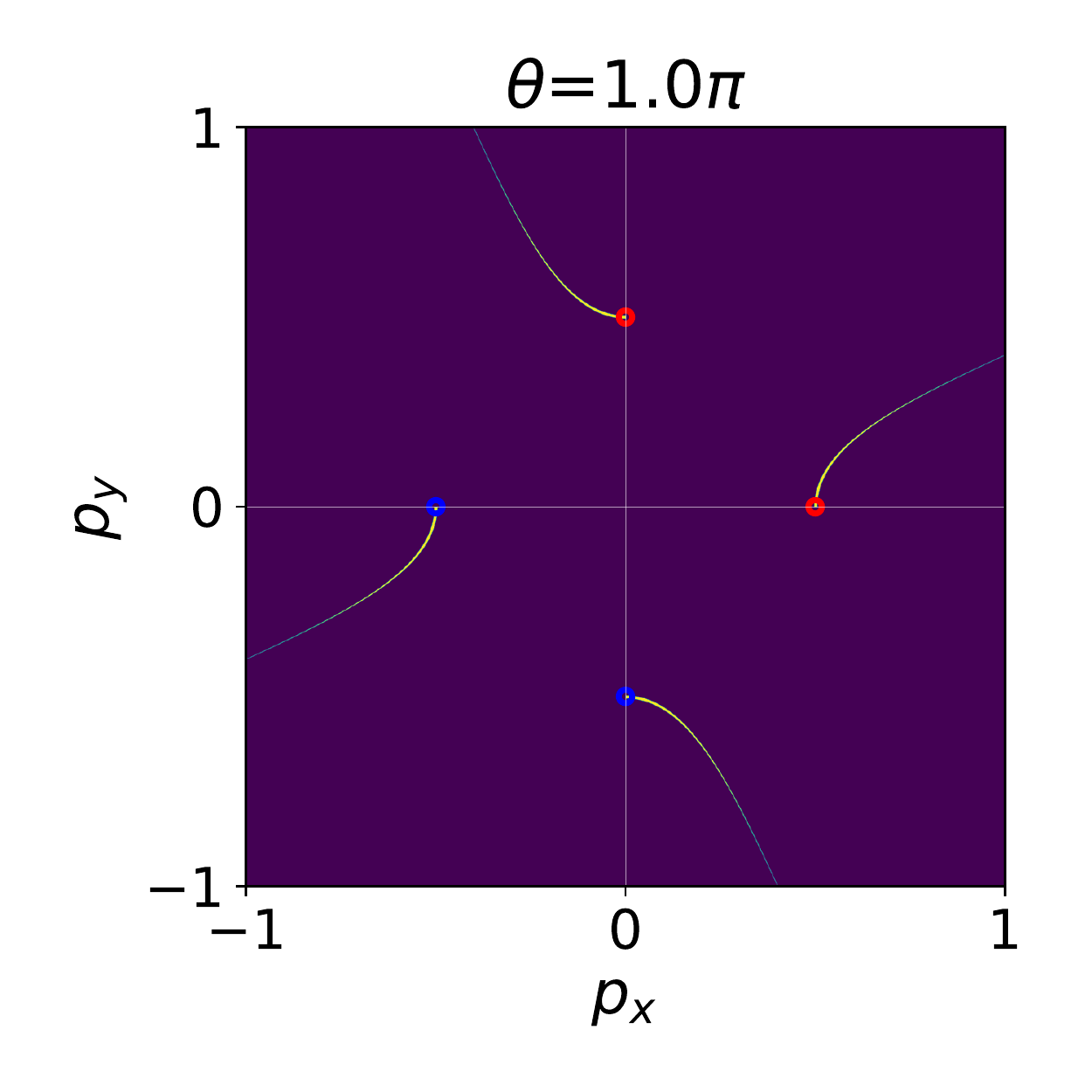}};
 \node (4) at (3.1,-4) {\includegraphics[width=14em]{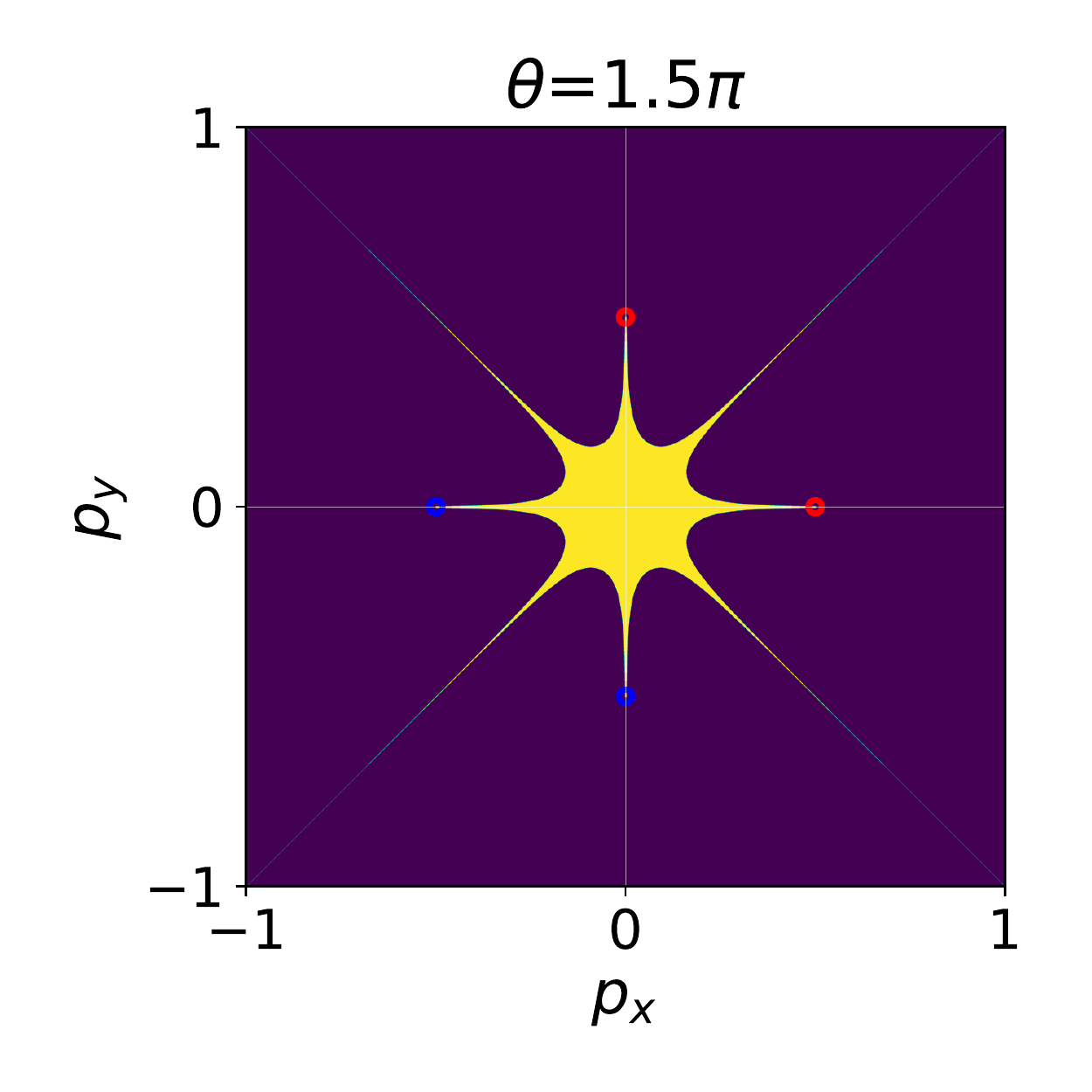}};
 \node (5) at (-1.2,-8) {\includegraphics[width=14em]{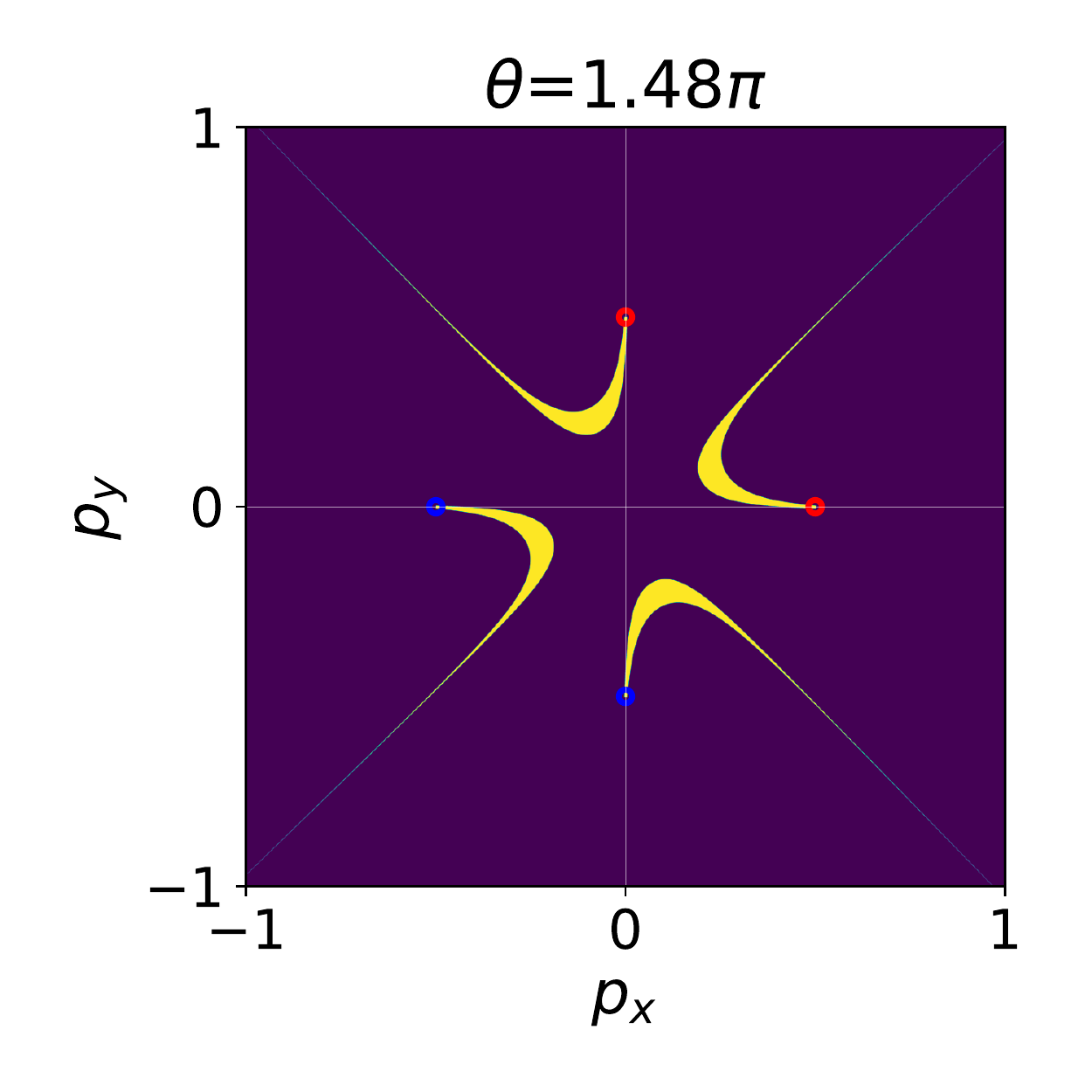}};
 \node (6) at (3.1,-8) {\includegraphics[width=14em]{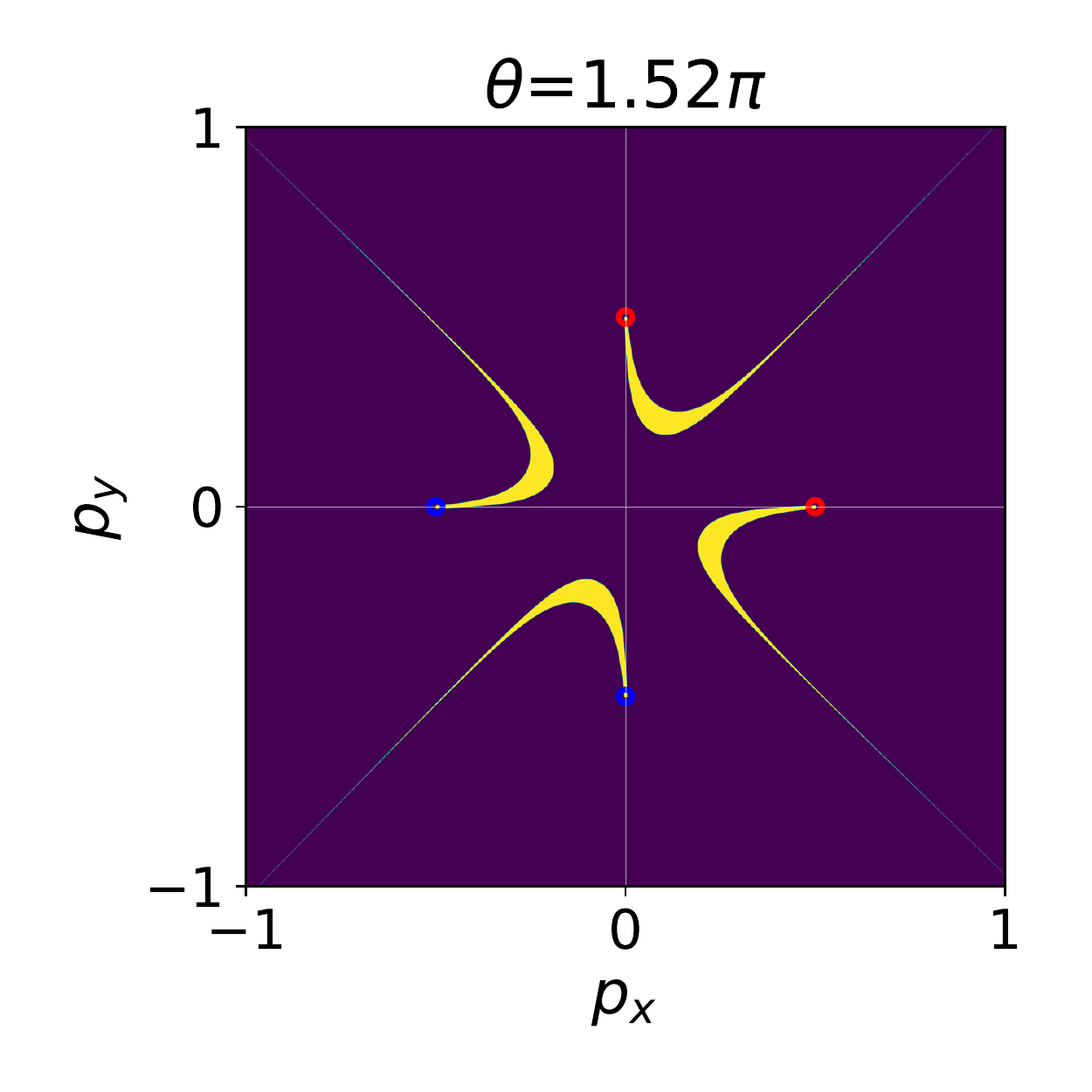}};

 \node at (-2.0,2) {(a)};
 \node at (2.2,2) {(b)};
 \node at (-2.0,-2.0) {(c)};
 \node at (2.2,-2.0) {(d)};
 \node at (-2.0,-6) {(e)};
 \node at (2.2,-6) {(f)};
 \end{tikzpicture}
 \end{center}
 \caption{The Fermi arcs of four Weyl points $Q=(1,-1,1,-1)$ (sitting at $(\pm 0.5,0)$ and $(0,\pm 0.5)$ indicated by small red and blue circles) for the boundary condition parameter $\theta= 0,0.5\pi,1.0\pi,1.5\pi,1.48\pi,1.52\pi$ are shown successively.}
 \label{fig:1111}
\end{figure}
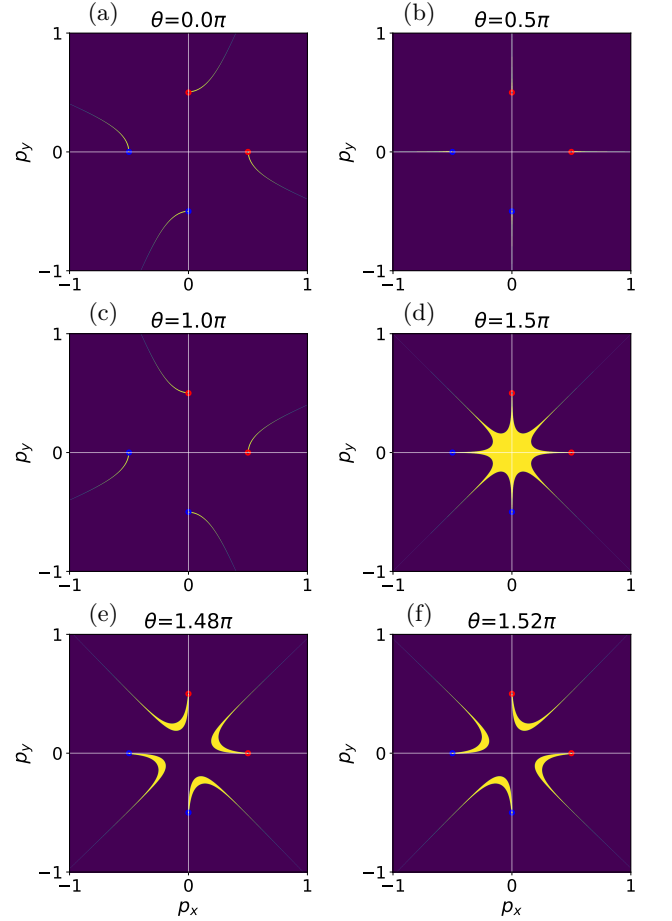
For two pairs of Weyl nodes locating at $(\pm a,0)$ and $(0,\pm a)$ with topological charge $\pm1$, its $g(p)=(p^2-a^2)(p^2+a^2)=p^4-a^4$, the real and imaginary parts of $g(p)$ are:
\begin{align}
&\Re{g(p)}=(p_1^2-p_2^2)^2-(2p_1\,p_2)^2-a^4=|p|^4\cos{4\beta}-a^4, \\
&\Im{g(p)}=4p_1\,p_2(p_1^2-p_2^2)=|p|^4\sin{4\beta},
\end{align}
where $p=|p|e^{i\beta}=\sqrt{p_1^2+p_2^2}e^{iarg(p)}$ is also used.
The bulk energy dispersion in this case is 
\begin{align}
E=\pm\sqrt{(|p|^4\cos{4\beta}-a^4)^2+(|p|^4\sin{4\beta})^2+p_3^2}.
\end{align}
The corresponding energy dispersion of surface states is
\begin{align}\label{EDR1111}
	\epsilon=-(|p|^4\cos{4\beta}-a^4)\cos{\theta}-|p|^4\sin{4\beta}\sin{\theta},
\\
	\alpha=(|p|^4\cos{4\beta}-a^4)\sin{\theta}-|p|^4\sin{4\beta}\cos{\theta}>0.	
\end{align}
The Fermi arcs are
\begin{align}\label{EDR1111b}
(|p|^4\cos{4\beta}-a^4)\cos{\theta}+|p|^4\sin{4\beta}\sin{\theta}=0,
 \\
(|p|^4\cos{4\beta}-a^4)\sin{\theta}-|p|^4\sin{4\beta}\cos{\theta}>0.	
\end{align}
To compare with the case $Q=(1,-1)$, we itemize four special values of $\theta$:
\begin{itemize}
\item $\theta=\pi/2$, $\Rightarrow(\cos{\theta}=0,\sin{\theta}=1)$\\
 Fermi arcs:
\begin{align}
 \sin{4\beta}=0\cap(|p|^4\cos{4\beta}-a^4)>0,
\end{align}
which are four rays along the $p_1$ and $p_2$ axes from the four Weyl nodes $(\pm a,0)$ and $(0,\pm a)$ to infinite, respectively. 
\item $\theta=3\pi/2$, $\Rightarrow(\cos{\theta}=0,\sin{\theta}=-1)$\\
 Fermi arcs: 
 \begin{align}
 \sin{4\beta}=0\cap(|p|^4\cos{4\beta}-a^4)<0, 
\end{align}
which constitutes two line segments and two direct lines both crossing at the Weyl points. In first situation, $\cos{4\beta}=1$, the corresponding Fermi arcs connect two pairs of Weyl points $(\pm a,0)$ and $(0,\pm a)$ along the $p_1$ and $p_2$ axes, respectively; in second case, $\cos{4\beta}=-1$, which corresponds to the two diagonal lines of the $p_1-p_2$ plane.
\item $\theta=0$, $\Rightarrow(\cos{\theta}=1,\sin{\theta}=0)$\\
 Fermi arcs:
\begin{align}
|p|^4=a^4\sec{(4\beta)}, \label{heterpola}\\
\sin{4\beta}<0,
\end{align}
where the equation represents four "compressed" hyperbolae along the $p_1$ and $p_2$ axes from the four Weyl nodes $(\pm a,0)$ and $(0,\pm a)$ to infinite, respectively, which one may call it as "quartibola" since it is quartic curves including four branches; while the inequality further excluding half of each branch. 
\item $\theta=\pi$, $\Rightarrow(\cos{\theta}=-1,\sin{\theta}=0)$\\
 Fermi arcs: 
 \begin{align}
|p|^4=a^4\sec{(4\beta)}\cap\sin{4\beta}>0,
\end{align}
which is the other half of the quartibola \eqref{heterpola}. 
\end{itemize}
The Fermi arc of $Q=(1,-1,1,-1)$ for the four special cases above are shown in FIG. \ref{fig:1111}(a-d). One can find that the Fermi arcs in double-pairs of Weyl nodes are two copies of that for $Q=(1,-1)$ along $p_x$ and $p_y$ axes except for $\theta=\pi/2$ and $3\pi/2$. The Fermi arcs for $\theta=1.5\pi$ should be two crosses, as demonstrated above. While the error of the Fermi arcs curves around the origin shown in FIG. \ref{fig:1111}(d) actually arises from the algorithm using in our plotting program. The Fermi arcs of $Q=(1,-1,1,-1)$ shown in FIG. \ref{fig:1111}(e,f) for $\theta=1.48\pi$ and $1.52\pi$ can help us to identify the connection variation of Fermi arcs crossing over $\theta=1.5\pi$ .

For general $\theta$ except for $\theta=\pi/2$ and $3\pi/2$, we find that Fermi arcs become
\begin{eqnarray}
|p|^4=a^4\cos\theta\sec{(4\beta-\theta)}, \label{HPPL41}\\
\cos{\theta}\tan{(4\beta-\theta)}+\sin{\theta}<0.\label{HPPL42}
\end{eqnarray}
The equation \eqref{HPPL41} represents two tilt quartibola rotating $\theta/4$ counter clockwisely from the two compressed conjugate hyperbolae $p_1^2-p_2^2=a^2$ with vertices $a$ shorten as $a\sqrt{|\cos\theta|}$, while the inequality \eqref{HPPL42} further selects the half of each branches of these hyperbolae. To be concretely, we item them in two situations:
\begin{itemize}
\item $-\pi/2<\theta<\pi/2$, $\Rightarrow(\cos{\theta}>0)$,\\
 Fermi arcs: 
\begin{align}
|p|^4=a^4\cos\theta\sec{(4\beta-\theta)}, \\
\cos{(4\beta-\theta)}>0\cap\sin{4\beta}<0,
\end{align} 
which is the rotating squeezed half quartibola from that of $(\theta=\dfrac{\pi}{2})$ by angle $\theta/4$ with $a^4\rightarrow (a^4\cos\theta)$.
\item $\pi/2<\theta<3\pi/2$, $\Rightarrow(\cos{\theta}<0)$\\
 Fermi arcs: 
 \begin{align}
|p|^4=a^4\cos\theta\sec{(4\beta-\theta)}, \\
\cos{(4\beta-\theta)}<0\cap\sin{4\beta}>0, 
\end{align}
which is the rotating squeezed half quartipola from that in $(\theta=\dfrac{3\pi}{2})$ case by angle $\theta/4$ with $a^4\rightarrow (a^4\cos\theta)$.
\end{itemize}

\subsubsection{$Q=(2,-2,2,-2)$}

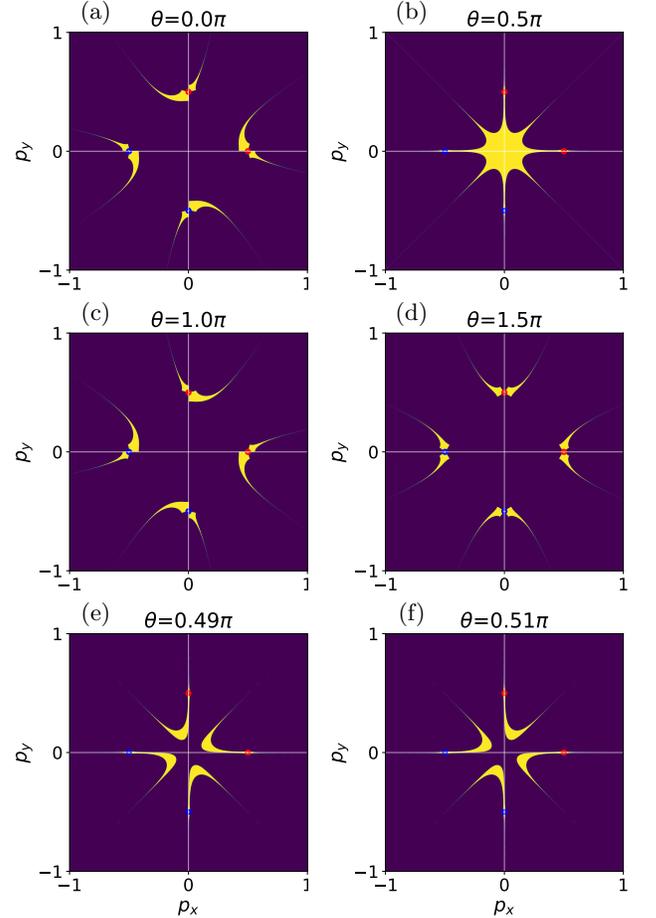
\begin{figure}[htb]
 \begin{center}
 \begin{tikzpicture}
 \node (1) at (-1.1,0) {\includegraphics[width=14em]{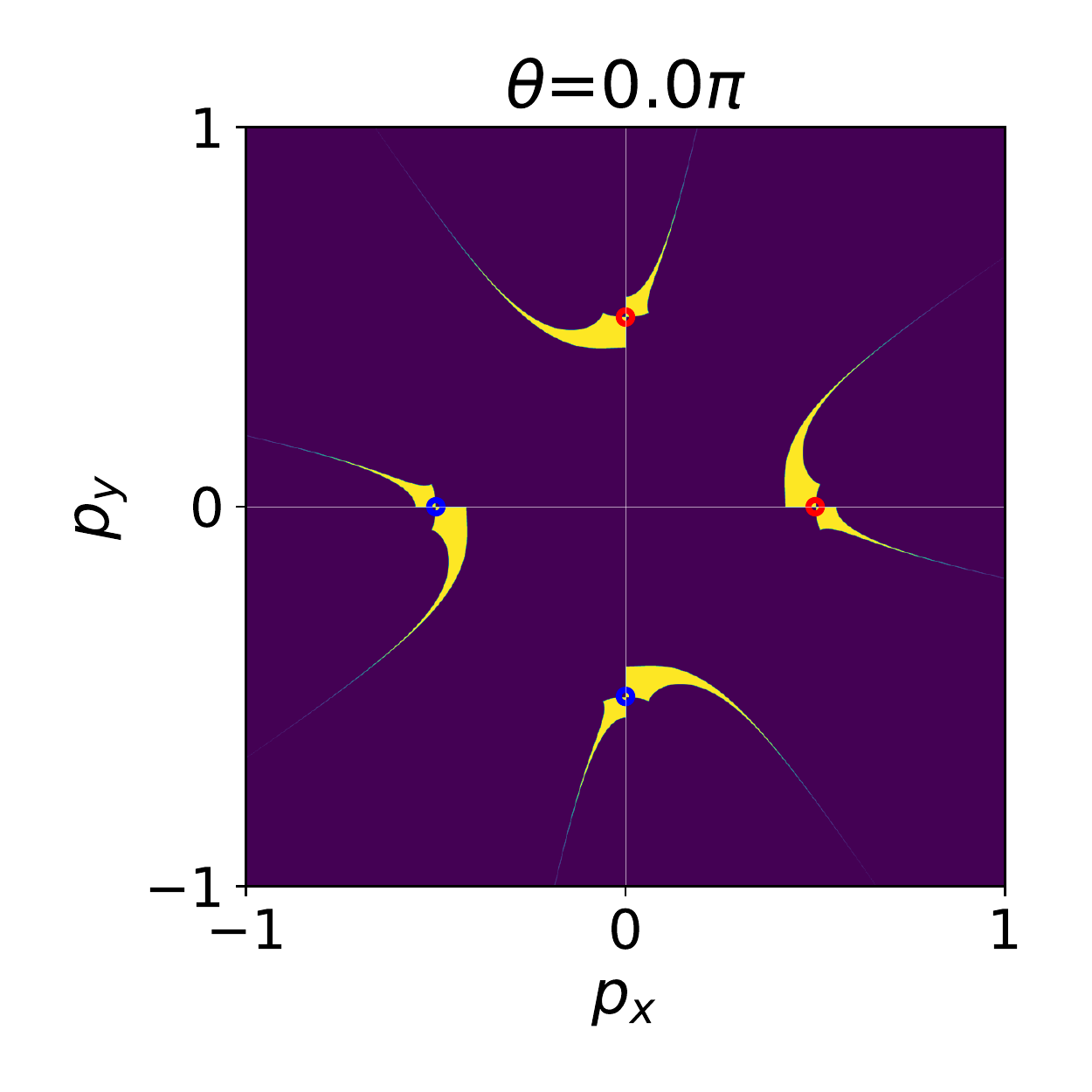}};
 \node (2) at (3.1,0) {\includegraphics[width=14em]{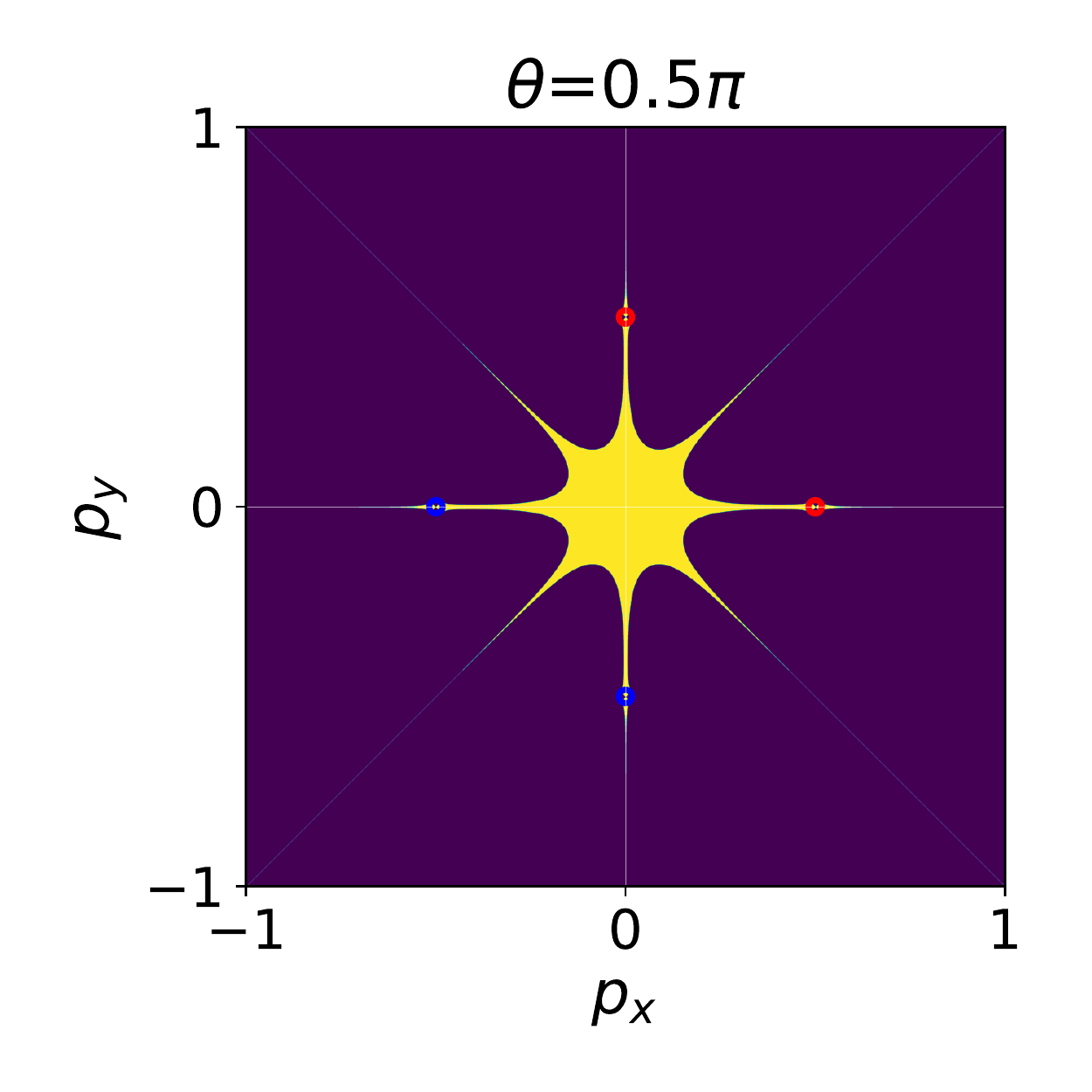}};
 \node (3) at (-1.1,-4) {\includegraphics[width=14em]{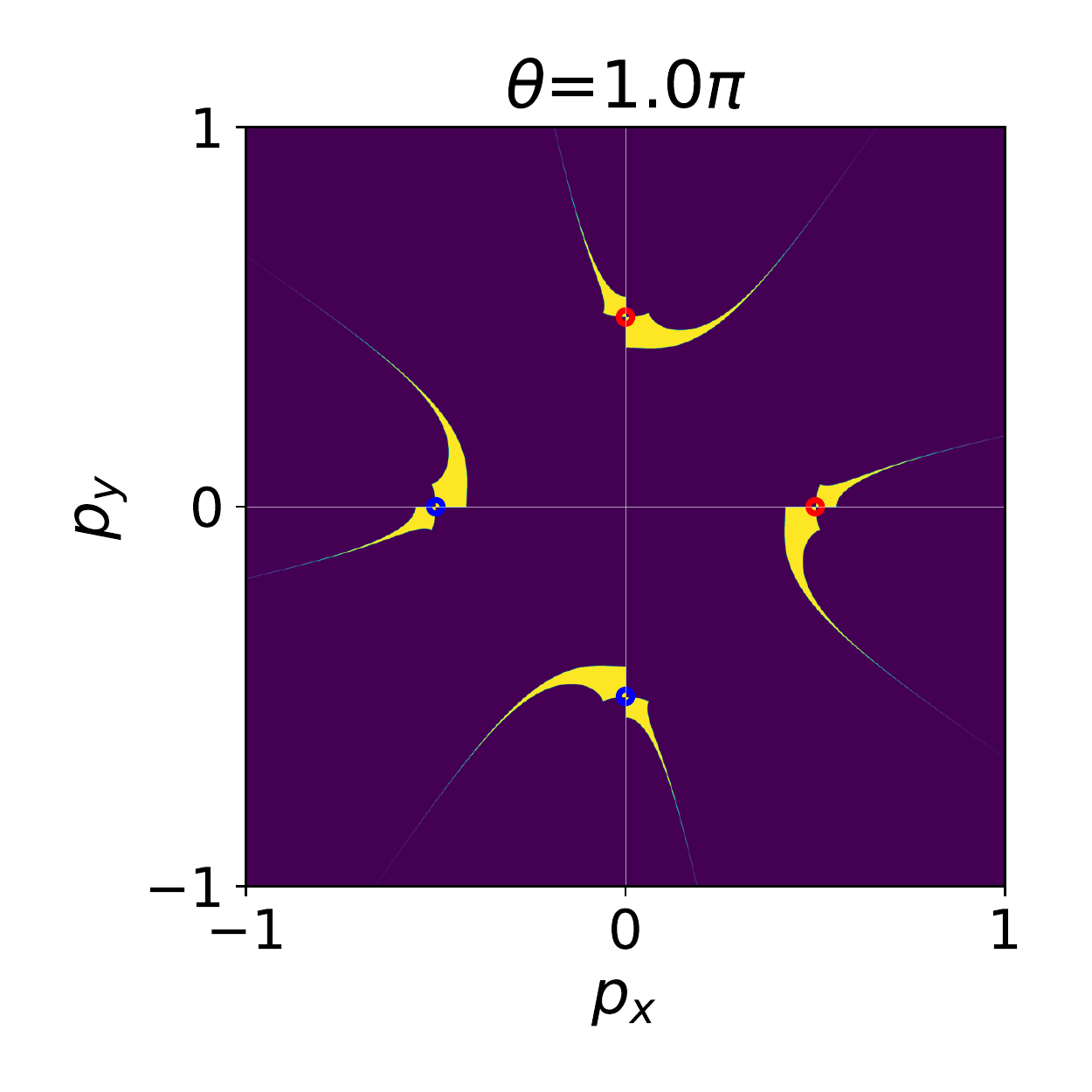}};
 \node (4) at (3.1,-4) {\includegraphics[width=14em]{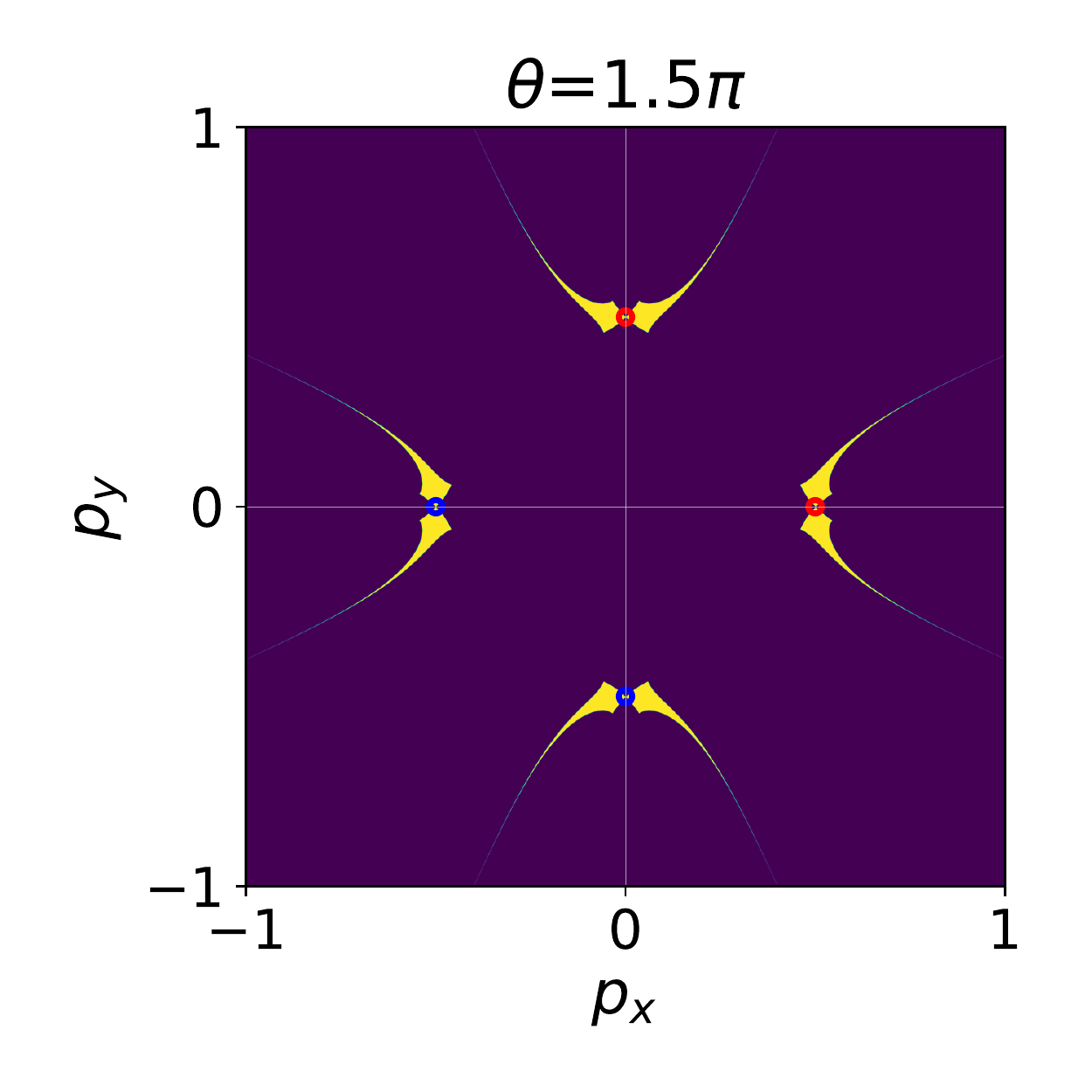}};
 \node (5) at (-1.1,-8) {\includegraphics[width=14em]{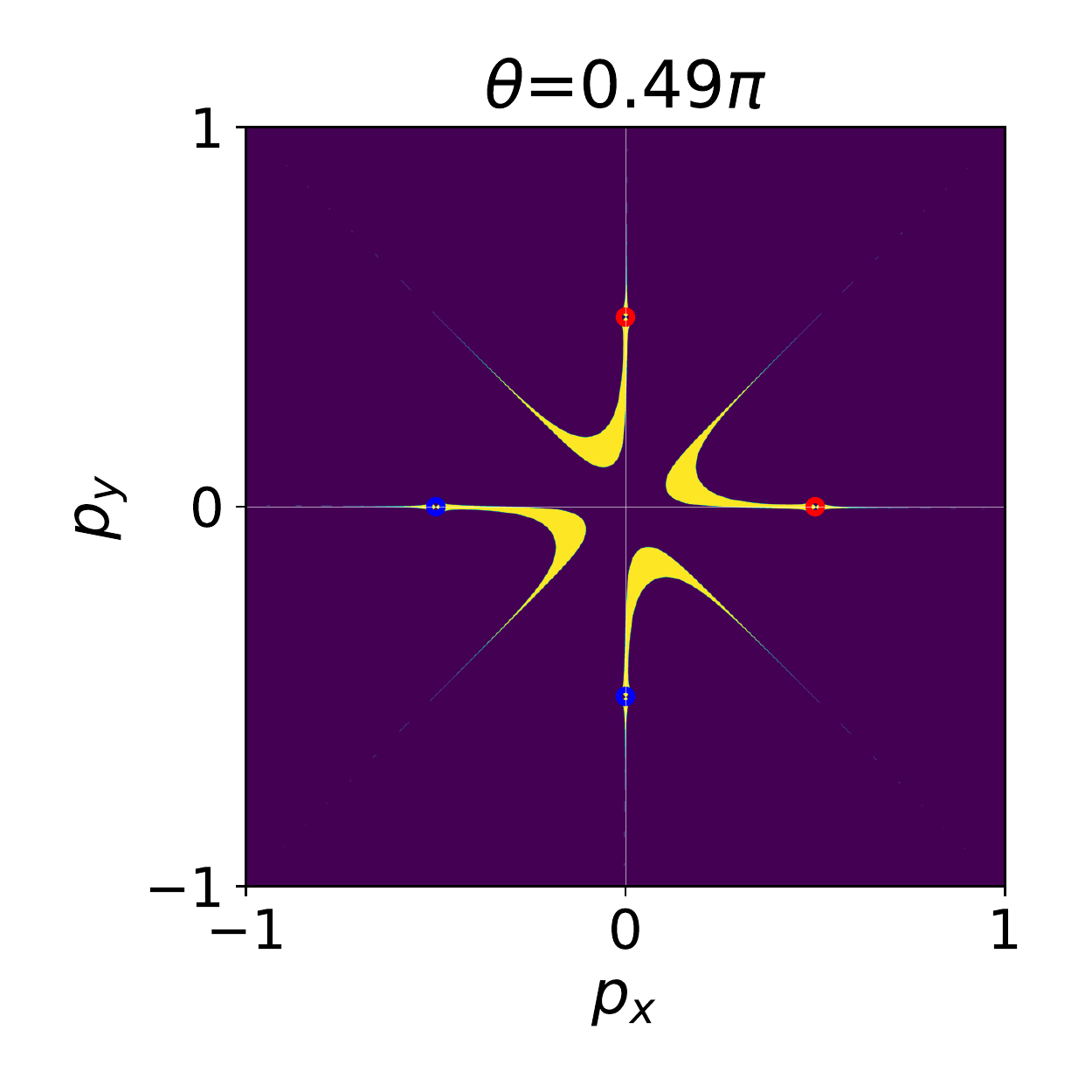}};
 \node (6) at (3.1,-8) {\includegraphics[width=14em]{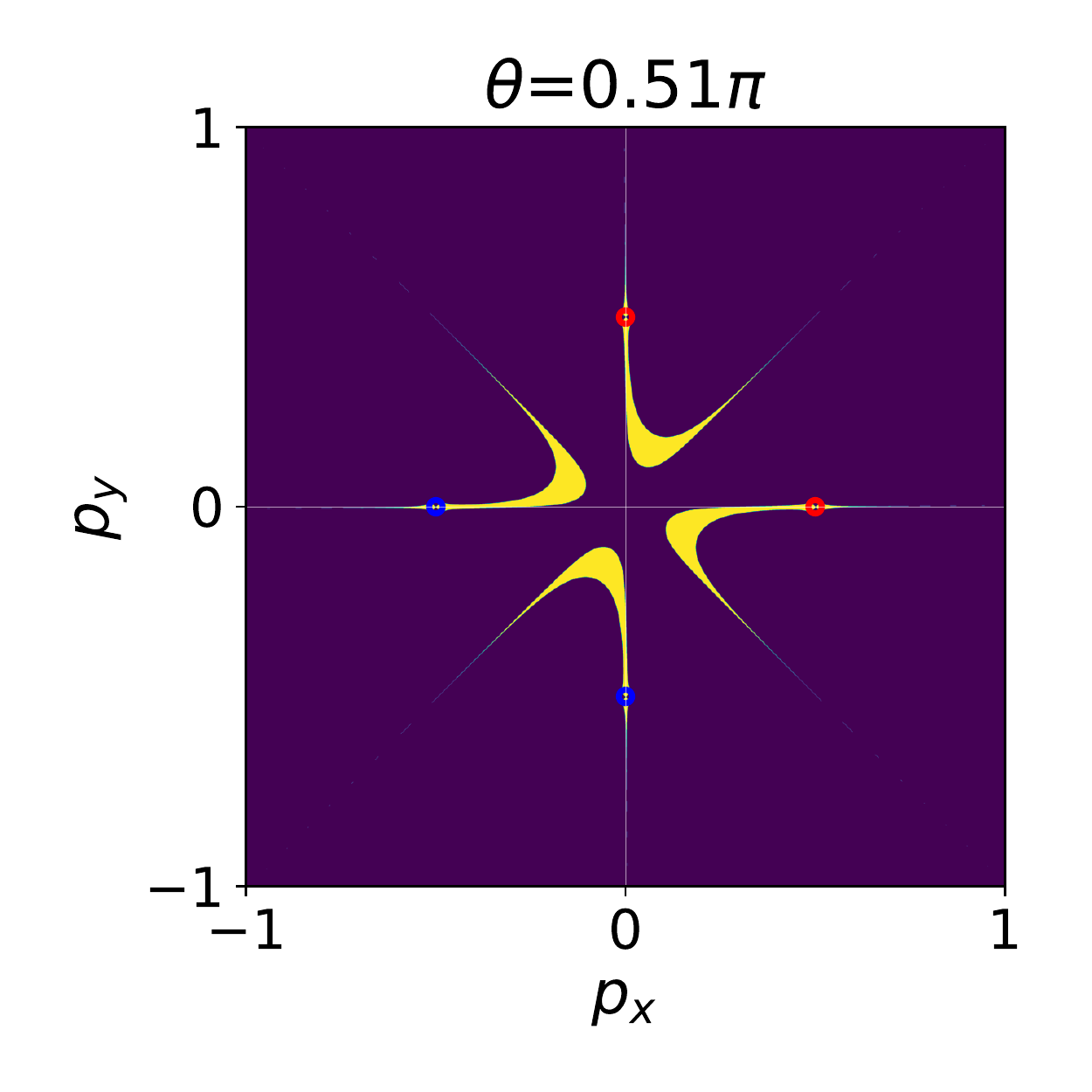}};

 \node at (-2.0,2) {(a)};
 \node at (2.2,2) {(b)};
 \node at (-2.0,-2.0) {(c)};
 \node at (2.2,-2.0) {(d)};
 \node at (-2.0,-6) {(e)};
 \node at (2.2,-6) {(f)};
 \end{tikzpicture}
 \end{center}
 \caption{The Fermi arcs of four Weyl points $Q=(2,-2,2,-2)$ (sitting at $(\pm 0.5,0)$ and $(0,\pm 0.5)$ indicated by small red and blue circles) for the boundary condition parameter $\theta= 0,0.5\pi,1.0\pi,1.5\pi,0.49\pi,0.51\pi$ are shown successively.}
 \label{fig:2222}
\end{figure}

For two pairs of Weyl nodes locating at $(\pm a,0)$ and $(0,\pm a)$ with topological charge $\pm2$, its $g(p)=(p^2-a^2)^2(p^2+a^2)^2=(p^4-a^4)^2$. The real and imaginary part of $g(p)$ are:
\begin{align}
&\Re{g(p)}=(|p|^4\cos{4\beta}-a^4)^2-(|p|^4\sin{4\beta})^2,\\
&\Im{g(p)}=2|p|^4\sin{4\beta}(|p|^4\cos{4\beta}-a^4).
\end{align}
The bulk energy dispersion in this case is
\begin{align}
E=\pm\sqrt{[(|p|^4\cos{4\beta}-a^4)^2+(|p|^4\sin{4\beta})^2]^2+p_3^2}.
\end{align}
The corresponding energy dispersion of surface states is
\begin{align}\label{EDR2222}
\epsilon=-[(|p|^4\cos{4\beta}-a^4)^2-(|p|^4\sin{4\beta})^2]\cos{\theta}\nonumber\\
-[2|p|^4\sin{4\beta}(|p|^4\cos{4\beta}-a^4)]\sin{\theta},
\\
\alpha=[(|p|^4\cos{4\beta}-a^4)^2-(|p|^4\sin{4\beta})^2]\sin{\theta}\nonumber\\
-[2|p|^4\sin{4\beta}(|p|^4\cos{4\beta}-a^4)]\cos{\theta}>0.
\end{align}
The Fermi arcs are
\begin{align}\label{EDR2222b}
[(|p|^4\cos{4\beta}-a^4)^2-(|p|^4\sin{4\beta})^2]\cos{\theta}\nonumber\\
+[2|p|^4\sin{4\beta}(|p|^4\cos{4\beta}-a^4)]\sin{\theta}=0,
 \\
	[(|p|^4\cos{4\beta}-a^4)^2-(|p|^4\sin{4\beta})^2]\sin{\theta}\nonumber\\
	-[2|p|^4\sin{4\beta}(|p|^4\cos{4\beta}-a^4)]\cos{\theta}>0.	
\end{align}
Comparing with the case $Q=(1,-1,1,-1)$, we itemize four special values of $\theta$:
\begin{itemize}
\item $\theta=\pi/2$, $\Rightarrow(\cos{\theta}=0,\sin{\theta}=1)$\\
 Fermi arcs:
\begin{align}
 \sin{4\beta}=0\cap(|p|^4\cos{4\beta}-a^4)^2>0,
\end{align}
which are two lines along the $p_1$ and $p_2$ axes except for the four Weyl nodes $(\pm a,0)$ and $(0,\pm a)$ and the whole two diagonal lines of the $p_1-p_2$ plane.. 
\item $\theta=3\pi/2$, $\Rightarrow(\cos{\theta}=0,\sin{\theta}=-1)$\\
 Fermi arcs: 
 \begin{align}
 |p|^4=a^4\sec{4\beta}, 
\end{align}
which is the whole quartibola that appeared in $Q=(1,-1,1,-1)$ representing two "compressed hyperbolas" along the $p_1$ and $p_2$ axes from the four Weyl nodes $(\pm a,0)$ and $(0,\pm a)$ to infinite, respectively.
\item $\theta=0$, $\Rightarrow(\cos{\theta}=1,\sin{\theta}=0)$\\
 Fermi arcs:
\begin{align}
 |p|^4=a^4\cos{(\dfrac{\pi}{4})}\sec{(4\beta-\dfrac{\pi}{4})},
\end{align}
which is the rotating quartibola from that of $(\theta=\dfrac{3\pi}{2})$ by angle $\pi/4$ squeezed from $a^4\rightarrow (a^4\cos\pi/4)$. 
\item $\theta=\pi$, $\Rightarrow(\cos{\theta}=-1,\sin{\theta}=0)$\\
 Fermi arcs: 
 \begin{align}
 |p|^4=a^4\cos{(\dfrac{\pi}{4})}\sec{(4\beta+\dfrac{\pi}{4})},
\end{align}
which is the rotating quartipola from that of $(\theta=\dfrac{3\pi}{2})$ by angle $-\pi/4$ squeezed from $a^4\rightarrow (a^4\cos\pi/4)$. 
\end{itemize}
For general $\theta$ in $[0,2\pi)$ except for $\pi/2$, we find that Fermi arcs are
\begin{align}
&|p|^4=a^4\cos{(\dfrac{\theta}{2}+\dfrac{\pi}{4}+k\pi)}\sec{(4\beta-\dfrac{\theta}{2}-\dfrac{\pi}{4}-k\pi)}\nonumber\\
&\cap \cos{(\dfrac{\theta}{2}+\dfrac{\pi}{4}+k\pi)}>0 ,k\in\{0,1\}.
\end{align}
The Fermi arcs of $Q=(2,-2,2,-2)$ for the four special cases above are shown in FIG. \ref{fig:2222}(a-d). One can find that the Fermi arcs in double-pairs of Weyl nodes are two copies of that for $Q=(2,-2)$ along $p_x$ and $p_y$ axes except for $\theta=\pi/2$ and $\theta=3\pi/2$. 
We find that the error of the Fermi arcs curves around the origin of $p_x-p_y$ plane and topological charges shown in FIG. \ref{fig:2222}, arising from the algorithm used in our plotting program, increases with the winding number of Weyl points. The Fermi arcs shown in FIG. \ref{fig:2222}(e,f) for $\theta=0.49\pi$ and $0.51\pi$ can help us to identify the variation of Fermi arcs connection when $\theta$ crosses the critical point$\theta=0.5\pi$.

\subsubsection{$Q=(3,-3,3,-3)$}

For two pairs of Weyl nodes locating at $(\pm a,0)$ and $(0,\pm a)$ with topological charge $\pm3$, its $g(p)=(p^4-a^4)^3$, the real and imaginary parts of $g(p)$ are:
\begin{align}
\Re{g(p)}&=(|p|^4\cos{4\beta}-a^4)[(|p|^4\cos{4\beta}-a^4)^2-3(|p|^4\sin{4\beta})^2], \\
\Im{g(p)}&=|p|^4\sin{4\beta}[3(|p|^4\cos{4\beta}-a^4)^2-(|p|^4\sin{4\beta})^2].
\end{align}
The bulk energy dispersion is
\begin{align}
E=\pm\sqrt{[(|p|^4\cos{4\beta}-a^4)^2+(|p|^4\sin{4\beta})^2]^3+p_3^2}.
\end{align}
The equation of Fermi arcs are
\begin{align}\label{EDR3333b}
(|p|^4\cos{4\beta}-a^4)[(|p|^4\cos{4\beta}-a^4)^2-3(|p|^4\sin{4\beta})^2]\cos{\theta}\nonumber\\ 
+|p|^4\sin{4\beta}[3(|p|^4\cos{4\beta}-a^4)^2-(|p|^4\sin{4\beta})^2]\sin{\theta}=0,
 \\
(|p|^4\cos{4\beta}-a^4)[(|p|^4\cos{4\beta}-a^4)^2-3(|p|^4\sin{4\beta})^2]\sin{\theta}\nonumber\\ 
-|p|^4\sin{4\beta}[3(|p|^4\cos{4\beta}-a^4)^2-(|p|^4\sin{4\beta})^2]\cos{\theta}>0.
\end{align}
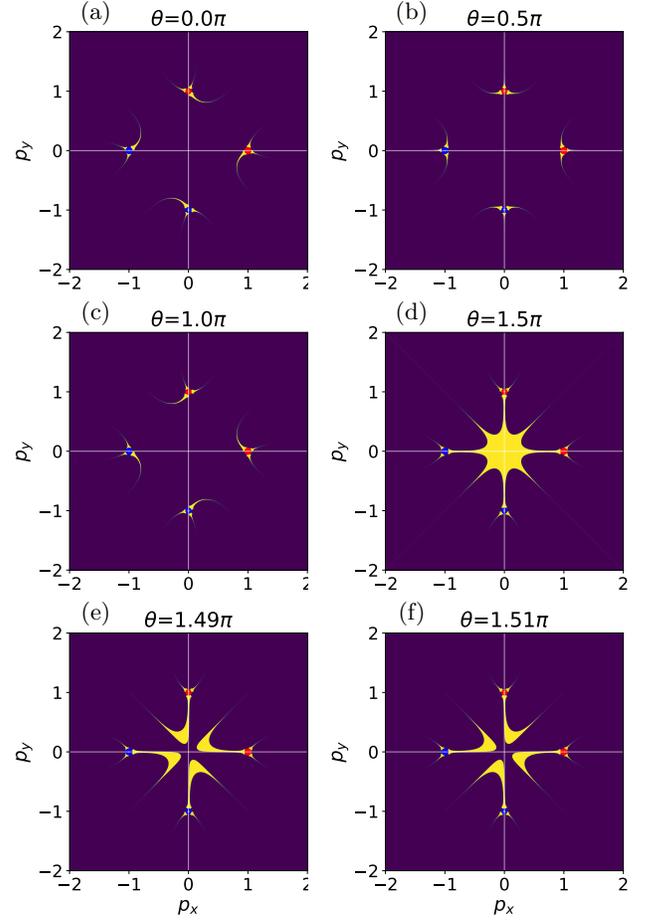
\begin{figure}[htb]
 \begin{center}
 \begin{tikzpicture}
 \node (1) at (-1.1,0) {\includegraphics[width=14em]{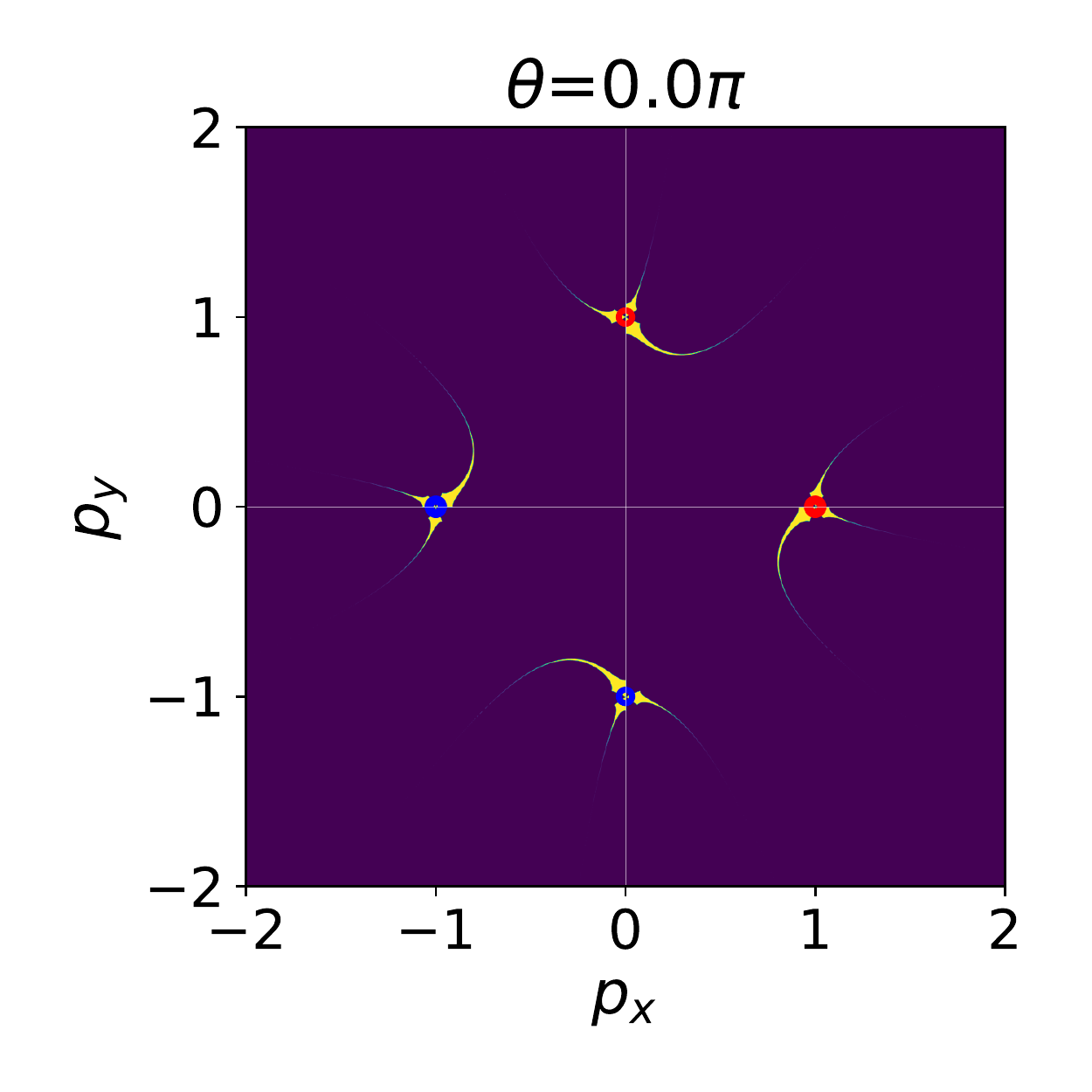}};
 \node (2) at (3.1,0) {\includegraphics[width=14em]{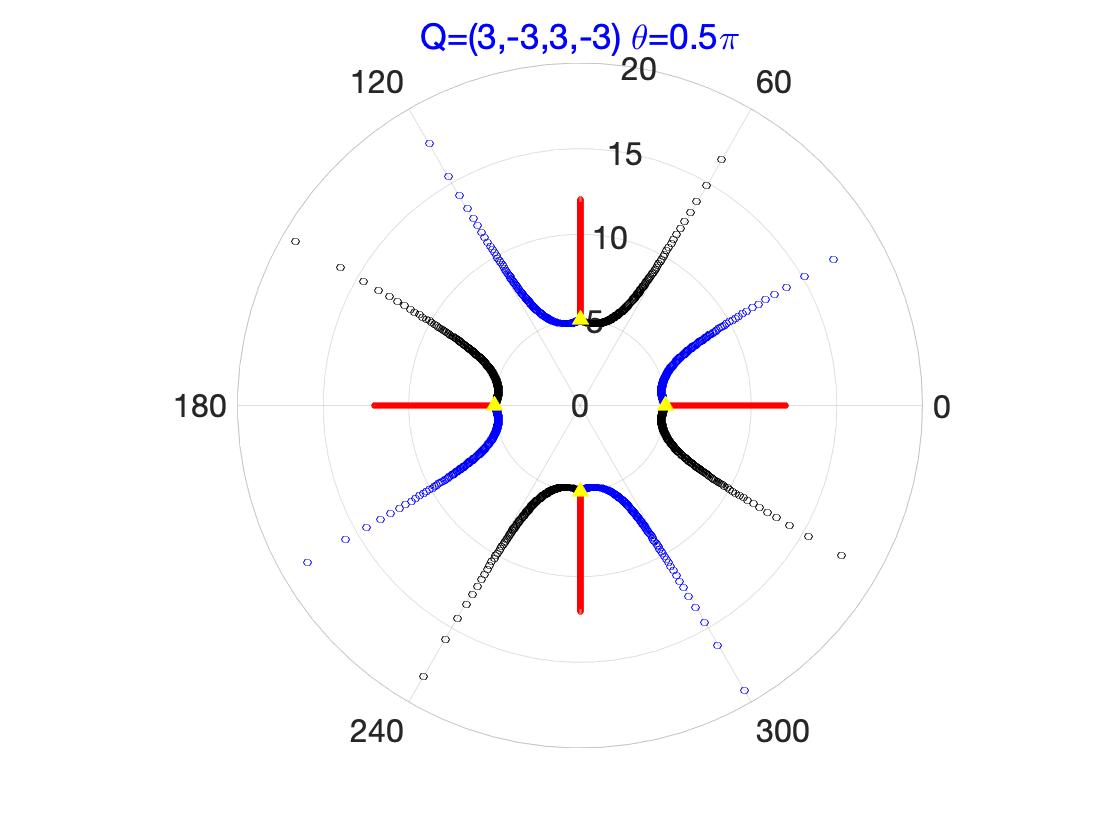}};
 \node (3) at (-1.1,-4) {\includegraphics[width=14em]{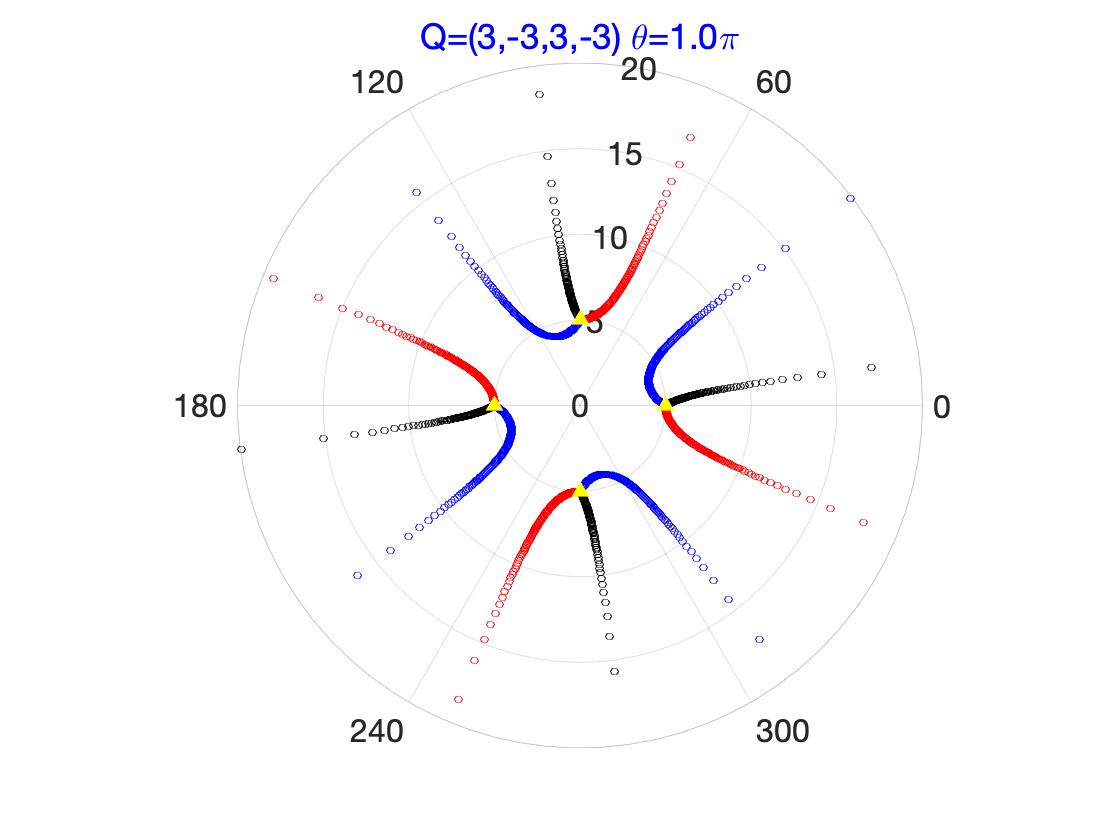}};
 \node (4) at (3.1,-4) {\includegraphics[width=14em]{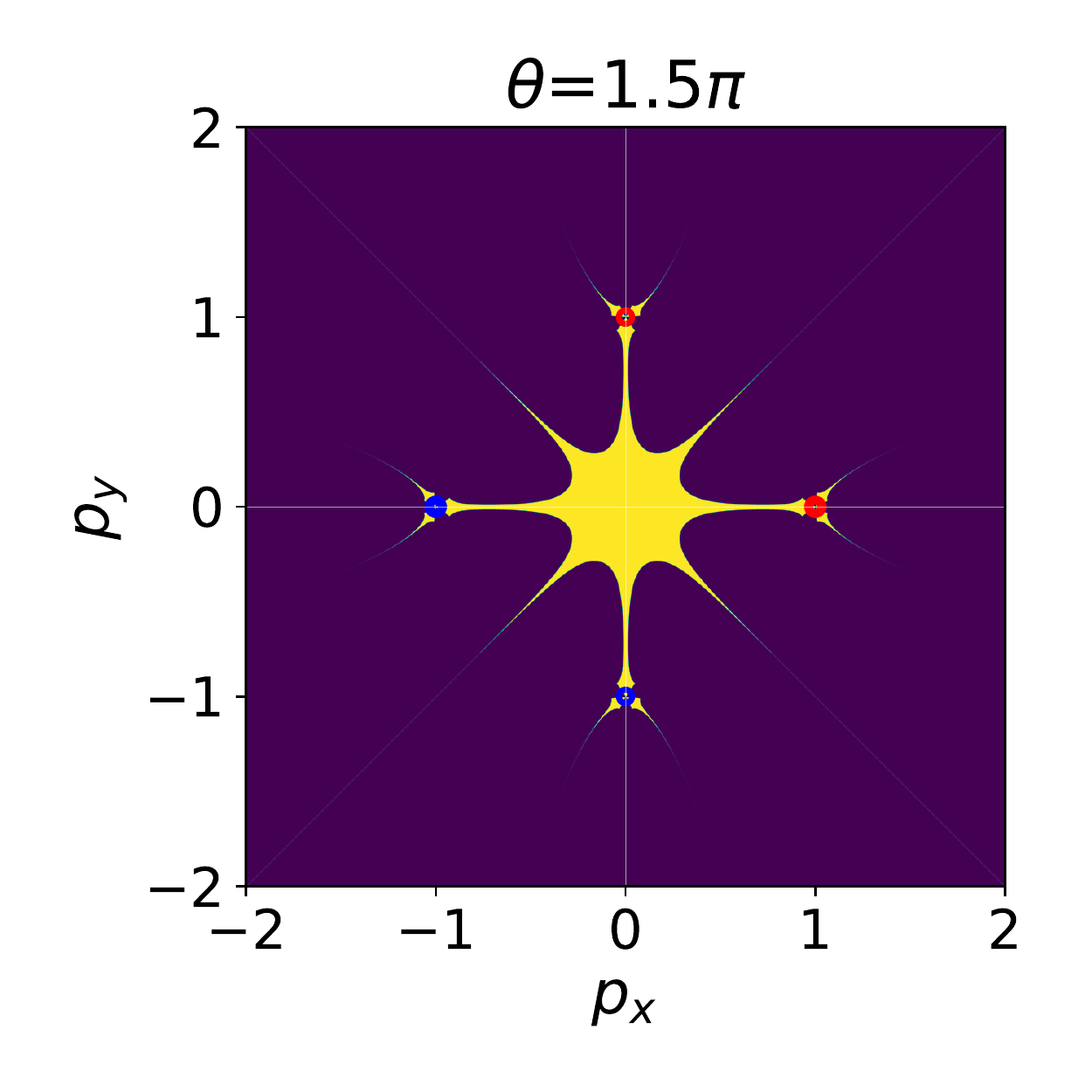}};
 \node (5) at (-1.1,-8) {\includegraphics[width=14em]{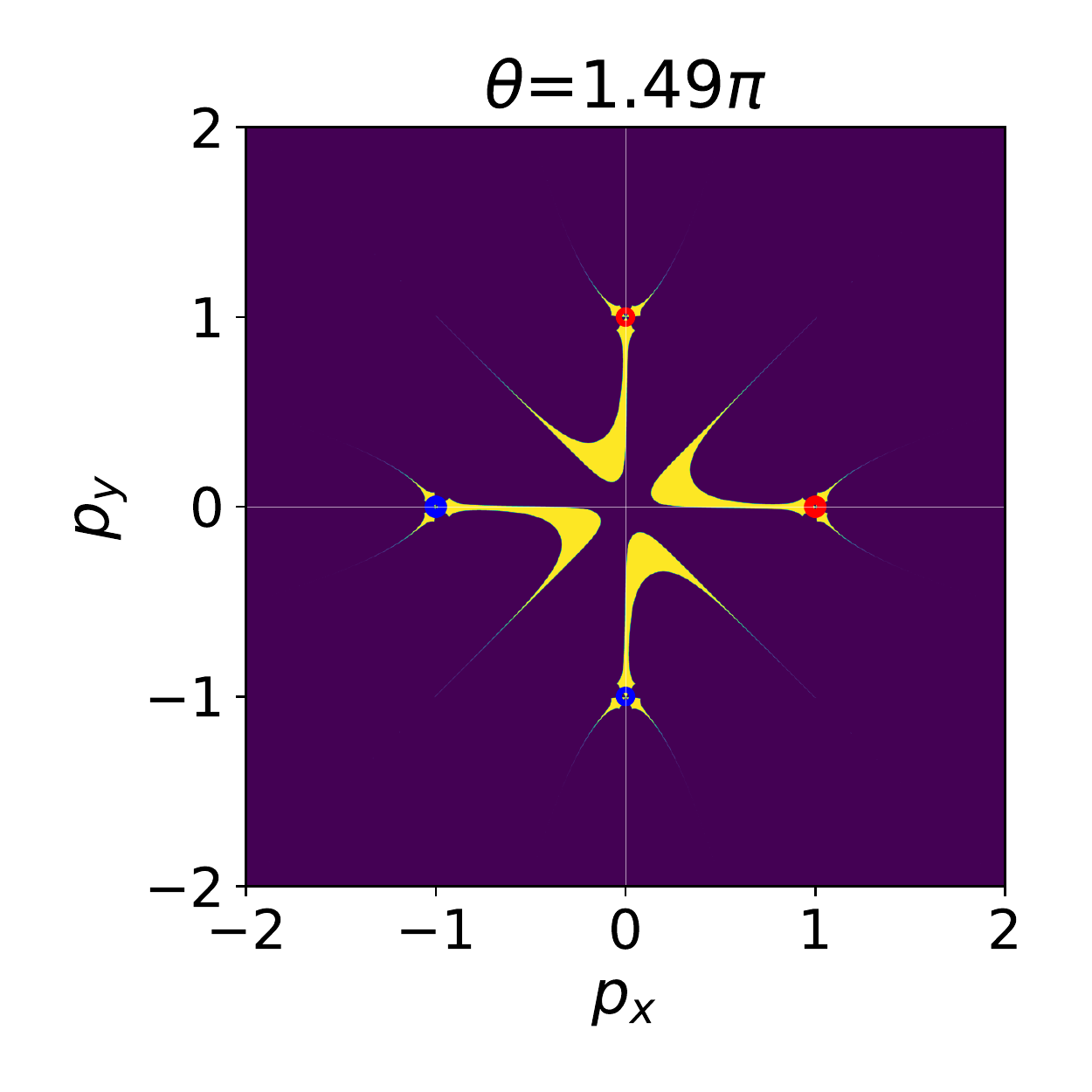}};
 \node (6) at (3.1,-8) {\includegraphics[width=14em]{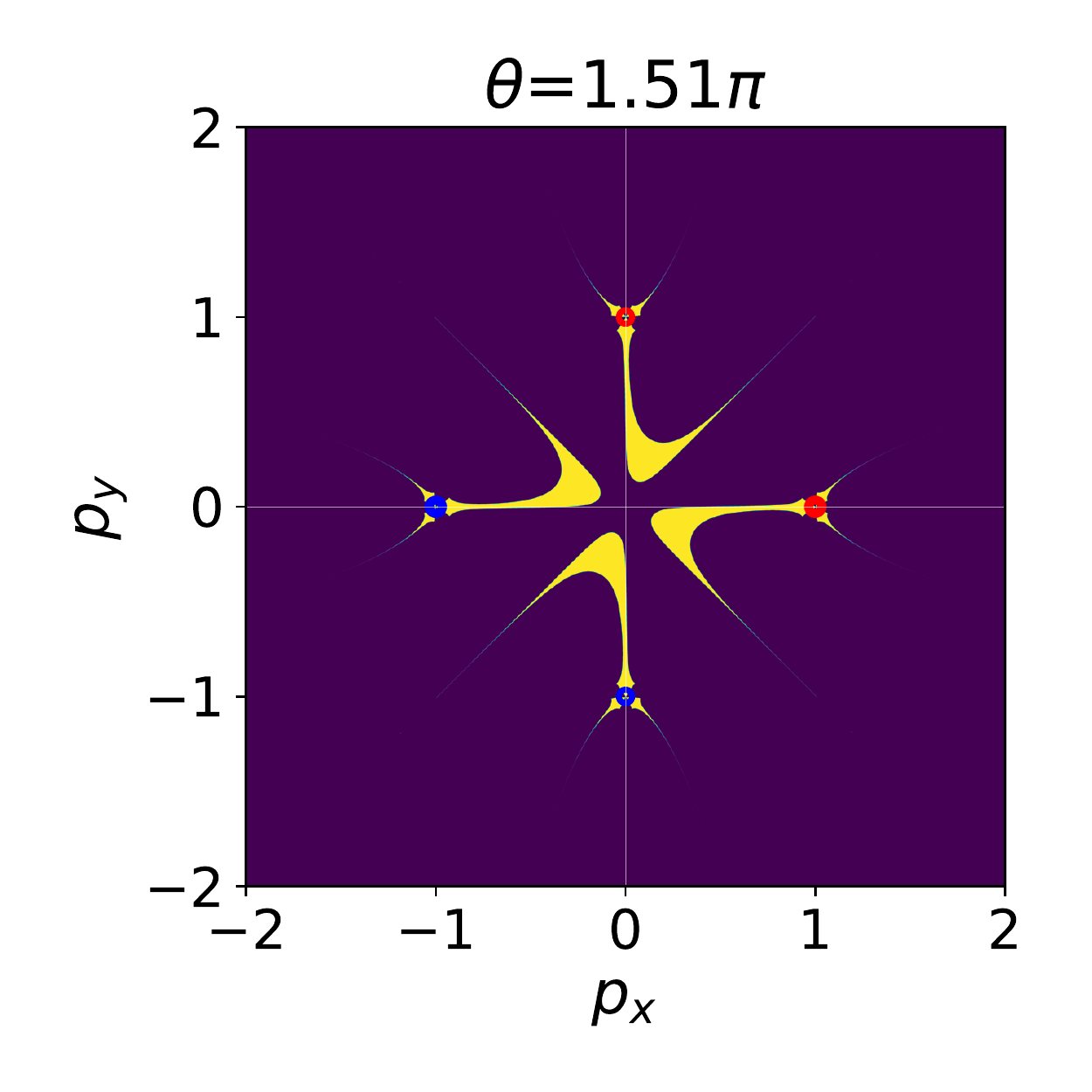}};

 \node at (-2.0,2) {(a)};
 \node at (2.2,2) {(b)};
 \node at (-2.0,-2.0) {(c)};
 \node at (2.2,-2.0) {(d)};
 \node at (-2.0,-6) {(e)};
 \node at (2.2,-6) {(f)};
 \end{tikzpicture}
 \end{center}
 \caption{The Fermi arcs of four Weyl points $Q=(3,-3,3,-3)$ (sitting at $(\pm 1,0)$ and $(0,\pm 1)$ indicated by small red or blue circles) for the boundary condition parameter $\theta= 0,0.5\pi,1.0\pi,1.5\pi,1.49\pi,1.51\pi$ are shown successively.}
 \label{fig:3333-a}
\end{figure}
\begin{figure}[bht]
 \begin{center}
 \begin{tikzpicture}
 \node (1) at (-1.2,0) {\includegraphics[width=15em]{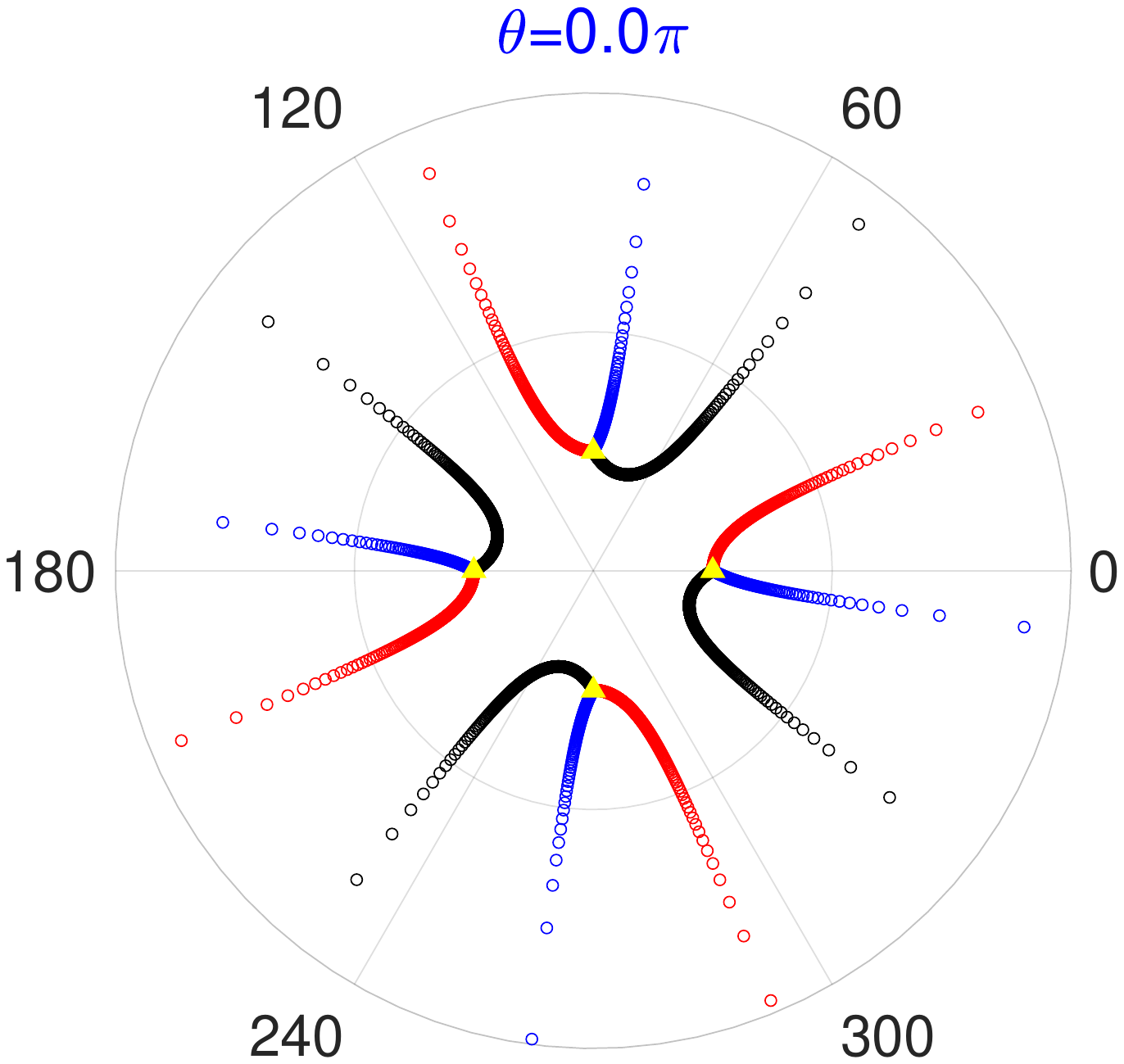}};
 \node (2) at (3,0) {\includegraphics[width=15em]{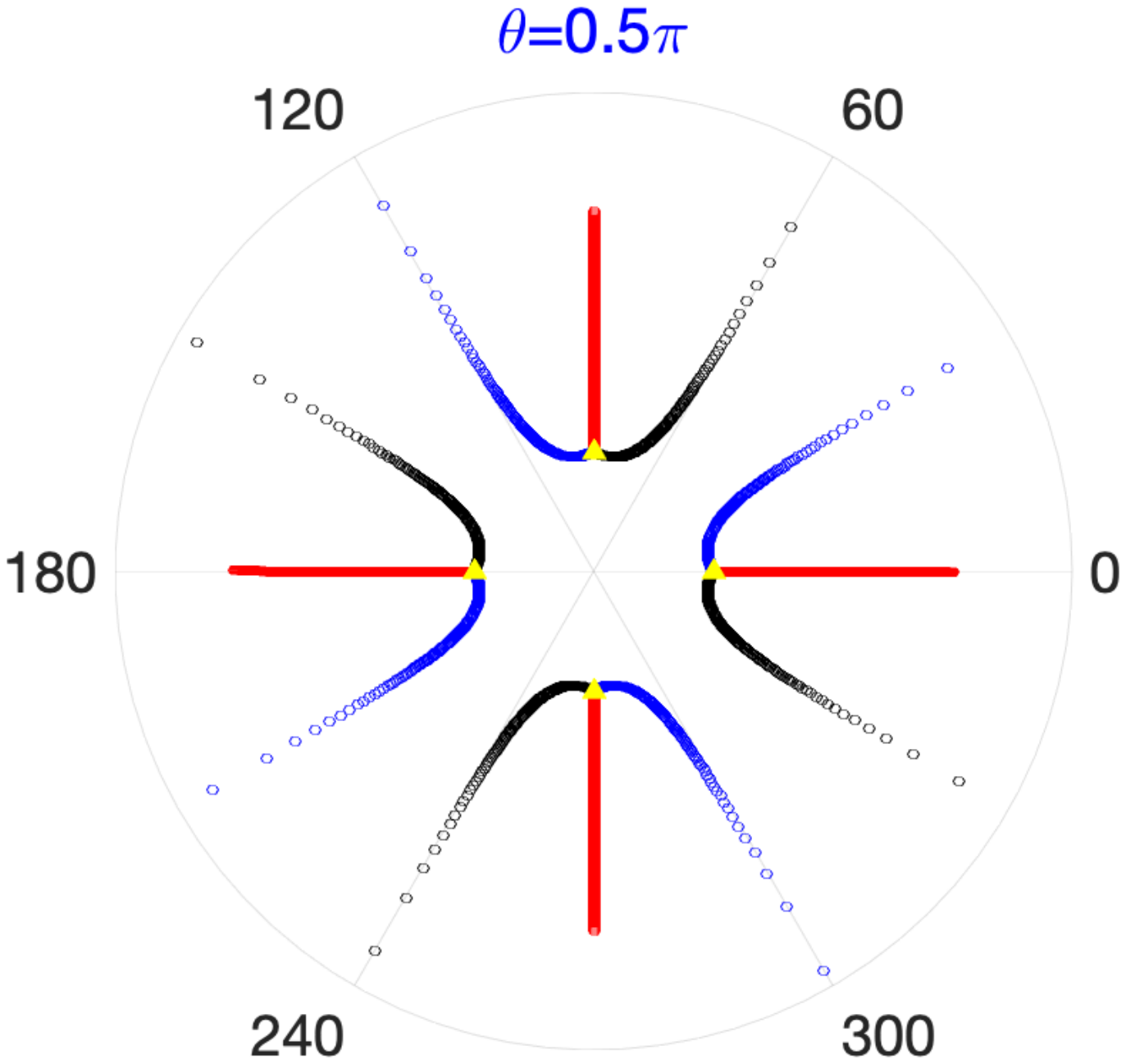}};
 \node (3) at (-1.2,-4) {\includegraphics[width=15em]{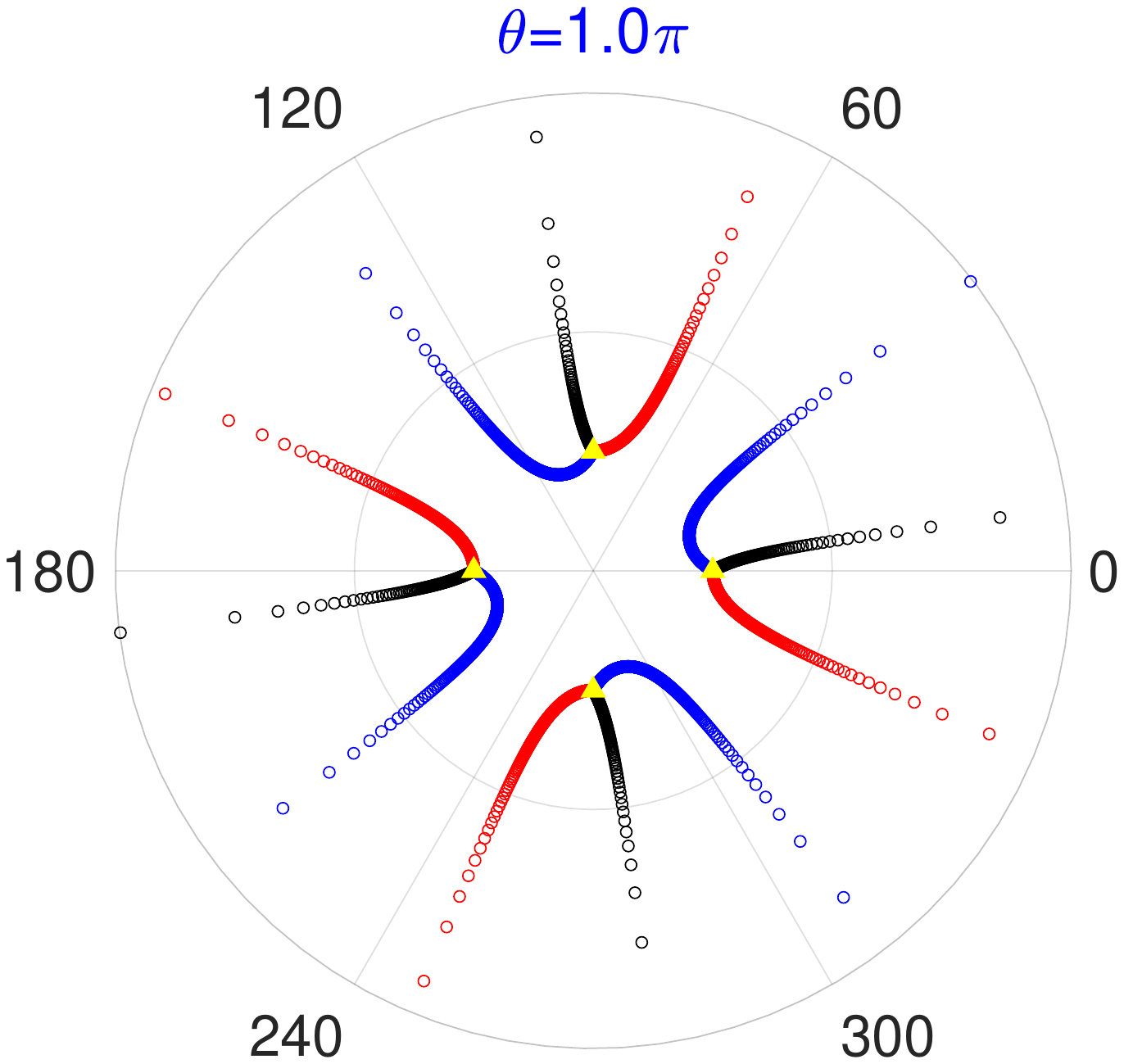}};
 \node (4) at (3,-4) {\includegraphics[width=15em]{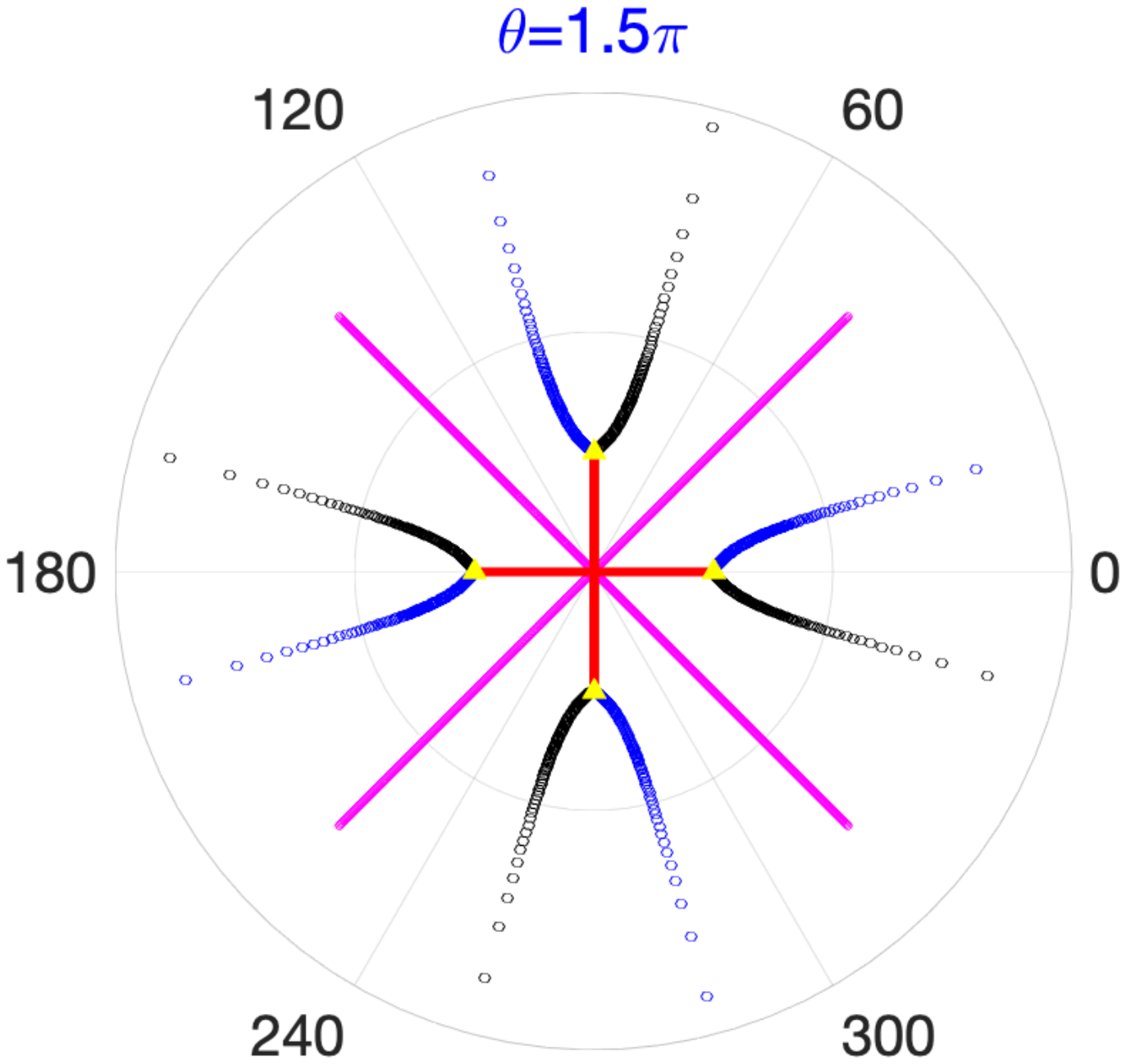}};
 \node at (-2.4,1.7) {(a)};
 \node at (1.8,1.7) {(b)};
 \node at (-2.4,-2.3) {(c)};
 \node at (1.8,-2.3) {(d)};
 \end{tikzpicture}
 \end{center}
 \caption{The Fermi arcs of four Weyl points $Q=(3,-3,3,-3)$ (sitting at $(\pm 5,0)$ and $(0,\pm 5)$ indicated by yellow triangle) for the boundary condition parameter $\theta= 0,0.5\pi,1.0\pi,1.5\pi$ are shown successively.}
 \label{fig:3333-b}
\end{figure}
Now discuss four special values of $\theta$.
\begin{itemize}
\item $\theta=\pi/2$, $\Rightarrow(\cos{\theta}=0,\sin{\theta}=1)$\\
 Fermi arcs:
\begin{align}
&(|p|^4=a^4\cos{(\dfrac{\pi}{6})}\sec{(4\beta\pm \dfrac{\pi}{6}})\cap(\pm\sin{(4\beta)}<0))\cup\nonumber\\
&(\cos{4\beta}=1 \cap |p|> a),
\end{align}
which are large part of $p_1$ and $p_2$ axes with $|p|> a$ and half of the two heterobolae $|p|^4=a^4\cos{(\dfrac{\pi}{6})}\sec{(4\beta\pm \dfrac{\pi}{6})}$. 
\item $\theta=3\pi/2$, $\Rightarrow(\cos{\theta}=0,\sin{\theta}=-1)$\\
 Fermi arcs: 
\begin{align}
&(|p|^4=a^4\cos{(\dfrac{\pi}{6})}\sec{(4\beta\pm \dfrac{\pi}{6}})\cap(\pm\sin{(4\beta)}>0))\cup\nonumber\\
&[(\cos{4\beta}=1 \cap |p|< a)\cup(\cos{4\beta}=-1)],
\end{align}
which are small part of $p_1$ and $p_2$ axes with $|p|< a$ and the whole diagonal lines as well as the other half of the two heterobolae $|p|^4=a^4\cos{(\dfrac{\pi}{6})}\sec{(4\beta\pm \dfrac{\pi}{6})}$. 
\item $\theta=0$, $\Rightarrow(\cos{\theta}=1,\sin{\theta}=0)$\\
 Fermi arcs:
\begin{align}
&(|p|^4=a^4\cos{(\dfrac{\pi}{3})}\sec{(4\beta\pm \dfrac{\pi}{3}})\cap(\sin{(4\beta)}<0))\cup\nonumber\\
&(|p|^4=a^4\sec{(4\beta)})\cap(\sin{(4\beta)}>0),
\end{align}
which are three half-branches of the hyperbolas. 
\item $\theta=\pi$, $\Rightarrow(\cos{\theta}=-1,\sin{\theta}=0)$\\
 Fermi arcs: 
\begin{align}
&(|p|^4=a^4\cos{(\dfrac{\pi}{3})}\sec{(4\beta\pm \dfrac{\pi}{3}})\cap(\sin{(4\beta)}>0))\cup\nonumber\\
&(|p|^4=a^4\sec{(4\beta)})\cap(\sin{(4\beta)}<0),
\end{align}
which are the other three half-branches of these hyperbolas. 
\end{itemize} 
For general $\theta$, we find that Fermi arcs are half of these three hyperbolas
\begin{align}
&|p|^4=a^4\cos{(\dfrac{\theta+(2k+1)\pi}{3})}\sec{(4\beta-\dfrac{\theta+(2k+1)\pi}{3})},\nonumber\\
&k\in\{0,1,2\},
\end{align}
with the condition
\begin{align}
\sin{4\beta}\lessgtr 0, for \cos{(\dfrac{\theta+(2k+1)\pi}{3})}\lessgtr 0.\\
\end{align}
When $\cos{(\dfrac{\theta+(2k+1)\pi}{3})}=0$, the equation above is not suitable and the Fermi arcs become direct lines
\begin{align}
&[(\cos{4\beta}=1 \cap |p|\lessgtr a), for \sin{(\dfrac{\theta+(2k+1)\pi}{3})}=\mp1]\nonumber\\
&\cup(\cos{4\beta}=-1, for \sin{(\dfrac{\theta+(2k+1)\pi}{3})}=-1).
\end{align}
The Fermi arc of $Q=(3,-3,3,-3)$ for the four special cases above are shown in FIG. \ref{fig:3333-a}(a-d). One can find that the Fermi arcs in double-pairs of Weyl points are two copies of that for $Q=(2,-2)$ along $p_x$ and $p_y$ axes except for $\theta=\pi/2$ and $3\pi/2$. 
The error of the Fermi arcs curves around the origin of $p_x-p_y$ plane and topological charges shown in FIG. \ref{fig:3333-a} increases significantly. The Fermi arcs shown in FIG. \ref{fig:3333-a}(e,f) for $\theta=1.49\pi$ and $1.51\pi$ demonstrate clearly the connection variation of Fermi arcs crossing $\theta=1.5\pi$. Besides, we provide the Fermi arcs of $Q=(3,-3,3,-3)$ for the four special cases according to the analytical resolution obtained above for comparison, which are shown in FIG. \ref{fig:3333-b}.

\subsubsection{$Q=(2,-2,1,-1)$ and $Q=(4,-4,3,-3)$}

For two pairs of Weyl nodes locating at $(\pm a,0)$ and $(0,\pm b)$ with different chirality $\pm w$ and $\pm x$, its $g(p)=(p^2-a^2)^{w}(p^2+b^2)^x$. The analytic solutions for these cases can not be simplified effectively, thereby we only provide the Fermi arcs of $Q=(2,-2,1,-1)$ and $Q=(4,-4,3,-3)$, which $g(p)$ are $(p^2-a^2)^{2}(p^2+b^2)$ and $(p^2-a^2)^{4}(p^2+b^2)^3$, respectively. The real and imaginary part of $g(p)$ for $Q=(2,-2,1,-1)$ are:
\begin{align}
\Re{g(p)}=(p_1^2-p_2^2-a^2)^2(p_1^2-p_2^2+b^2)\nonumber\\
-4p_1^2p_2^2[(b^2-2a^2)+3(p_1^2-p_2^2)], \\
\Im{g(p)}=2p_1 p_2(p_1^2-p_2^2-a^2)[(2b^2-a^2)\nonumber\\
+3(p_1^2-p_2^2)]-8p_1^3p_2^3.
\end{align}
The Fermi arcs for different $\theta$ are shown in FIG. \ref{fig:2211}. Obviously, this is the combination of the Fermi arcs of $Q=(1,-1)$ at points$(0,\pm b)$ and $Q=(2,-2)$ at points$(\pm a,0)$. 

For comparison, the Fermi arcs for $Q=(4,-4,3,-3)$ are shown in FIG. \ref{fig:4433}. FIG. \ref{fig:2211}(a-d) and FIG. \ref{fig:4433}(a-d) show that the relation between the number of Fermi arcs and the topological charge of Weyl points remains as before. However, FIG. \ref{fig:2211}(e,f) and FIG. \ref{fig:4433}(e,f) show the new structure with more bifurcations appeared in the Fermi arcs near the $\theta=0.5\pi$, where one of the two diagonal line is no longer the asymptote of Fermi arcs.
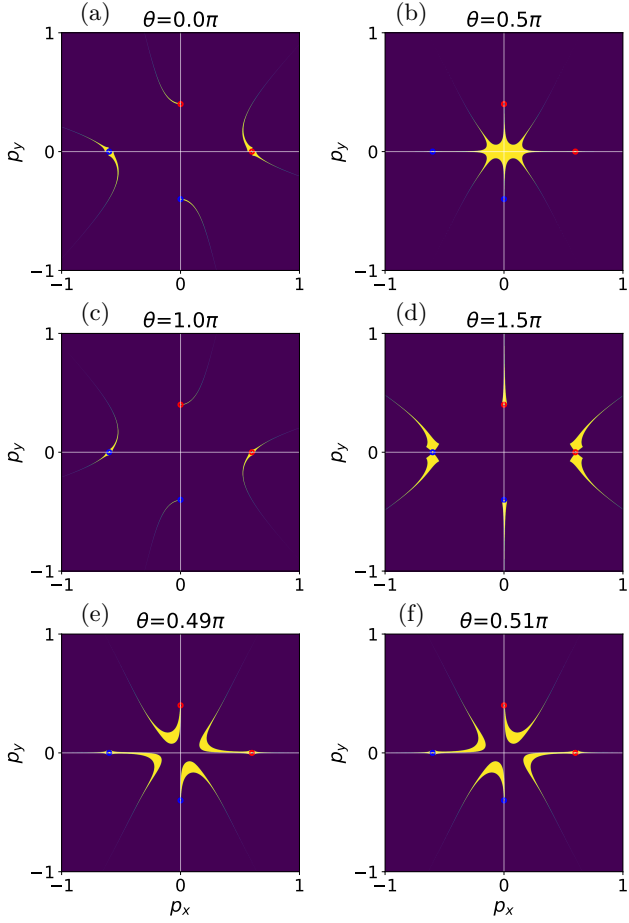
\begin{figure}[htbp]
 \begin{center}

 \begin{tikzpicture}
 \node (1) at (-1.2,0) {\includegraphics[width=14em]{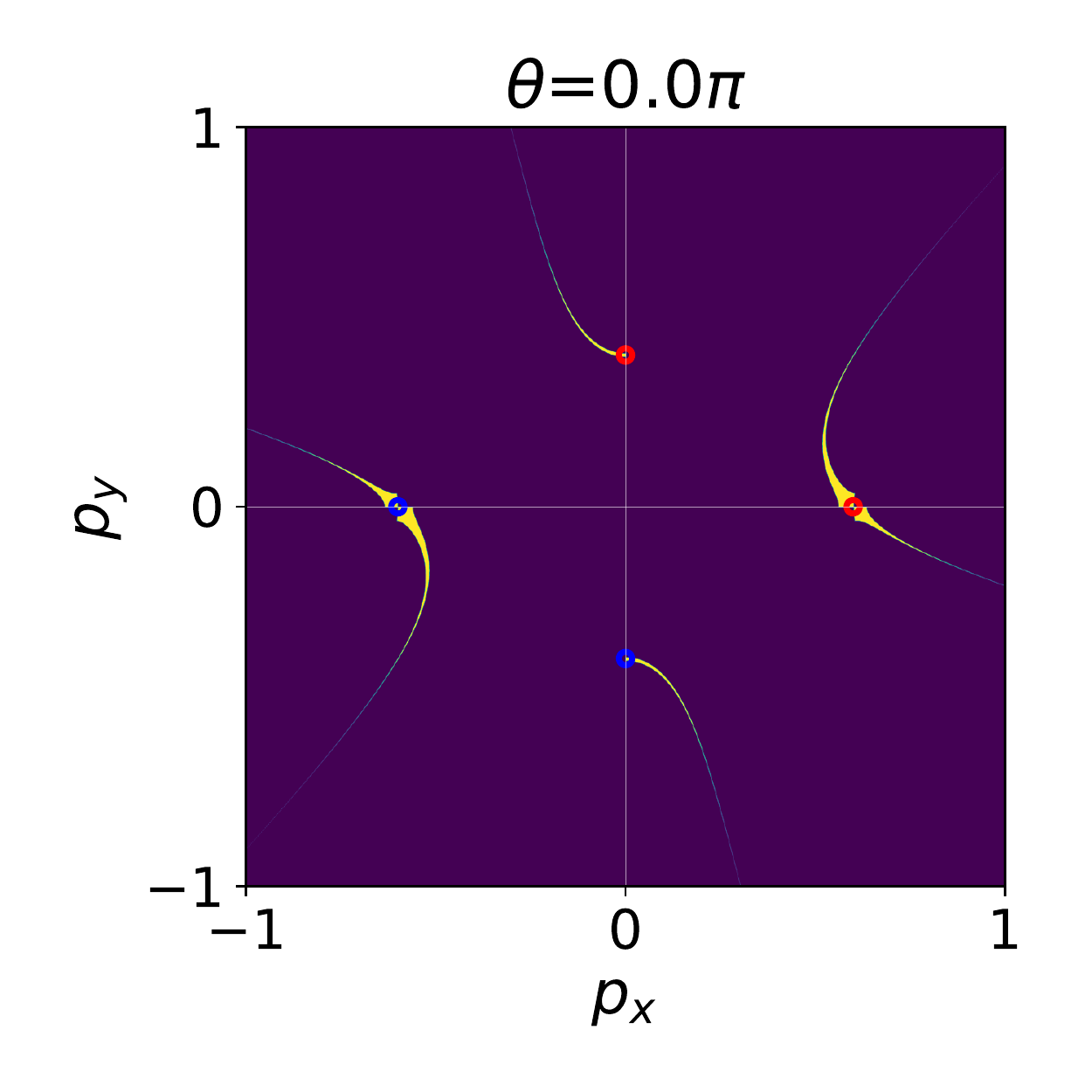}};
 \node (2) at (3.1,0) {\includegraphics[width=14em]{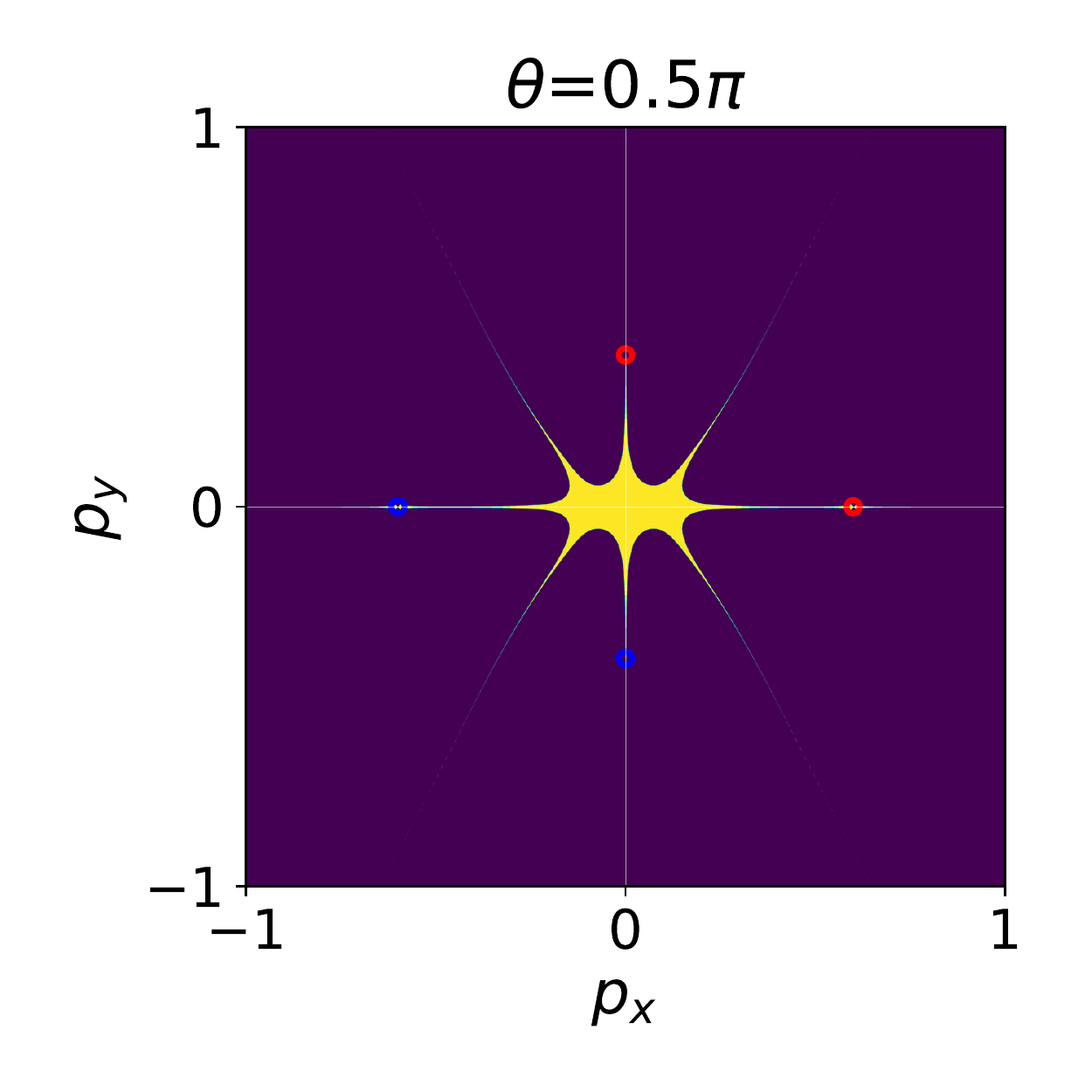}};
 \node (3) at (-1.2,-4.0) {\includegraphics[width=14em]{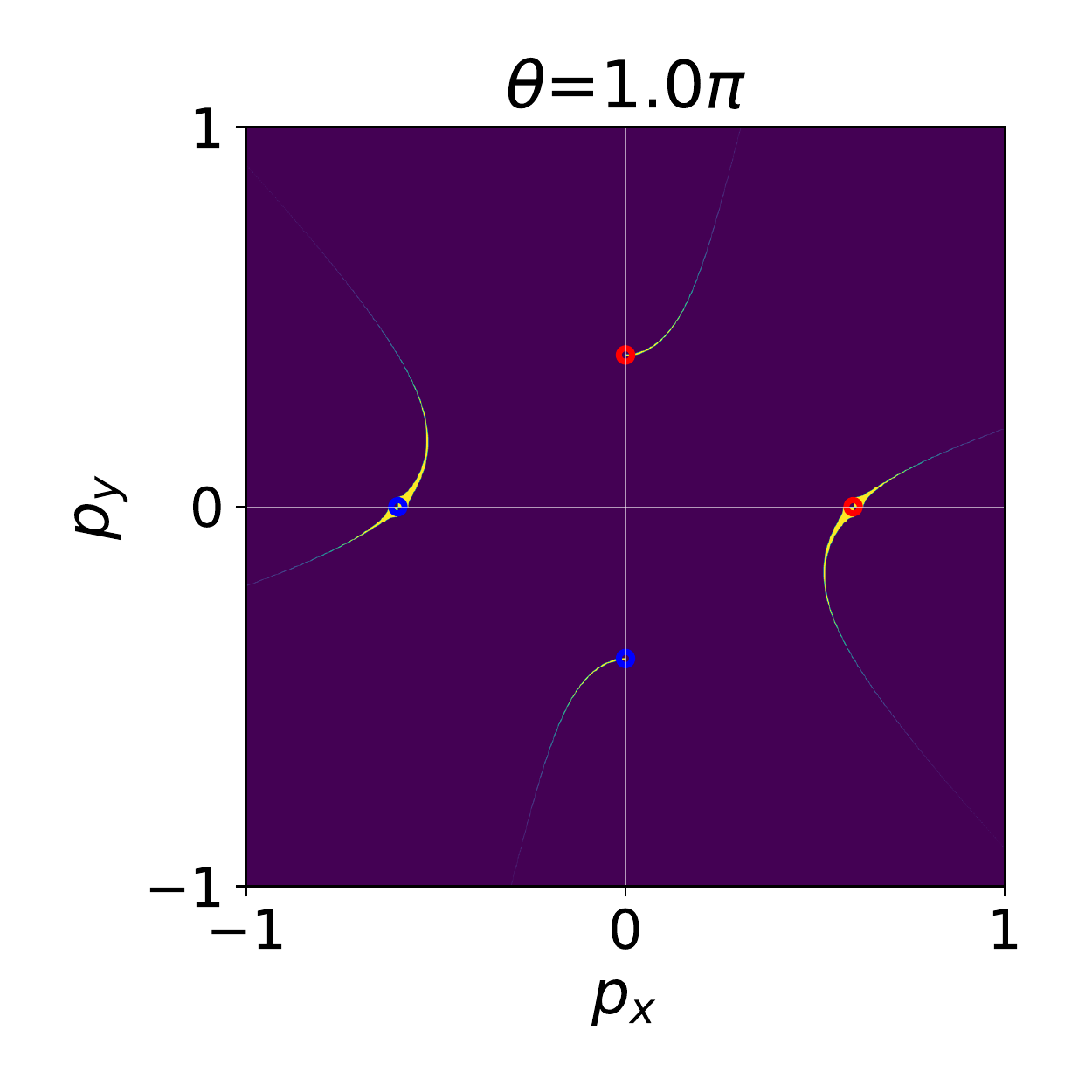}};
 \node (4) at (3.1,-4.0) {\includegraphics[width=14em]{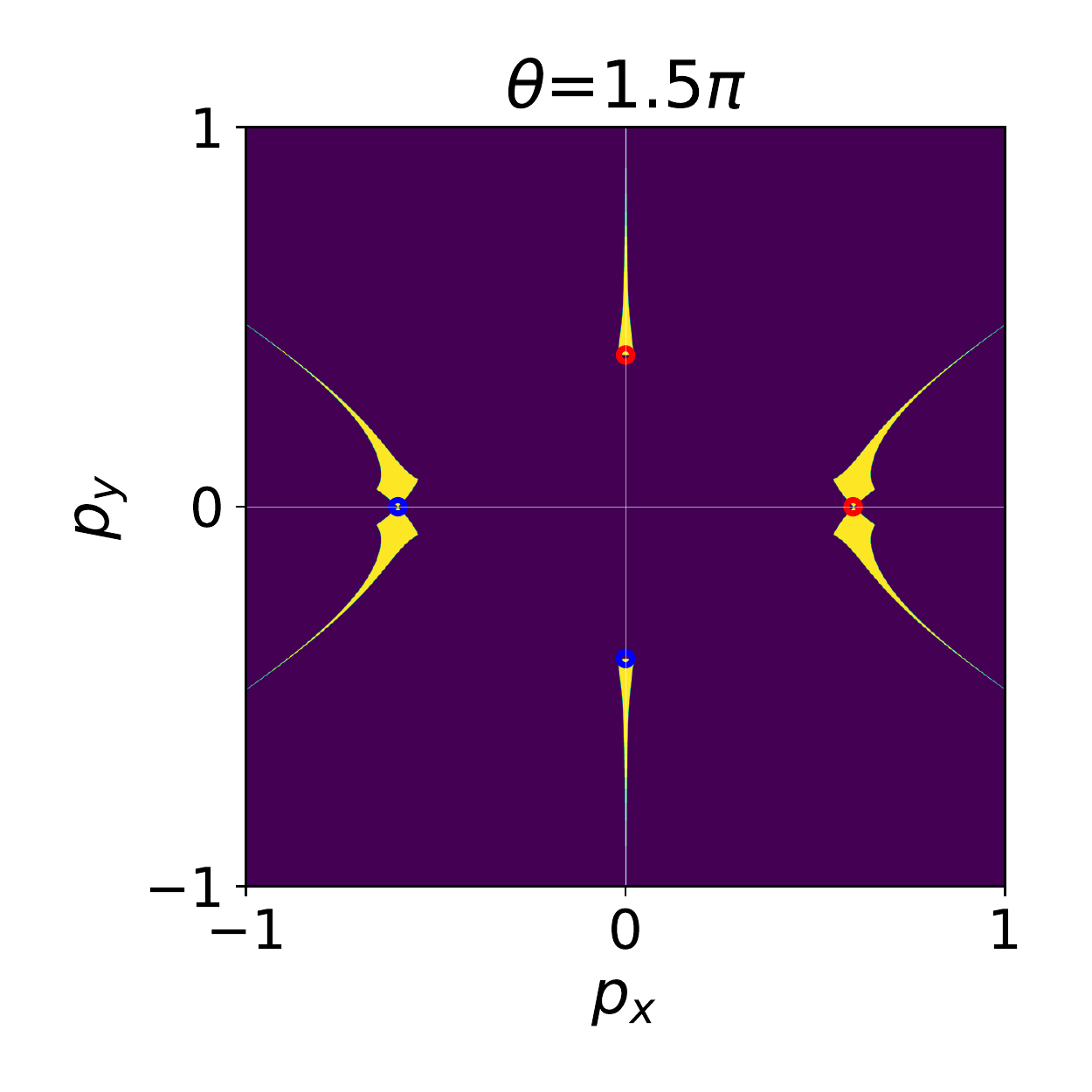}};
 \node (5) at (-1.2,-8.0) {\includegraphics[width=14em]{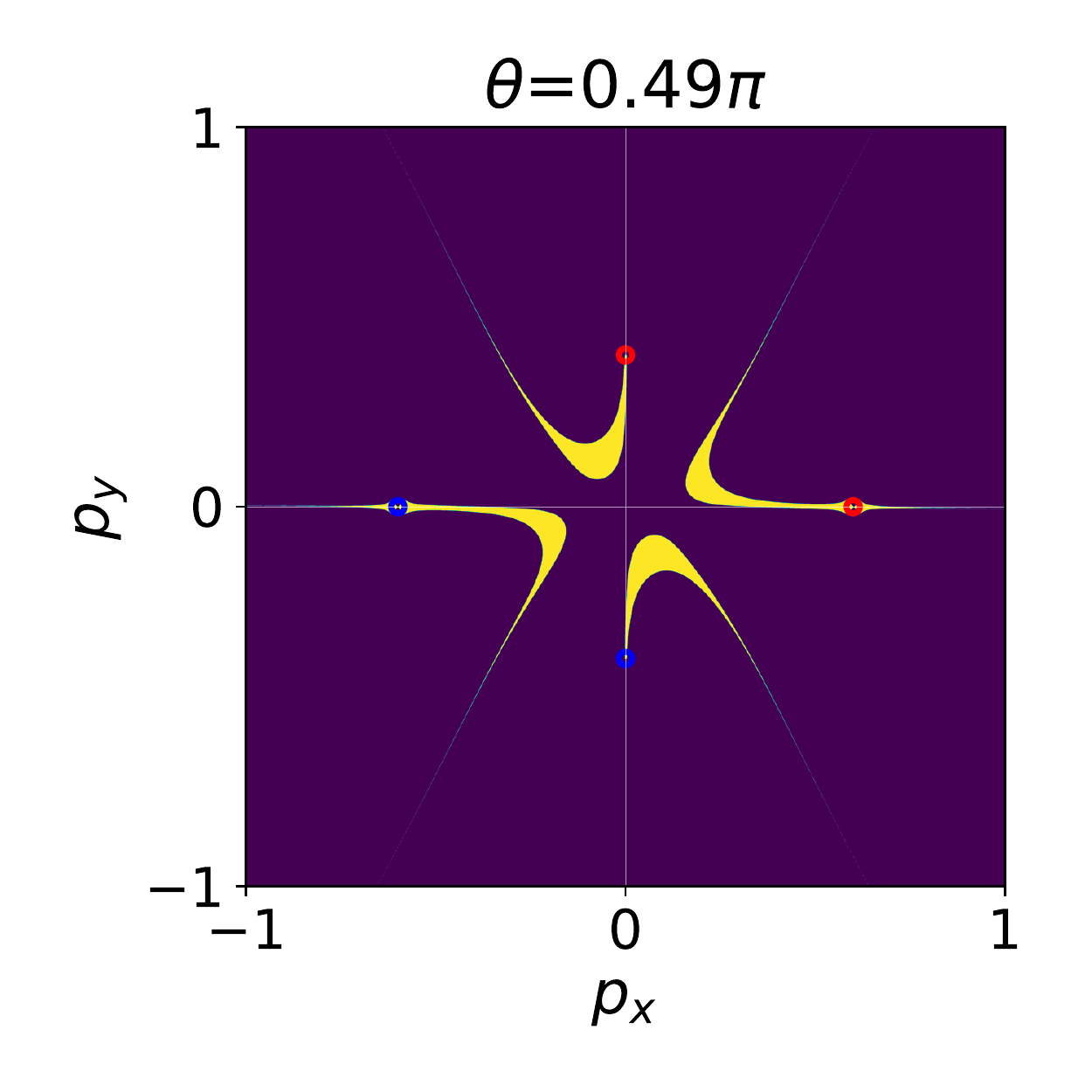}};
 \node (6) at (3.1,-8.0) {\includegraphics[width=14em]{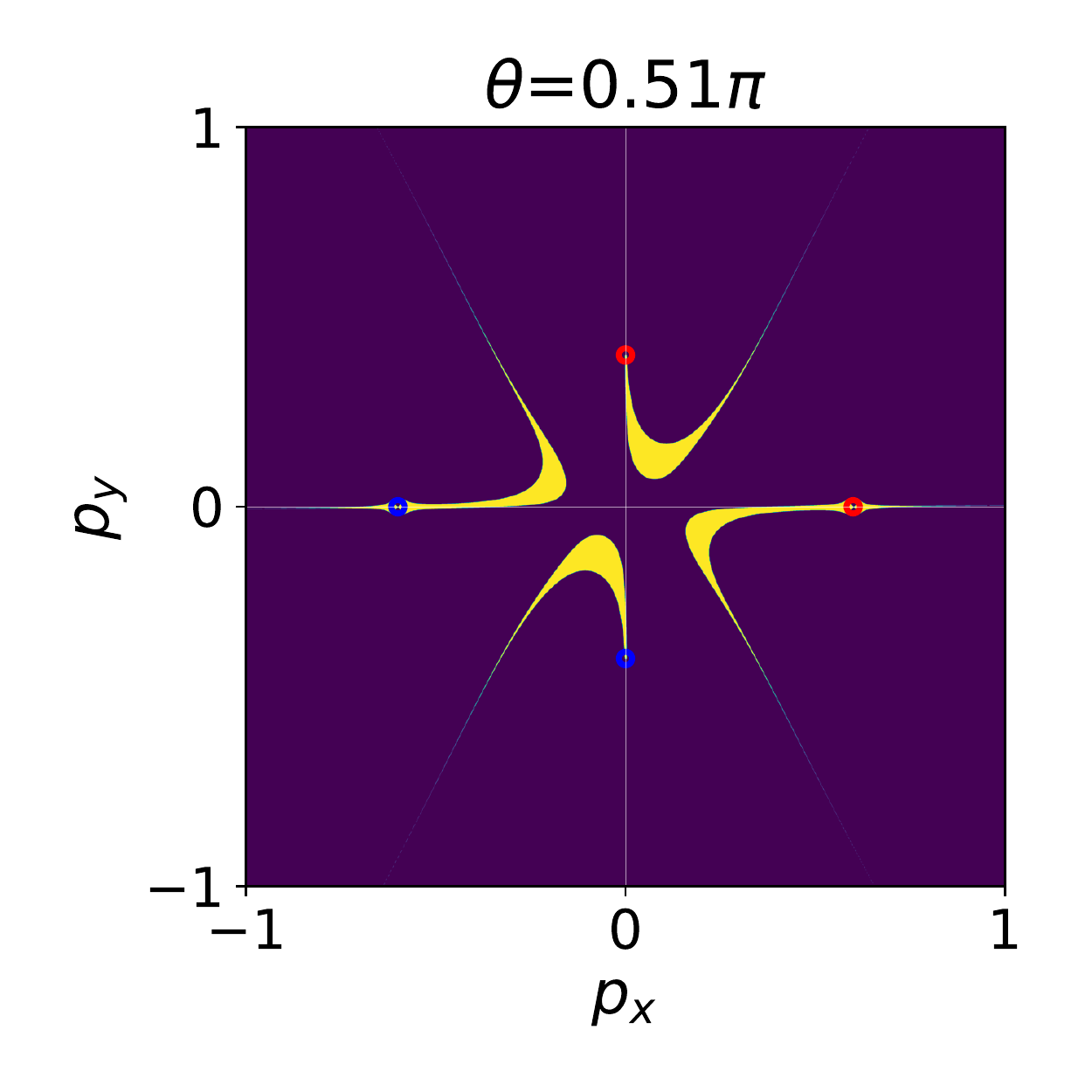}};

 \node at (-2.0,2.0) {(a)};
 \node at (2.2,2.0) {(b)};
 \node at (-2.0,-2.0) {(c)};
 \node at (2.2,-2.0) {(d)};
 \node at (-2.0,-6.0) {(e)};
 \node at (2.2,-6.0) {(f)};
 \end{tikzpicture}
 \end{center}
 \caption{The Fermi arcs of two pairs of Weyl points $Q=(2,-2,1,-1)$ (sitting at $(\pm 0.6,0)$ and $(0,\pm 0.4)$ indicated by small red or blue circles) for the boundary condition parameter $\theta= 0,0.5\pi,1.0\pi,1.5\pi,0.49\pi,0.51\pi$ are shown successively.}
 \label{fig:2211}
\end{figure}
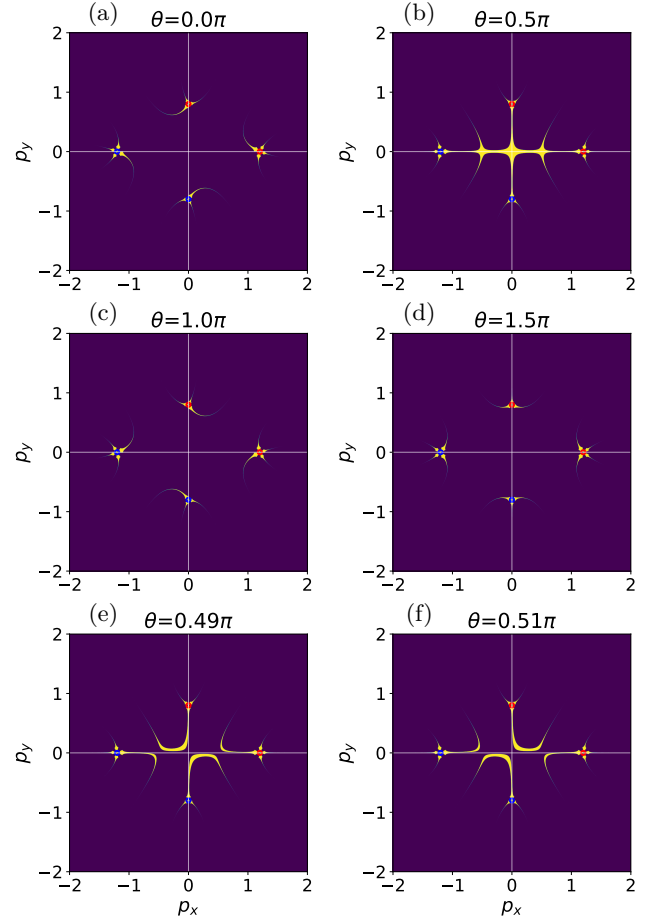
\begin{figure}[htb]
 \begin{center}
 \begin{tikzpicture}
 \node (1) at (-1.2,0) {\includegraphics[width=14em]{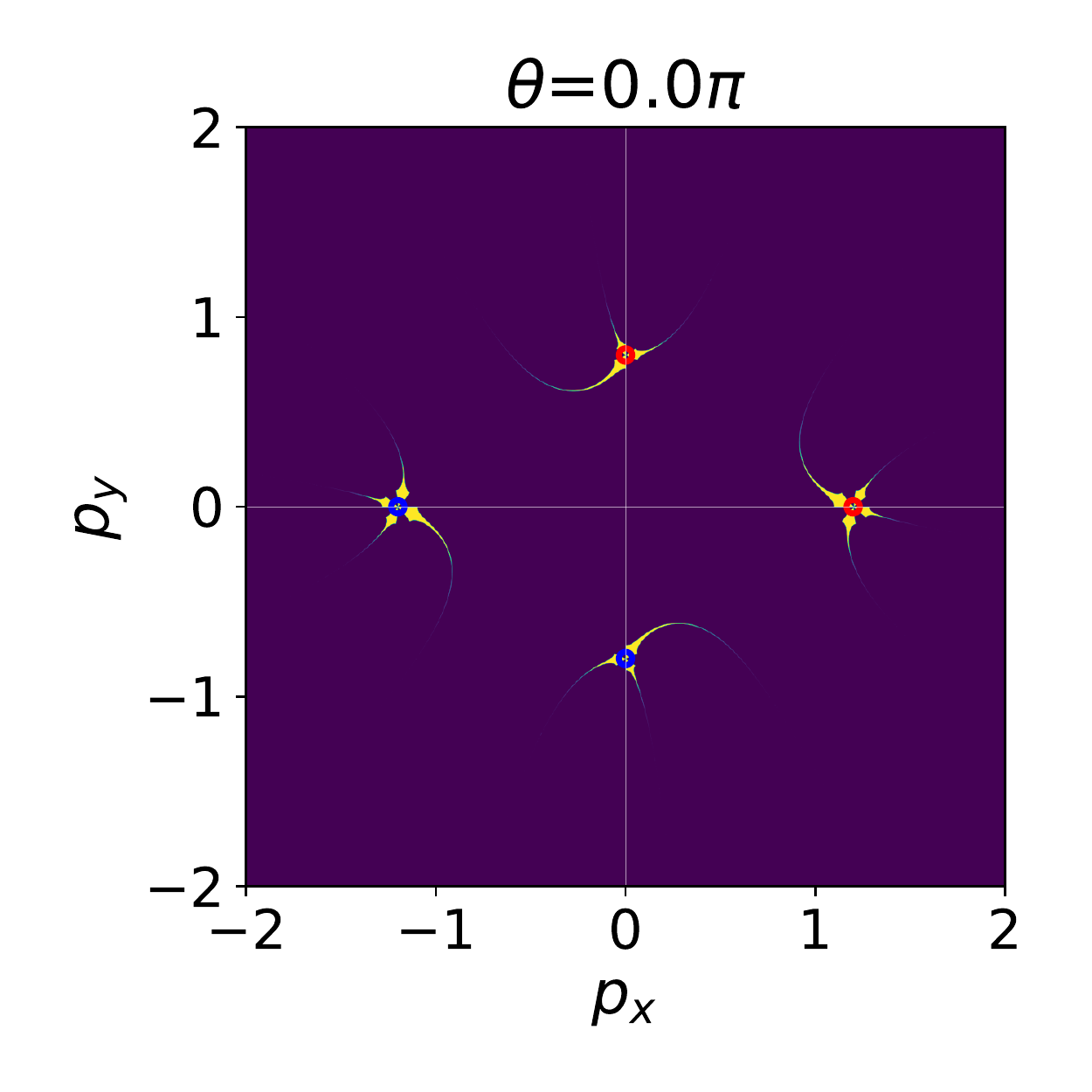}};
 \node (2) at (3.1,0) {\includegraphics[width=14em]{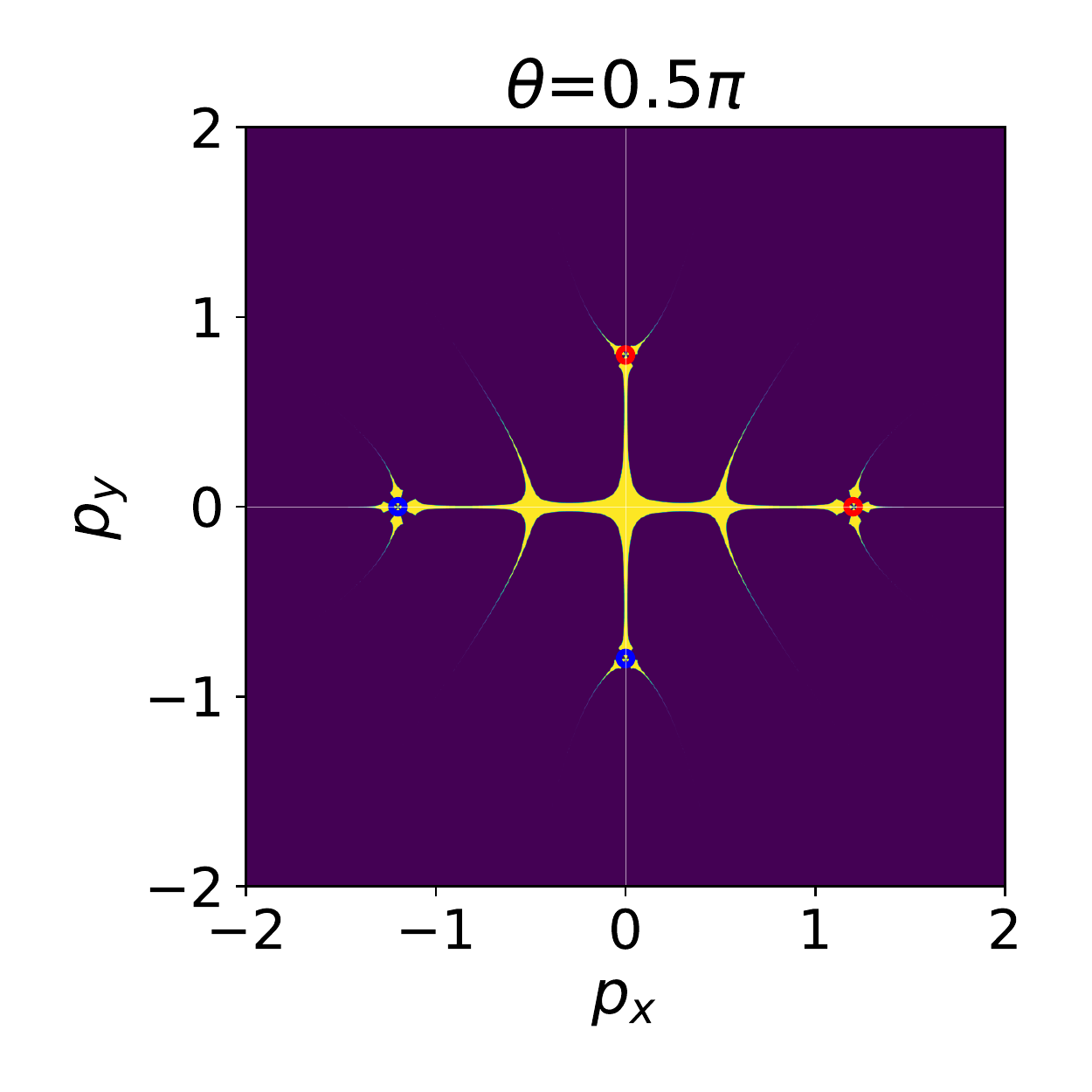}};
 \node (3) at (-1.2,-4.0) {\includegraphics[width=14em]{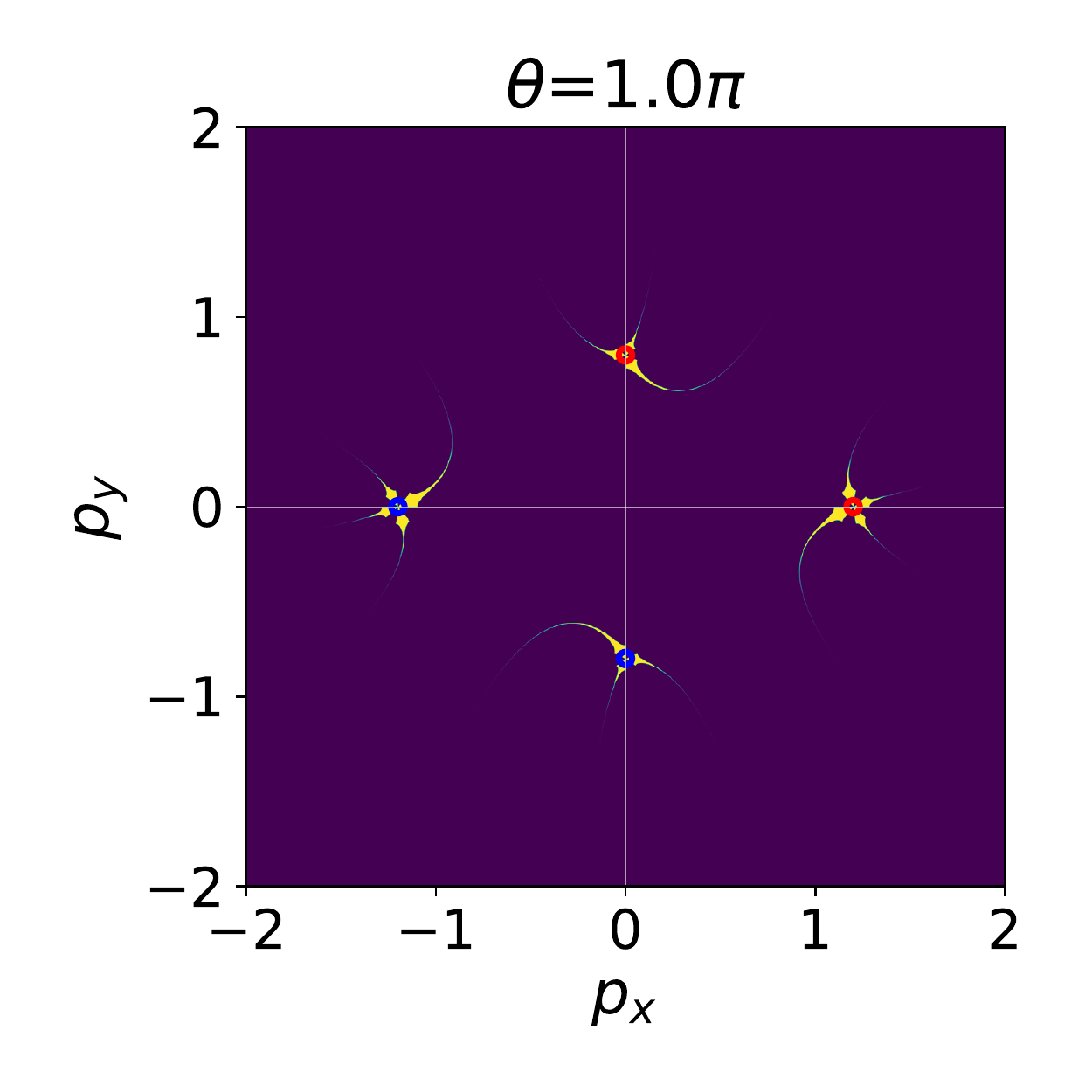}};
 \node (4) at (3.1,-4.0) {\includegraphics[width=14em]{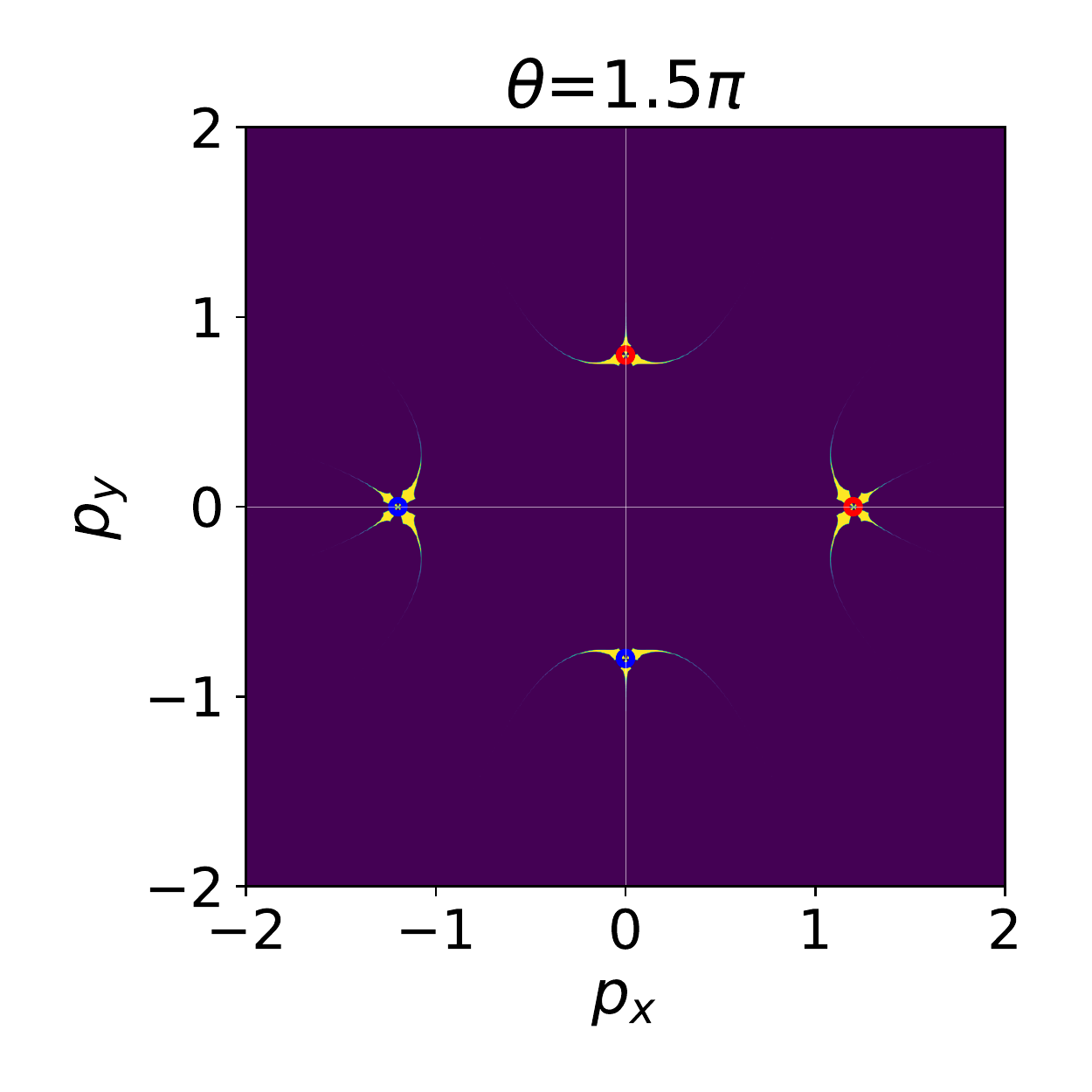}};
 \node (5) at (-1.2,-8.0) {\includegraphics[width=14em]{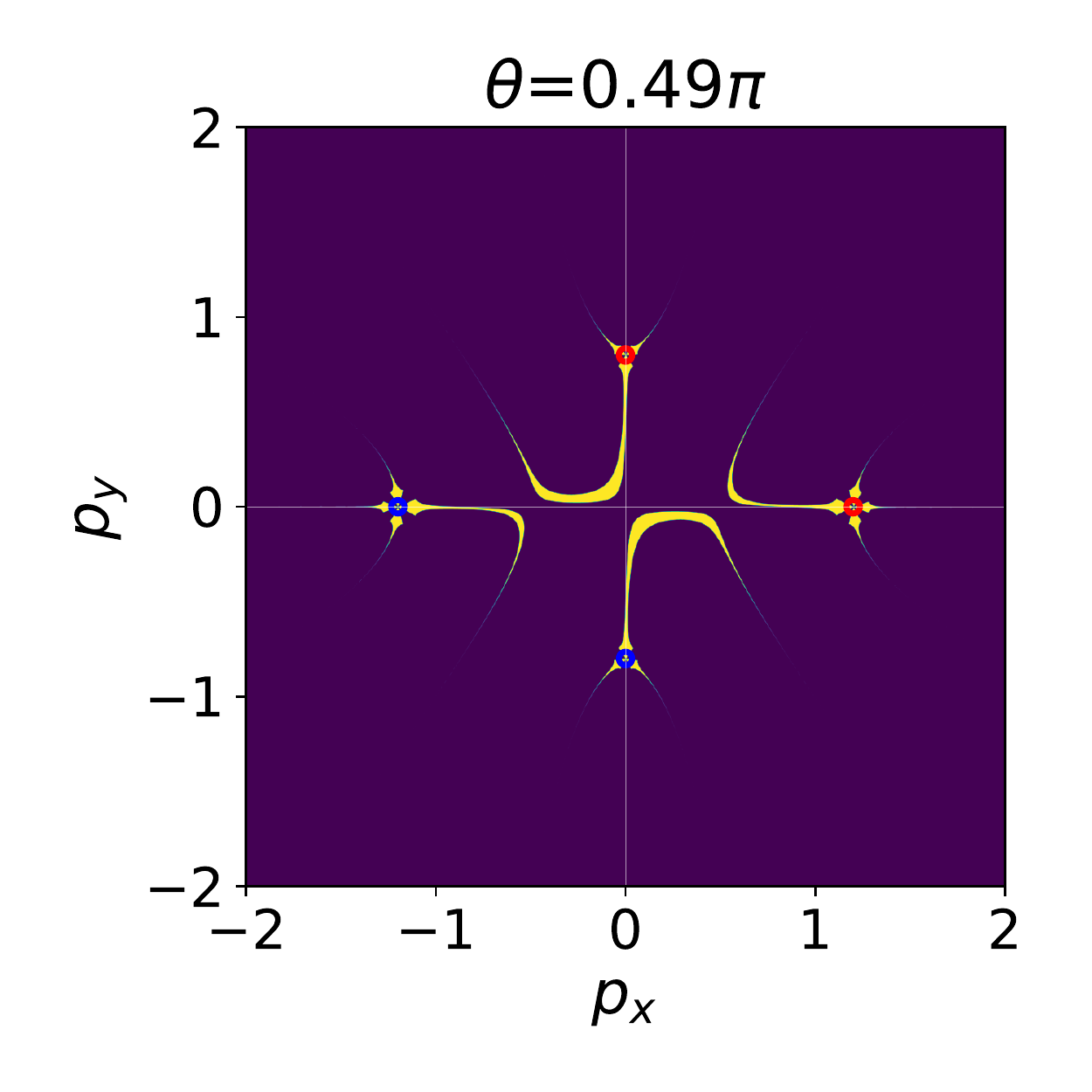}};
 \node (6) at (3.1,-8.0) {\includegraphics[width=14em]{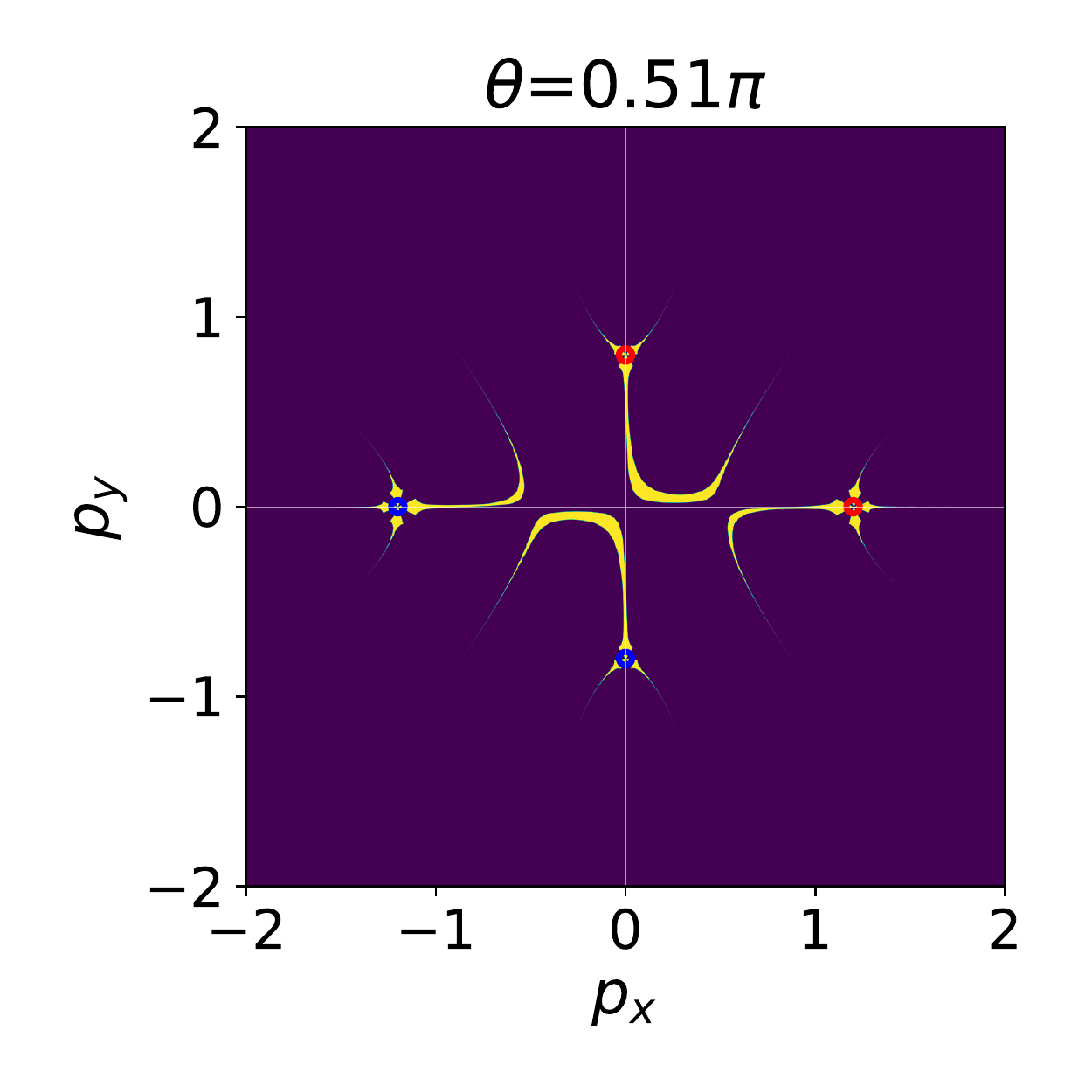}};
 
 \node at (-2.0,2.0) {(a)};
 \node at (2.2,2.0) {(b)};
 \node at (-2.0,-2.0) {(c)};
 \node at (2.2,-2.0) {(d)};
 \node at (-2.0,-6.0) {(e)};
 \node at (2.2,-6.0) {(f)};
 \end{tikzpicture}
 \end{center}
 \caption{The Fermi arcs of two pairs of Weyl points $Q=(4,-4,3,-3)$ (sitting at $(\pm 1.2,0)$ and $(0,\pm 0.8)$ indicated by red or blue small circles) for the boundary condition parameter $\theta= 0,0.5\pi,1.0\pi,1.5\pi,0.49\pi,0.51\pi$ are shown successively.}
 \label{fig:4433}
\end{figure}

\section{Discussion and Summary}

In this paper, we have systematically investigated the Fermi arcs of topological surface states in the 3D multi-Weyl semimetals by a continuum model. The generic boundary condition for linear-Weyl semimetals dictated by a single real parameter $\theta$ within $0<\theta\leq 2\pi$ can also be applied to multi-Weyl semimetals. The Lifshitz phase transition of Fermi arcs relative to boundary condition parameter $\theta$ has demonstrated distinctly. 

First, the generic boundary condition for Weyl semimetals with single flat surface boundary has been re-derived by both Hamiltonian and Lagrangian formulations compactly. We have also derived the the boundary condition for the double parallel identical flat surfaces boundary and find that the two boundary condition should with the same boundary operator $M$ but with different eigenvalues. Then we have analytically calculated the wave functions and energy spectra for the bulk and surface states in linear-Weyl semimetal. These results for topological surface states can be expressed in a compact complex function formalism especially for the Fermi arc of the topological surface states. The point is that this complex function formalism can be generalized to the multi-Weyl cases much more readily than the vector formalism.

Based on the generalized complex function formalism, we first calculated the Fermi arc of topological surface states analytically in multi-Weyl semimetals. The Fermi arcs for multi-Weyl semimetals are discussed in three cases: single Weyl point, single-pair Weyl points and double-pairs Weyl points. In every case, we have analyzed them in the different situations with the different topological charges (or winding numbers). In single Weyl node case, the Fermi arcs for Weyl node with chirality $w$ are just $w$ rays emitting symmetrically from the Weyl point, and the emission angles are determined both by chirality $w$ and boundary parameter $\theta$. In single-pair Weyl points case, the Fermi arcs for Weyl points with topological charges $Q=(w,-w)$ are generally $w$ half or $w/2$ whole hyperbolas through Weyl points, depending on the parity of the winding number $w$ of Weyl points. However, in special case with $\theta=\pi/2$($\theta=3\pi/2$), the Fermi arcs become rays emitting from the origin of momentum plane with $w=2m$($w=2m-1$). In double-pair Weyl points case, the Fermi arcs for Weyl point with topological charge $Q=(w,-w,w,-w)$ are generally $w$ half or $w/2$ whole quartibolas through Weyl points, depending on the parity of the winding number $w$ of Weyl nodes. In special cases with $\theta=\pi/2$($\theta=3\pi/2$), the Fermi arcs become rays emitting from the origin of momentum plane jumping over or terminating at the Weyl points with $w=2m$($w=2m-1$). Besides, the Fermi arcs and their evolution in the more complicated cases with $Q=(2,-2,1,-1)$ and $Q=(4,-4,3,-3)$ have also been displayed, where the new structure appeared in the Fermi arcs for $\theta=\pi/2$.

It is found that in general case the number of the Fermi arcs emitting from every Weyl point is always equal to its chirality $w$. The extra Fermi arcs without passing through Weyl points seem present in special case when $\theta=\pi/2$ or $\theta=3\pi/2$, but the Fermi arcs structures for $\theta$ near these special points demonstrate explicitly that there is no extra Fermi arc actually. In addition, these Fermi arcs connection at special points also indicate clearly that the Lifshitz phase transition of Fermi arcs occurs indeed at $\theta=\pi/2$ or $\theta=3\pi/2$. In general case for $p_{x}\neq0$, there is no Fermi arc connecting two Weyl points. This may be due the absence of valley degree of freedom in our continuum model. 

In summary, we have obtained analytically the Fermi arcs pattern of topological Fermi surface states in multi-Weyl semimetals and clear demonstrated the topological Lifshitz phase transition of Fermi arcs relative to boundary condition parameter. Our continuum model and analytic solutions provide explicitly the structure and phase transition of the Fermi arcs of topological Fermi surface states in multi-Weyl semimetal. which may inspire some new insights to further investigation of multi-Weyl semimetals. Our future work will focus on the deduction of the boundary condition parameter for the reconstruction of boundary and metal atom decoration on boundary surface.

\appendix

\section{The orthogonal boundary condition for spinor in lattice models}\label{sec:lattice}

The effective model above indicate the significant dependence of the edge state on the boundary condition. In this appendix, the same boundary condition is realized in lattice models with tight-binding Hamiltonian, which is a revised version of Hashimoto et.al.\cite{Hashimoto2017} with important modification.

Following Hashimoto et.al., consider a discrete model defined on a finite one-dimensional lattice labeled by $k = 1, \ldots, N$, which is easy to generalize to three-dimensional case. The self-conjugate operator is $\mathcal{H} = - i \sigma \nabla$ with $\sigma$ a Hermitian matrix to be taken as a Pauli matrix. The difference operator for discrete model is defined by
\begin{align}
 \nabla \psi_k & = \psi_{k+1} - \psi_k
 \, , \\
 \nabla^\dag \psi_k & = \psi_{k-1} - \psi_k
 \, .
\end{align}
This difference operator can be reduced to the differential operator in the continuum limit, then the self-conjugate operator becomes the standard Dirac Hamiltonian $\mathcal{H} \to -i \sigma \partial_x$. Since they are related to each other, $i\nabla^\dag \psi_{k+1} = -i\nabla \psi_k$, this is locally self-conjugate. However, the discrete Dirac Hamiltonian is self-conjugate up to the boundary term
\begin{align}
 \sum_{k=1}^N \psi^\dag_n \left( -i \sigma \nabla \psi_k \right) 
 & =
 \sum_{k=1}^N \left( i \sigma \nabla^\dag \psi_k \right)^\dag \psi_k
 \nonumber \\
 & \quad
 + \psi^\dag_0 (i\sigma) \psi_1
 - \psi^\dag_N (i\sigma) \psi_{N+1}
\end{align}
where $\psi_0$ and $\psi_{N+1}$ are auxiliary fields to describe the effect of neighboring environment at boundary. The second line shows the surface term in this case, and the self-conjugacy of the Hamiltonian requires that this part should vanish:
\begin{align}\label{DBC}
 \psi^\dag_0 (i\sigma) \psi_1
 - \psi^\dag_N (i\sigma) \psi_{N+1} = 0 \, 
\end{align}

For scalar wave function, there are two possibilities to solve this condition. The first situation is the periodic boundary condition demanding that $\psi_k = \psi_{k+N}$ for $\forall k \in \{1, \ldots, N\}$, then these two terms cancel each other. The second situation demands that $\psi_0=\psi_{N+1}=0$, which corresponds to open boundary condition usually used in topological insulators.
For spinor wave function, however, there exists the third situation: 
\begin{align}\label{PBC3}
\psi_0\perp\sigma\psi_{1},\; 
\sigma\psi_N\perp\psi_{N+1},
\end{align}
which may be called as the orthogonal boundary condition. This is the key modification of our deduction. In this situation, the two terms in \eqref{DBC} vanish independently. 

Let us show that this orthogonal boundary condition is equivalent to \eqref{BC} considered in continuum theory if we assume $\sigma=\sigma_3$. 
$\psi_0\perp\psi_1$ means that they are orthogonal wave functions of a certain operator. Without loss any generality for Weyl spinor, we can assume that $\psi_0$ is the eigenfunction of operator $M$ with eigenvalue $+1$, i.e., 
\begin{align}
M\psi_0=+1\psi_0.
\end{align}
To satisfy the boundary condition, $\sigma_3\psi_1$ must be the other eigenfunction of $M$ with eigenvalue $-1$, i.e.,
\begin{align}
M\sigma_3\psi_1=-1\sigma_3\psi_1.
\end{align}
Thus we get 
\begin{align}\label{SP1}
\sigma_3\,M\,\sigma_3\psi_1=-1\psi_1.
\end{align}
On the other hand, since the translation invariance of spinor, $\psi_0$ must be parallel to $\psi_1$, which means that
\begin{align}\label{P1}
M\psi_1=+1\psi_1.
\end{align}
The consistence of \eqref{P1} and \eqref{SP1} demands that
\begin{align}
\sigma_3\,M\,\sigma_3=-M,
\end{align}
which is identical to 
\begin{align}
M\,\sigma_3+\sigma_3\,M=\{M,\sigma_3\}=0.
\end{align}
It is the same condition for $M$ deriving form the hermiticity of the Hamiltonian in continuum model.

\begin{acknowledgments}
Liu Yachao acknowledge the financial support of the National Public Visiting Scholar Program from China Scholarship Council (File No.201908610030) and the hospitality of the First-Principles Simulation Group at the International Center for Materials Nanoarchitectonics of National Institute of Materials Science (NIMS) in Japan. The work of Liu Yachao was also supported by Doctoral research start-up funds of Teacher in Xi'an University of Technology (Grant No.109-451119001) and in part by the Natural Science Research Program of the Science Program of Shaanxi Province (Grant No. 2019JQ-317).

\end{acknowledgments}
\newcommand{\PR}{ Phys.\ Rev.}
\bibliographystyle{aipnum4-1}
\bibliography{Liu-bib}

\end{document}